\documentclass[sn-mathphys]{sn-jnl}

\jyear{2024}%

\usepackage{amsmath,amssymb}
\usepackage{graphicx}
\usepackage{subfloat}
\usepackage{subfig}
\usepackage{enumerate}
\usepackage{url}

\begin{document}

\title{Strong-field QED effects on polarization states in dipole and quadrudipole pulsar emissions}


\author[1,2]{\fnm{Dong-Hoon} \sur{Kim}}\email{ki13130@gmail.com}

\author[3]{\fnm{Chul Min} \sur{Kim}}\email{chulmin@gist.ac.kr}

\author[4,5]{\fnm{Sang Pyo} \sur{Kim}}\email{sangkim@kunsan.ac.kr}


\affil[1]{\orgdiv{The Research Institute of Basic Science}, \orgname{Seoul National University}, \orgaddress{\street{1 Gwanak-ro}, \city{Seoul}, \postcode{08826},  \country{Republic of Korea}}}

\affil[2]{\orgdiv{Department of Physics and Astronomy}, \orgname{Seoul National University}, \orgaddress{\street{1 Gwanak-ro}, \city{Seoul}, \postcode{08826}, \country{Republic of Korea}}}

\affil[3]{\orgdiv{Advanced Photonics Research Institute}, \orgname{Gwangju Institute of Science and Technology}, \orgaddress{\street{123 Cheomdangwagi-ro}, \city{Gwangju}, \postcode{61005}, \country{Republic of Korea}}}

\affil[4]{\orgdiv{Department of Physics}, \orgname{Kunsan National University}, \orgaddress{\street{558 Daehak-ro}, \city{Gunsan}, \postcode{54150}, \country{Republic of Korea}}}

\affil[5]{\orgdiv{Asia Pacific Center for Theoretical Physics}, \orgname{POSTECH}, \orgaddress{\street{77 Cheongam-ro}, \city{Pohang}, \postcode{37673}, \country{Republic of Korea}}}

\abstract{Highly magnetized neutron stars have quantum refraction effects on pulsar emission due to the non-linearity of the quantum electrodynamics (QED) action. 
In this paper, we investigate the evolution of the polarization states of pulsar emission under the quantum refraction effects, combined with the dependence on the emission frequency, 
for dipole and quadrudipole pulsar models; we solve a system of evolution equations of the Stokes vector, where the birefringent vector, in which such effects are encoded, acts on the Stokes vector. 
At a fixed emission frequency, depending on the magnitude of the birefringent vector, dominated mostly by the magnetic field strength, 
the evolution of the Stokes vector largely exhibits three different patterns: (i) monotonic, or (ii) half-oscillatory, or (iii) highly oscillatory behaviors. 
These features are understood and confirmed by means of approximate analytical solutions to the evolution equations. 
Also, the evolution patterns are shown to differ between dipole and quadrudipole pulsar models, depending on the magnetic field strength.}


\maketitle

\section{Introduction}
\label{intro}
Strong fields may open a window for testing fundamental physics. Even before quantum electrodynamics (QED) was fully developed and precisely tested in the weak field regime, 
Heisenberg and Euler showed that a strong electromagnetic field can polarize the Dirac vacuum \cite{Heisenberg1936Folgerungen}. Schwinger introduced the proper-time integral 
method to obtain the one-loop effective QED action of the vacuum under a uniform electromagnetic field \cite{Schwinger1951Gauge}. The so-called Heisenberg-Euler-Schwinger (HES) 
action provides an effective theory of electrodynamics in strong fields, in which the linear Maxwell vacuum is turned into a dielectric medium with electric, magnetic, and magneto-electric 
responses. Consequently, a photon propagating in a region of strong electromagnetic fields can experience vacuum birefringence, i.e., a quantum refraction effect \cite{Meszaros1992High,Kim2023Vacuum}. 
Furthermore, when the electric field is sufficiently strong to be comparable to the critical electric field strength ($E_{\rm{c}}=m_{\mathrm{e}}^{2}c^{3}/ \left(e\hbar \right) \approx 1.3\times10^{16}\, \rm{V/cm}$), 
electron-positron pairs can be created spontaneously out of the vacuum, which is called Sauter-Schwinger pair production \cite{Ruffini_2010,Fedotov:2022ely,Hattori:2023egw}. 
Observing these effects will validate the quantum vacuum model in the strong-field regime; however, vacuum birefringence is practically far more likely to be implemented or observed 
than Sauter-Schwinger pair production.  

Although the relevant field strength is too high to attain by terrestrial means, an experiment and several proposals for testing strong-field QED have been reported. 
In the PVLAS (Polarizzazione del Vuoto con Laser, i.e., polarization of vacuum with laser) project, permanent, superconducting magnets have been used with laser as a probe, 
and a limit on vacuum birefringence has been reported for a field strength of $2.5\times10^{4}\, \mathrm{G}$ \cite{Ejlli2020PVLASa}. Recently, several proposals have appeared, 
in which ultra-intense laser fields are used with X-rays as a probe \cite{Karbstein2021Vacuum, Shen2018Exploring, Yu2023X}; the current ultra-intense laser can provide 
a magnetic field strength of $10^{10}\, \rm{G}$ \cite{Danson2019Petawatt}, which corresponds to millisecond pulsars, but not to young pulsars. 
Although the fields from such lasers are still weaker than the critical magnetic field strength, 
$B_\mathrm{c} = m_\mathrm{e}^2 c^3/e\hbar \approx 4.414\times 10^{13} \,\mathrm{G}$ by three orders, these proposals are promising for the observation of vacuum birefringence 
in the relatively weak-field regime. Thus, they are seriously considered to be conducted at upcoming ultra-intense laser facilities \cite{Danson2019Petawatt}.

However, the observation of vacuum birefringence in the strong-field regime requires a field strength comparable to the critical value. It has been predicted that such extreme fields 
are available from astrophysical compact objects. For instance, highly magnetized neutron stars have magnetospheres whose field strength approaches up to $\sim 2\times10^{15}\, \rm{G}$ 
(about 50 times as high as the critical field strength) \cite{Meszaros1992High,Harding2006Physics}. In this regard, several space telescope missions are being conducted or proposed 
to observe the X-rays from neutron stars for vacuum birefringence: the Imaging X-ray Polarimetry Explorer (IXPE) \cite{IXPE}, X-ray Polarimeter Satellite (XPoSat) \cite{XPoSat},
the enhanced X-ray Timing and Polarimetry (eXTP) \cite{Santangelo_2019} and the Compton Telescope project \cite{Wadiasingh2019Magnetars}. 
The X-rays from a neutron star contain information about vacuum birefringence in its magnetosphere, and the birefringence effect accumulates over the magnetospheric size. 
Such accumulation is a great advantage compared to terrestrial laser experiments, not to mention the available field strength. This way, astrophysical compact objects can be used 
as a laboratory to test fundamental physics in the strong-field regime \cite{Raffelt1996Stars,Meszaros1992High,Kim2024Magnetars}.

The HES action is well approximated by the post-Maxwellian action, even up to the strength one order lower than the critical magnetic field $B_\mathrm{c}$, 
which keeps up to the quadratic terms of the Maxwell scalar and pseudo-scalar. Therefore, the post-Maxwellian action exhibits non-linear characteristics of vacuum polarization, 
such as quantum refraction \cite{Adler1971Photon,Ni2013Foundations, Denisov2016Pulsar}. Previously, we have studied the quantum refraction effects 
on the propagation of a probe photon in the magnetic dipole field background of a pulsar model \cite{10.1093/mnras/stae1304}. The study is non-trivial 
in comparison with other similar studies wherein the background magnetic field is assumed to be uniform, in that we have to deal with a dipole magnetic field, 
the strength and direction of which vary over space.

In this work, we investigate the evolution of the polarization states of pulsar emission under the quantum refraction effects, combined with the dependence on the emission frequency, 
for dipole and quadrudipole (for the first time, to our knowledge) pulsar models; with growing theoretical and observational concerns for beyond-dipole effects, this study extends 
the scope of previous works by taking into account the multipolar magnetic field structure, the importance of which has been illuminated in different contexts of pulsar astronomy 
by a number of studies (see \cite{Gralla_2016,10.1093/mnras/stz2524,Kazmierczak2019,Kalapotharakos_2021}, \cite{Petri2016Theory} and references therein). 
To this end, we employ the evolution equations of the Stokes vector, where such effects are encoded into the birefringent vector that acts on the Stokes vector. 
The Stokes vector has a crucial advantage over the polarization vector in representing polarization states: it can be directly determined from experimentally measurable quantities 
and accommodate depolarization effects due to incomplete coherence and random processes during the photon propagation. Solutions of the evolution equations describe 
how the polarization states change along the photon propagation path from the emission point towards an observer. It turns out that the evolution of the Stokes vector, 
at a fixed frequency of emission, largely exhibits three different patterns, depending on the magnitudes of the birefringent vector, dominated mostly by the magnetic field strength: 
(i) fractionally oscillatory - monotonic, or (ii) half-oscillatory, or (iii) highly oscillatory behaviors, which are found by numerical solutions and also confirmed by approximate analytical solutions. 
These are novel features rarely illuminated in previous studies on the same topic. In addition, it is investigated how the aforementioned features regarding the evolution of the Stokes vector 
change as we replace a dipole field with a quadrudipole field to modify the pulsar magnetic field structure. Throughout our analysis, X-ray emission from pulsars, with frequency 
$\sim 10^{18}\,\mathrm{Hz}$, is considered; in this regime, the vacuum contribution to the birefringence dominates that of the plasma \cite{Heyl2000,Wang2007Wave}. 
Also, our analysis is sufficiently rigorous in solving the evolution equations of the Stokes vector, in that we feed into the equations the precise information of photon propagation 
under the pulsar rotation effect, through the magnetic field geometries of oblique dipole and quadrudipole rotators, with all the quantities involved fully affine-parameterized; 
then, the equations are solved solely in terms of an affine parameter. 

The paper is organized as follows. In Sect. \ref{eveq}, we introduce a system of evolution equations of the Stokes vector and apply this formalism to our pulsar emission model 
for an oblique dipole rotator. In Sect. \ref{soleveq}, the evolution equations are solved for some known rotation-powered pulsars (RPPs) in three ways: fully numerically, 
via perturbation analysis, and using an analytical approximation. Also, we discuss the evolution patterns of the Stokes vectors resulting from the solutions. 
In Sect. \ref{eveq2}, we consider a magnetic quadrudipole model for pulsar emission and look into the evolution equations under this model. 
In Sect. \ref{soleveq2}, we solve the evolution equations for the same RPPs fully numerically, and compare the results with those for the dipole case in Sect. \ref{soleveq}.
Then finally, we conclude the paper with discussions on other similar studies and future follow-up studies.

\section{Evolution of polarization states in strong magnetic field -- dipole pulsars}
\label{evol}

\subsection{Evolution equations of Stokes vector}
\label{eveq}

\begin{figure*}
\centering
\includegraphics[width=\textwidth]{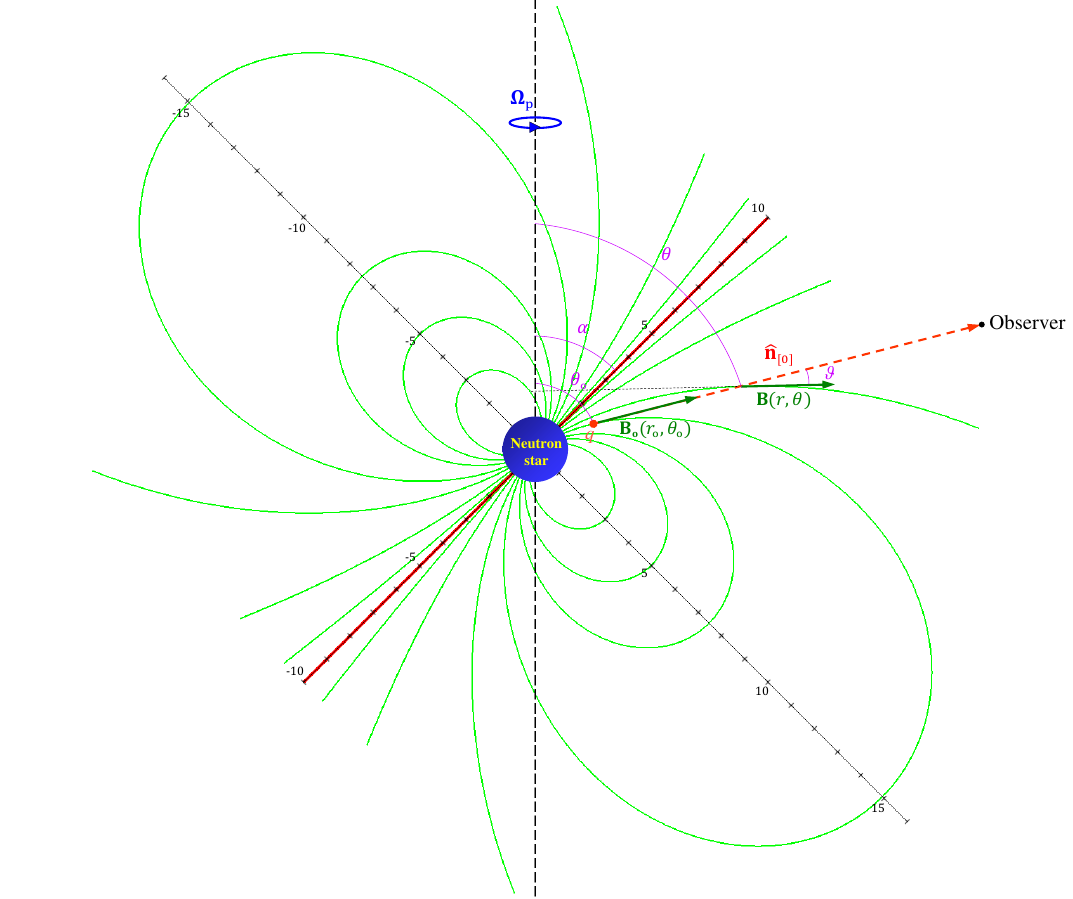}
\caption{A cross-sectional view of a pulsar magnetosphere with the dipole
magnetic field lines (green) around a neutron star. The vertical dashed line (black)
and the inclined solid line (red) represent the rotation axis and the
magnetic axis, respectively. $\alpha$ between these axes denotes
the inclination angle. The scale of the unity in this graph is
equivalent to the neutron star radius $\sim 10^{6}\,\mathrm{cm}$. The red
dashed line represents the trajectory curve of the light ray traced by the propagation vector $\mathbf{%
\hat{n}}_{\left[ 0\right] }$ as projected onto the $xz$-plane. (Credit: 
\cite{Kim2021Generala}, reproduced with modifications.) }
\label{fig1}
\end{figure*}

Classically, polarization properties of pulsar emission are described by the Stokes parameters 
$\left\{I,Q,U,V \right\}$, where $I$ is a measure of the total intensity, $Q$ and $U$ 
jointly describe the linear polarization, and $V$ describes the circular polarization of 
pulsar emission (for more details, see Appendix \ref{appA}). 
However, in the presence of a strong magnetic field in the background of the emission, 
the polarization evolves along the photon propagation path from the emission point towards an observer. 
The evolution of the polarization can be investigated systematically using the formalism
initiated by \cite{Kubo1981,Kubo1983,Kubo1985}, and further developed 
by \cite{Heyl2000,Heyl2003highenergy,Heyl2018Strongly,Novak2018}, namely, a system of evolution equations 
of the Stokes vector, described as
\begin{equation}
\frac{\mathrm{d}\mathbf{S}}{\mathrm{d}s}=k\mathbf{\hat{\Omega}}\times 
\mathbf{S},  \label{ev}
\end{equation}%
where $k\equiv \omega /c$ denotes the wave number for the electromagnetic
radiation and $s$ is an affine parameter to measure the length of the photon
trajectory, and $\mathbf{S}$ is the normalized Stokes vector, defined out of 
the Stokes parameters as 
$\mathbf{S}=\left( S_{1},S_{2},S_{3}\right) \allowbreak \equiv \left(Q/I,U/I,V/I\right)$,\footnote{The classical Stokes vector 
can be expressed via pulse profiles of pulsar curvature emission, as illustrated in Appendix \ref{appA}.} 
and $\mathbf{\hat{\Omega}}$ is the \textit{dimensionless} birefringent vector, defined as\footnote{Note that our $k\mathbf{\hat{\Omega}}$ 
is equivalent to the birefringent vector as defined in the references above.}  
\begin{equation}
\mathbf{\hat{\Omega}}\equiv \frac{\alpha_{\mathrm{e}}}{30\pi }\left( B/B_{\mathrm{c}%
}\right) ^{2}\sin ^{2}\vartheta \left( \mathcal{E}_{\mathrm{I}}^{2}-\mathcal{%
E}_{\mathrm{II}}^{2},2\mathcal{E}_{\mathrm{I}}\mathcal{E}_{\mathrm{II}%
},0\right) , \label{br}
\end{equation}%
where $\alpha_{\mathrm{e}}$ denotes the fine-structure constant and 
$\alpha_{\mathrm{e}} /\left( 30\pi \right) \approx 7.743\times
10^{-5}$ and $B_{\mathrm{c}}\approx 4.414\times
10^{13}\,\mathrm{G}$ is the critical magnetic field, and $\vartheta $ denotes 
the angle between the photon trajectory and the local magnetic field line (see Fig. \ref{fig1}), i.e.,%
\begin{equation}
\vartheta =\cos ^{-1}\left( \mathbf{\hat{n}}_{\left[ 0\right] }\cdot \mathbf{%
\hat{B}}\right) ,  \label{th}
\end{equation}%
with $\mathbf{\hat{n}}_{\left[ 0\right] }$ being the classical propagation
vector and $\mathbf{\hat{B}}\equiv \mathbf{B}/\left\vert \mathbf{B}%
\right\vert $, and 
\begin{equation}
\mathcal{E}_{i}\equiv -\mathbf{\hat{B}\cdot }\left( \mathbf{\hat{n}}_{\left[
0\right] }\times \boldsymbol{\varepsilon} _{i\left[
0\right]}\right) ,~~ i=\mathrm{I},\mathrm{II}, \label{Eps}
\end{equation}%
with $\boldsymbol{\varepsilon }_{\mathrm{I}\left[ 0\right] }$ and $\boldsymbol{%
\varepsilon }_{\mathrm{II}\left[ 0\right] }$ being the two classical mode
polarization vectors, orthogonal to each other and to $\mathbf{\hat{n}}_{%
\left[ 0\right] }$; the specific forms of $\mathbf{\hat{n}}_{\left[ 0\right] }$, 
$\boldsymbol{\varepsilon }_{\mathrm{I}\left[ 0\right] }$ and $\boldsymbol{%
\varepsilon }_{\mathrm{II}\left[ 0\right] }$ are later given by Eqs. (\ref{no}), 
(\ref{ei0}) and (\ref{eii0}), respectively, for the magnetic field of an oblique dipole rotator 
as described by Eq. (\ref{B}) and Fig. \ref{fig1}.

In our pulsar emission model, we consider curvature radiation produced along
the magnetic field lines of an oblique dipole rotator as illustrated in Fig. \ref{fig1}: 
\begin{align}
\mathbf{B}\left( r,\theta ,\phi \right) &=\frac{2\mu \left( \cos \alpha \cos
\theta +\sin \alpha \sin \theta \cos \phi \right) }{r^{3}}\mathbf{e}_{\hat{r}%
}+\frac{\mu \left( \cos \alpha \sin \theta -\sin \alpha \cos \theta \cos
\phi \right) }{r^{3}}\mathbf{e}_{\hat{\theta}} \notag \\
& \hspace{12pt} +\frac{\mu \sin \alpha \sin
\phi \,}{r^{3}}\mathbf{e}_{\hat{\phi}},  \label{B}
\end{align}%
where $\mu $ is the magnetic dipole moment and $\alpha$ denotes the
inclination angle between the rotation axis and the magnetic axis.\footnote{Here 
the symbol $\alpha$ must be distinguished from the fine-structure constant $\alpha_{\mathrm{e}}$.}
The photon beam from curvature radiation is tangent to the field line at the emission point $\left( x_{%
\mathrm{o}},y_{\mathrm{o}},z_{\mathrm{o}}\right) =\left( r_{\mathrm{o}}\sin
\theta _{\mathrm{o}},0,r_{\mathrm{o}}\cos \theta _{\mathrm{o}}\right) $. 

However, at the same time, our pulsar magnetosphere rotates, and therefore 
the field lines get twisted due to the magneto-centrifugal acceleration on the 
plasma particles moving along the field lines \cite{Blandford1982}. Then, taking 
into consideration this magneto-hydrodynamic (MHD) effect, the direction of the 
classical photon propagation, which must line up with the particle velocity in order 
for an observer to receive the radiation, can be described as \cite{Gangadhara2005}%
\begin{equation}
\mathbf{\hat{n}}_{\left[ 0\right] }=\beta \mathbf{\hat{B}}+ \frac{\mathbf{\Omega }_{\mathrm{p}}\times \mathbf{r}}{c}, 
\label{mhd}
\end{equation}%
where on the right-hand side 
\begin{equation}
\beta \equiv \left[ 1-\left( \frac{\Omega _{\mathrm{p}}r}{c}\right) ^{2}\sin
^{2}\theta \left( 1-\frac{\sin ^{2}\alpha \sin ^{2}\phi }{3\cos ^{2}\theta
^{\prime }+1}\right) \right] ^{1/2}-\frac{\Omega _{\mathrm{p}}r}{c}\frac{%
\sin \alpha \sin \theta \sin \phi }{\left( 3\cos ^{2}\theta ^{\prime
}+1\right) ^{1/2}},  \label{beta}
\end{equation}%
with $c$ being the speed of light and $\cos \theta ^{\prime }\equiv \cos
\alpha \cos \theta +\sin \alpha \sin \theta \cos \phi $, and the second term
accounts for the centrifugal acceleration, with $\mathbf{\Omega }_{\mathrm{p}%
}\equiv \Omega _{\mathrm{p}}\mathbf{e}_{z}$\footnote{Here the symbol $\mathbf{\Omega }_{\mathrm{p}}$ 
must be distinguished from the birefringent vector $\mathbf{\hat{\Omega}}$.} and $\Omega _{\mathrm{p}}=2\pi/P$ being
a pulsar rotation (angular) frequency, as given in terms of the rotation period $P$.

During the rotation the azimuthal phase changes by $\phi \sim \Omega _{%
\mathrm{p}}t$, while our photon has propagated a distance by $s\sim ct$. In
our analysis, the photon propagation is described with the consideration of
the MHD effect above, assuming $\phi $ to be very small; e.g., $\phi
\lesssim 10^{-1}$ is considered for a millisecond pulsar with $\Omega _{%
\mathrm{p}}\sim 10^{2}\,\mathrm{Hz}$, during the time of rotation $t\lesssim
10^{-3}\,\mathrm{s}$, such that $s\lesssim 10^{7}\,\mathrm{cm}$, which
corresponds to the propagation distance within about $10$ times the neutron
star radius. For Eq. (\ref{mhd}) we take only the leading order expansions
of $\mathbf{\hat{B}}\left( r_{\mathrm{o}},\theta _{\mathrm{o}},\phi \right) $
and $\beta \left( r_{\mathrm{o}},\theta _{\mathrm{o}},\phi \right) $ in $%
\phi $ from Eqs. (\ref{B}) and (\ref{beta}), respectively, and can express 
the classical propagation vector $\mathbf{\hat{n}}_{\left[ 0\right] }$ 
in Cartesian coordinates as
\begin{equation}
\mathbf{\hat{n}}_{\left[ 0\right] }=\hat{n}_{x\left[ 0\right] }\mathbf{e}%
_{x}+\hat{n}_{y\left[ 0\right] }\mathbf{e}_{y}+\hat{n}_{z\left[ 0\right] }%
\mathbf{e}_{z} \label{no} \\
\end{equation}%
with
\begin{align}
\hat{n}_{x\left[ 0\right] }&\approx \frac{2\cos \left( \theta _{\mathrm{o}%
}-\alpha \right) \sin \theta _{\mathrm{o}}+\sin \left( \theta _{\mathrm{o}%
}-\alpha \right) \cos \theta _{\mathrm{o}}}{\left( 3\cos ^{2}\left( \theta _{%
\mathrm{o}}-\alpha \right) +1\right) ^{1/2}}+\mathcal{O}\left( \phi
^{2},\left( \Omega _{\mathrm{p}}r_{\mathrm{o}}/c\right) ^{2},\phi \left(
\Omega _{\mathrm{p}}r_{\mathrm{o}}/c\right) \right) ,  \label{nox} \\
\hat{n}_{z\left[ 0\right] }&\approx \frac{2\cos \left( \theta _{\mathrm{o}%
}-\alpha \right) \cos \theta _{\mathrm{o}}-\sin \left( \theta _{\mathrm{o}%
}-\alpha \right) \sin \theta _{\mathrm{o}}}{\left( 3\cos ^{2}\left( \theta _{%
\mathrm{o}}-\alpha \right) +1\right) ^{1/2}}+\mathcal{O}\left( \phi
^{2},\left( \Omega _{\mathrm{p}}r_{\mathrm{o}}/c\right) ^{2},\phi \left(
\Omega _{\mathrm{p}}r_{\mathrm{o}}/c\right) \right) ,  \label{noz}
\end{align}%
and 
\begin{equation}
\hat{n}_{y\left[ 0\right] }\approx \frac{\Omega _{\mathrm{p}}}{c}\left[ \frac{\sin\alpha \,s}
{\left( 3\cos ^{2}\left( \theta _{\mathrm{o}}-\alpha \right) +1\right)
^{1/2}}+r_{\mathrm{o}}\sin \theta _{\mathrm{o}} \right] +\mathcal{O}\left( \phi ^{2},
\left( \Omega _{\mathrm{p}}r_{\mathrm{o}}/c\right) ^{2},
\phi \left( \Omega _{\mathrm{p}}r_{\mathrm{o}}/c\right) \right) ,  \label{noy}
\end{equation}%
where we have considered $\Omega _{\mathrm{p}}r_{\mathrm{o}}/c\lesssim \phi $%
, e.g., for a millisecond pulsar with $\Omega _{\mathrm{p}}\sim 10^{2}\,%
\mathrm{Hz}$ and $r_{\mathrm{o}}\sim 10^{6}\,\mathrm{cm}$, such that $\left(
\Omega _{\mathrm{p}}r_{\mathrm{o}}/c\right) ^{2}\lesssim \phi \left( \Omega
_{\mathrm{p}}r_{\mathrm{o}}/c\right) \lesssim \phi ^{2}$, all to be ignored
in our analysis, and have substituted $\phi =\Omega _{\mathrm{p}}s/c$ in Eq.
(\ref{noy}), the leading order rotational effect to be considered in our
analysis.

The orthogonal pair of classical mode polarization vectors, $\boldsymbol{\varepsilon }_{\mathrm{I}\left[ 0\right] }
$ and $\boldsymbol{\varepsilon }_{\mathrm{II}\left[ 0\right] }$, both being also
orthogonal to $\mathbf{\hat{n}}_{\left[ 0\right] }$ as given by Eq. (\ref{no}) above, are
determined as%
\begin{align}
\boldsymbol{\varepsilon }_{\mathrm{I}\left[ 0\right] }&=\hat{n}_{z\left[ 0\right]
}\mathbf{e}_{x}+\hat{n}_{y\left[ 0\right] }\mathbf{e}_{y}-\hat{n}_{x\left[ 0%
\right] }\mathbf{e}_{z},  \label{ei0} \\
\boldsymbol{\varepsilon }_{\mathrm{II}\left[ 0\right] }&=-\left( \hat{n}_{x\left[
0\right] }+\hat{n}_{z\left[ 0\right] }\right) \hat{n}_{y\left[ 0\right] }%
\mathbf{e}_{x}+\mathbf{e}_{y}+\left( \hat{n}_{x\left[ 0\right] }-\hat{n}_{z%
\left[ 0\right] }\right) \hat{n}_{y\left[ 0\right] }\mathbf{e}_{z}, \label{eii0}
\end{align}%
such that the three vectors, $\mathbf{\hat{n}}_{\left[ 0\right] }$, $\boldsymbol{%
\varepsilon }_{\mathrm{I}\left[ 0\right] }$ and $\boldsymbol{\varepsilon }_{%
\mathrm{II}\left[ 0\right] }$ form an orthonormal basis.\footnote{%
It can be checked out that $\mathbf{\hat{n}}_{\left[ 0\right] }\cdot \boldsymbol{%
\varepsilon }_{\mathrm{I}\left[ 0\right] }\approx 0+\mathcal{O}\left( \left(
\Omega _{\mathrm{p}}r_{\mathrm{o}}/c\right) ^{2}\right) $, $\mathbf{\hat{n}}%
_{\left[ 0\right] }\cdot \boldsymbol{\varepsilon }_{\mathrm{II}\left[ 0\right]
}=0$ and $\boldsymbol{\varepsilon }_{\,\mathrm{I}\left[ 0\right] }\cdot \boldsymbol{%
\ \varepsilon }_{\mathrm{II}\left[ 0\right] }=0$\ while $\mathbf{\hat{n}}_{%
\left[ 0\right] }^{2}\approx 1+\mathcal{O}\left( \left( \Omega _{\mathrm{p}%
}r_{\mathrm{o}}/c\right) ^{2}\right) $ and $\boldsymbol{\ \varepsilon }_{\mathrm{I%
},\mathrm{II}\left[ 0\right] }^{2}\approx 1+\mathcal{O}\left( \left( \Omega
_{\mathrm{p}}r_{\mathrm{o}}/c\right) ^{2}\right) $.} Using these for Eq.
(\ref{Eps}), we obtain 
\begin{align}
\mathcal{E}_{\mathrm{I}}&=-\mathbf{\hat{B}\cdot }\left( \mathbf{\hat{n}}_{%
\left[ 0\right] }\times \boldsymbol{\varepsilon }_{\mathrm{I}\left[ 0\right]
}\right)   \notag \\
&\approx \frac{4\cos \left( \theta _{\mathrm{o}}-\alpha \right) \cos \left(
\theta -\alpha \right) +\sin \left( \theta _{\mathrm{o}}-\alpha \right) \sin
\left( \theta -\alpha \right) +2\sin \left( \theta -\theta _{\mathrm{o}}\right) }
{\left( 3\cos ^{2}\left( \theta _{\mathrm{o}}-\alpha \right)
+1\right) ^{1/2}\left( 3\cos ^{2}\left( \theta -\alpha \right) +1\right)
^{1/2}}\hat{n}_{y\left[ 0\right] } \notag \\
&\hspace{12pt} -\frac{\Omega _{\mathrm{p}}\sin\alpha \,s}{c\left( 3\cos ^{2}\left( \theta
-\alpha \right) +1\right) ^{1/2}} +\mathcal{O}\left(\phi ^{2},\left( \Omega _{\mathrm{p}}r_{\mathrm{o}}/c\right) ^{2},\phi
\left( \Omega _{\mathrm{p}}r_{\mathrm{o}}/c\right) \right) ,  \label{EI} \\
\mathcal{E}_{\mathrm{II}}&=-\mathbf{\hat{B}\cdot }\left( \mathbf{\hat{n}}_{%
\left[ 0\right] }\times \boldsymbol{\varepsilon }_{\mathrm{II}\left[ 0\right]
}\right)   \notag \\
&\approx -\frac{2\sin \left( \theta -\theta _{\mathrm{o}}\right) }{\left(
3\cos ^{2}\left( \theta _{\mathrm{o}}-\alpha \right) +1\right) ^{1/2}\left(
3\cos ^{2}\left( \theta -\alpha \right) +1\right) ^{1/2}} \notag \\
&\hspace{12pt}+\mathcal{O}\left( \phi ^{2},\left( \Omega
_{\mathrm{p}}r_{\mathrm{o}}/c\right) ^{2},
\phi \left( \Omega _{\mathrm{p}}r_{\mathrm{o}}/c\right) \right) .  \label{EII}
\end{align}%

By means of Eqs. (\ref{th}), (\ref{B}) and (\ref{no}) one can express 
\begin{align}
\cos \vartheta &\approx \frac{4\cos \left( \theta _{\mathrm{o}}-\alpha
\right) \cos \left( \theta -\alpha \right) +\sin \left( \theta _{\mathrm{o}%
}-\alpha \right) \sin \left( \theta -\alpha \right) }{\left( 3\cos
^{2}\left( \theta _{\mathrm{o}}-\alpha \right) +1\right) ^{1/2}\left( 3\cos
^{2}\left( \theta -\alpha \right) +1\right) ^{1/2}} \notag \\
&\hspace{12pt}+\mathcal{O}\left( \phi
^{2},\left( \Omega _{\mathrm{p}}r_{\mathrm{o}}/c\right) ^{2},\phi \left(
\Omega _{\mathrm{p}}r_{\mathrm{o}}/c\right) \right) ,  \label{cos} \\
\sin \vartheta &\approx \frac{2\sin \left( \theta -\theta _{\mathrm{o}%
}\right) }{\left( 3\cos ^{2}\left( \theta _{\mathrm{o}}-\alpha \right)
+1\right) ^{1/2}\left( 3\cos ^{2}\left( \theta -\alpha \right) +1\right)
^{1/2}} \notag \\
&\hspace{12pt}+\mathcal{O}\left( \phi ^{2},\left( \Omega _{\mathrm{p}}r_{\mathrm{o}%
}/c\right) ^{2},\phi \left( \Omega _{\mathrm{p}}r_{\mathrm{o}}/c\right)
\right) .  \label{sin}
\end{align}%
Now, using the relations between Eqs. (\ref{EI})-(\ref{sin}),
the birefringent vector can finally be specified from Eq. (\ref{br}): 
\begin{align}
\hat{\Omega}_{1} &\approx -\eta B^{2} \sin ^{4}\vartheta +\mathcal{O}\left( \phi ^{2},\left( \Omega _{%
\mathrm{p}}r_{\mathrm{o}}/c\right) ^{2},\phi \left( \Omega _{\mathrm{p}}r_{%
\mathrm{o}}/c\right) \right) ,  \label{br1} \\
\hat{\Omega}_{2} &\approx -2\eta B^{2} \sin ^{3}\vartheta \left[ \left( \cos \vartheta +\sin \vartheta
\right) \hat{n}_{y\left[ 0\right] }-\frac{\Omega _{\mathrm{p}}\sin\alpha \,s}{%
c\left( 3\cos ^{2}\left( \theta -\alpha \right) +1\right) ^{1/2}}\right] \notag \\
&\hspace{12pt}+\mathcal{O}\left( \phi ^{2},\left( \Omega _{\mathrm{p}}r_{\mathrm{o}%
}/c\right) ^{2},\phi \left( \Omega _{\mathrm{p}}r_{\mathrm{o}}/c\right)
\right) ,  \label{br2} 
\end{align}%
where $\eta \equiv \alpha_\mathrm{e} /\left( 30\pi B_{\mathrm{c}}^{2}\right) $\footnote{
$\eta = \eta_{2}-\eta_{1}$, where $\eta_{1}$ and $\eta_{2}$ are parameters defined via 
$\eta_{1}/4=\eta_{2}/7=\alpha_\mathrm{e} /\left( 90\pi B_{\mathrm{c}}^{2}\right) \sim 10^{-31}\,\mathrm{g}^{-1}\,\mathrm{cm}\,\mathrm{s}^{2}$,
from the post-Maxwellian Lagrangian $\mathcal{L}_\mathrm{PM}=-\left( \mathbf{B}^{2}-\mathbf{E}^{2} \right)/2 
+\eta_{1}\left( \mathbf{B}^{2}-\mathbf{E}^{2} \right)^{2}/4 + \eta_{2}\left( \mathbf{E}\cdot\mathbf{B} \right)^{2}$ \cite{Euler1935Ueber}.} and
\begin{equation}
B=\frac{B_{\mathrm{max}}r_{*}^{3}\ \left( 3\cos ^{2}\left( \theta
-\alpha \right) +1\right) ^{1/2}}{2\left( x^{2}+z^{2}\right) ^{3/2}},
\label{B1}
\end{equation}%
with $B_{\mathrm{max}}$ being the maximum magnetic field intensity at the polar cap\footnote{From Eq. (\ref{B}) 
$B_{\mathrm{max}}=\left\vert B\left(r=r_{*},\theta=\alpha\right) \right\vert$.} 
and $r_{*}$ being the neutron star radius ($\approx10^{6}\,\mathrm{cm}$), 
and $\hat{n}_{y\left[0\right] }$, $\cos \vartheta $ and $\sin \vartheta $ are 
given by Eqs. (\ref{noy}), (\ref{cos}) and (\ref{sin}), respectively. 

To facilitate solving the evolution equation (\ref{ev}) in the next subsection, 
we substitute the following identities,
\begin{align}
&\cos \left( \theta -\alpha \right) =\frac{\sin\alpha \,x+\cos\alpha \,z}{\left( x^{2}+z^{2}\right) ^{1/2}},~
\sin \left( \theta -\alpha \right) =\frac{\cos\alpha \,x-\sin\alpha \,z}{\left( x^{2}+z^{2}\right) ^{1/2}}, \notag \\
&\sin \left(\theta -\theta _{\mathrm{o}}\right) =\frac{\cos \theta _{\mathrm{o}}\,x-\sin\theta _{\mathrm{o}}\,z}{\left( x^{2}+z^{2}\right) ^{1/2}},  \label{sb}
\end{align}%
together with 
\begin{equation}
x =\hat{n}_{x\left[ 0\right] }s +r_{\mathrm{o}}\sin \theta _{\mathrm{o}},~
z =\hat{n}_{z\left[ 0\right] }s +r_{\mathrm{o}}\cos \theta _{\mathrm{o}}  \label{xz}
\end{equation}%
into (\ref{cos}) and (\ref{sin}). Then our solutions for the Stokes vector $\mathbf{S}$ 
will be parameterized solely by $s$.

\subsection{Solving the evolution equations}
\label{soleveq}

From Eq. (\ref{ev}) we write down a system of first-order ordinary differential equations to solve:
\begin{align}
\dot{S}_{1}\left( s\right)  &=k\hat{\Omega}_{2}\left( s\right) S_{3}\left(
s\right) ,  \label{ode1} \\
\dot{S}_{2}\left( s\right)  &=-k\hat{\Omega}_{1}\left( s\right) S_{3}\left(
s\right) ,  \label{ode2} \\
\dot{S}_{3}\left( s\right)  &=k\left[ \hat{\Omega}_{1}\left( s\right)
S_{2}\left( s\right) -\hat{\Omega}_{2}\left( s\right) S_{1}\left( s\right) %
\right] ,  \label{ode3}
\end{align}%
where an over-dot ${\dot{}}$ denotes differentiation with respect to $s$, 
and $\hat{\Omega}_{1}\left( s\right)$ and $\hat{\Omega}_{2}\left( s\right)$ 
are given by (\ref{br1}) and (\ref{br2}), respectively. By solving these equations numerically, 
we find out how the photon polarization evolves through the strong magnetic field 
in the background of our pulsar emission.

However, in case $\left\vert k \hat{\Omega}_{1,2}\left( s\right)s \right\vert_{\mathrm{max}} \ll 1$, 
one can obtain a solution to Eq. (\ref{ev}) via perturbation: 
\begin{equation}
\mathbf{S}=\mathbf{S}_{\left[0\right]}+\delta \mathbf{S}_{\left[ 1\right] }
=\mathbf{S}_{\left[0\right]}+k\int \mathbf{\hat{\Omega}}\times 
\mathbf{S}_{\left[ 0\right] }\,\mathrm{d}s,  \label{dS}
\end{equation}%
where $\delta \mathbf{S}_{\left[ 1\right] }$ means the leading order quantum
correction to the \textit{unperturbed }(initial) Stokes vector $\mathbf{S}_{\left[ 0\right] }$. 
Here the correction can be treated as the leading order perturbation with 
$\alpha_{\mathrm{e}} /\left( 30\pi \right) \left( B/B_{\mathrm{c}}\right) ^{2}\sim
10^{-5}\left( B/B_{\mathrm{c}}\right) ^{2}$ being a perturbation parameter.
Upon inspection of Eqs. (\ref{br1}) and (\ref{br2}) for Eq. (\ref{dS}),
we can further write down our solution in terms of its components:
\begin{align}
S_{1} &\approx S_{1\left[0\right]}-2k\eta S_{3\left[0\right]} \int B^{2}\sin ^{3}\vartheta \left[ \left( \cos
\vartheta +\sin \vartheta \right) \hat{n}_{y\left[ 0\right] }-\frac{\Omega _{\mathrm{p}}\sin\alpha \,s}{c\left( 3\cos ^{2}\left( \theta -\alpha \right)
+1\right) ^{1/2}}\right] \mathrm{d}s, \label{dS1} \\
S_{2} &\approx S_{2\left[0\right]}+k\eta S_{3\left[0\right]} \int B^{2}\sin ^{4}\vartheta \mathrm{d}s, \label{dS2} \\
S_{3} &\approx S_{3\left[0\right]}+k\eta \left\{2S_{1\left[0\right]} \int B^{2}\sin ^{3}\vartheta \left[ \left( \cos
\vartheta +\sin \vartheta \right) \hat{n}_{y\left[ 0\right] }-\frac{\Omega _{\mathrm{p}}\sin\alpha \,s}{c\left( 3\cos ^{2}\left( \theta -\alpha \right)
+1\right) ^{1/2}}\right] \mathrm{d}s \right. \notag \\
& \!\left. \left. \hspace{12pt} -S_{2\left[0\right]} \int B^{2}\sin ^{4}\vartheta \mathrm{d}s \right.
_{\,_{\,_{\,_{{}}}}}^{\,^{\,^{\,^{{}}}}}\!\!\!\!\!\right\}. \label{dS3}
\end{align}%

\subsubsection{Examples}
\label{ex}
\begin{figure*}[!ht]
\centering
\includegraphics[width=\textwidth]{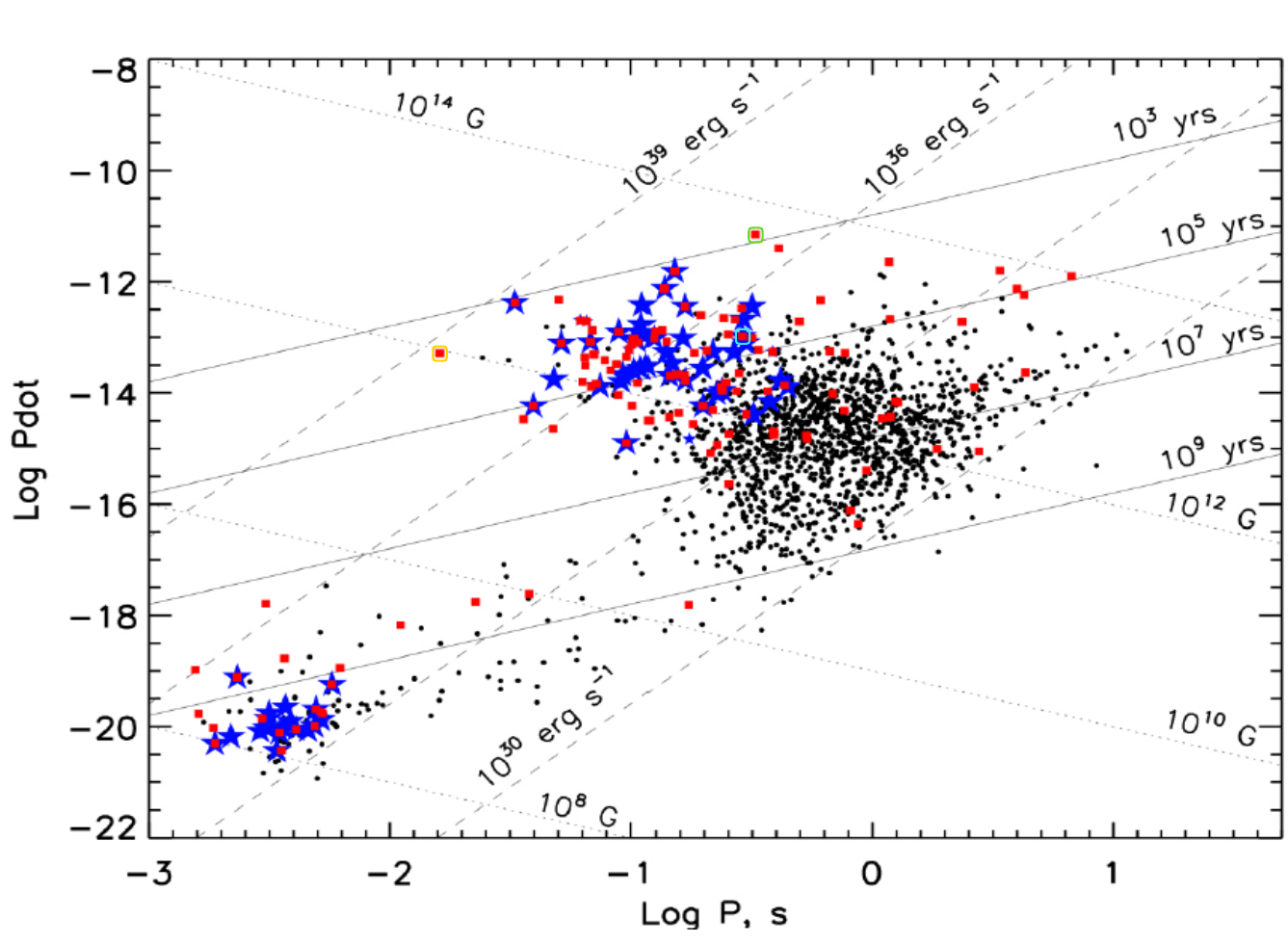}
\caption{The population of detected rotation-powered pulsars (RPPs) plotted against their rotation period. 
RPPs detected in X-rays and gamma-rays are represented by red dots and blue stars, respectively. 
Three RPPs chosen from the X-ray group, (i) one with $B_{\mathrm{max}}\approx10^{12}\,\mathrm{G}$ 
and $\Omega _{\mathrm{p}}\approx392.7\,\mathrm{Hz}$ ($P\approx0.016\,\mathrm{s}$), (ii) another with 
$B_{\mathrm{max}}\approx5.6\times10^{12}\,\mathrm{G}$ and $\Omega _{\mathrm{p}}\approx22.28\,\mathrm{Hz}$ 
($P\approx0.282\,\mathrm{s}$), (iii) the third with $B_{\mathrm{max}}\approx5.0\times10^{13}\,\mathrm{G}$ and 
$\Omega _{\mathrm{p}}\approx19.6\,\mathrm{Hz}$ ($P\approx0.32\,\mathrm{s}$), 
are encircled in orange, cyan and green colors, respectively. (Credit: \cite{Pavlov2013}, reproduced with modifications.) }
\label{fig2}
\end{figure*}

We consider X-ray emissions from three neutron stars: (i) one with $B_{\mathrm{max}}\approx10^{12}\,\mathrm{G}$ 
and $\Omega _{\mathrm{p}}\approx392.7\,\mathrm{Hz}$ ($P\approx0.016\,\mathrm{s}$), 
(ii) another with $B_{\mathrm{max}}\approx5.6\times10^{12}\,\mathrm{G}$ and $\Omega _{\mathrm{p}}\approx22.28\,\mathrm{Hz}$ 
($P\approx0.282\,\mathrm{s}$), (iii) the third with $B_{\mathrm{max}}\approx5.0\times10^{13}\,\mathrm{G}$ and 
$\Omega _{\mathrm{p}}\approx19.6\,\mathrm{Hz}$ ($P\approx0.32\,\mathrm{s}$). For all three, we assume 
$r_{\mathrm{o}}=2r_{*}\approx 2\times 10^{6}\,\mathrm{cm}$, $\theta _{\mathrm{o}}=60^{\circ }$, $\alpha=45^{\circ }$, 
$\eta\approx3.97\times 10^{-32}$ and $\omega\approx2\pi\times 10^{18}\,\mathrm{Hz}$ ($k\approx2.0958\times10^{8}\,\mathrm{cm^{-1}}$).\footnote{For the emission location 
$\left(r_{\mathrm{o}},\theta_{\mathrm{o}} \right)$ and the inclination angle $\alpha$ are given the same values for the three stars; 
the values are not based on actual observations. This is intended for comparing the QED effects from the three different sources under the same conditions.}  
These stars belong to `rotation-powered pulsars' (RPPs) \cite{Pavlov2013}.\footnote{RPPs refer to neutron stars whose radiation is powered 
by loss of their rotation energy, via creation and acceleration of $e^{+}e^{-}$ pairs in the strong magnetic field, $B_{\mathrm{max}}\sim10^{11}-10^{13}\,\mathrm{G}$. 
The number of detected RPPs are known to be about $\sim4000$ in radio, $\sim10$ in optical (including NIR and UV), $\sim100$ in X-ray and $\sim300$ 
in gamma-ray emissions \cite{Pavlov2013,ANTF2024,Smith_2023}.} In Fig. \ref{fig2} the three RPPs chosen from the X-ray group are encircled: (i) the one in orange, (ii) another in cyan, 
(iii) the third in green. Given the X-ray emissions from these, we solve the evolution equations (\ref{ev}) for the following cases.\\

\noindent
\textbf{Example (i)}

\begin{figure*}[!ht]
\centering

\subfloat[With $\mathbf{S}\left(0\right)=\left(1,0,0\right)$]
{\includegraphics[clip,width=0.86\textwidth]{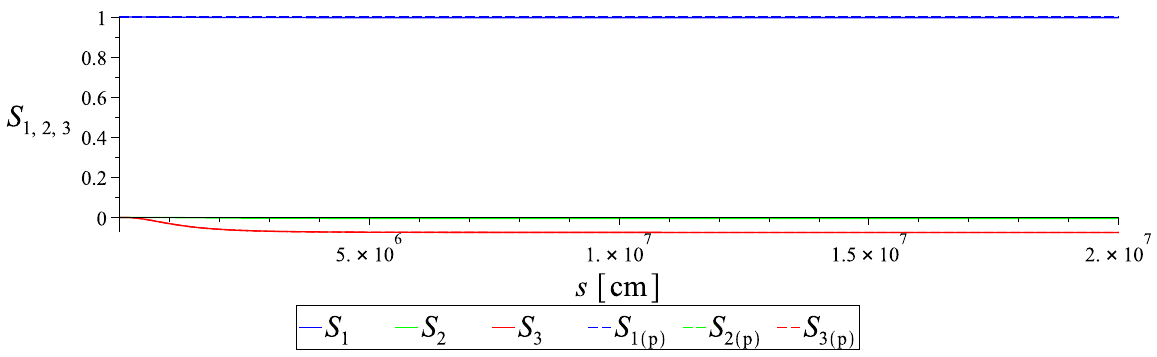}
\label{fig:subfig1}}

\subfloat[With $\mathbf{S}\left(0\right)=\left(0.8,0,0.6\right)$]
{\includegraphics[clip,width=0.86\textwidth]{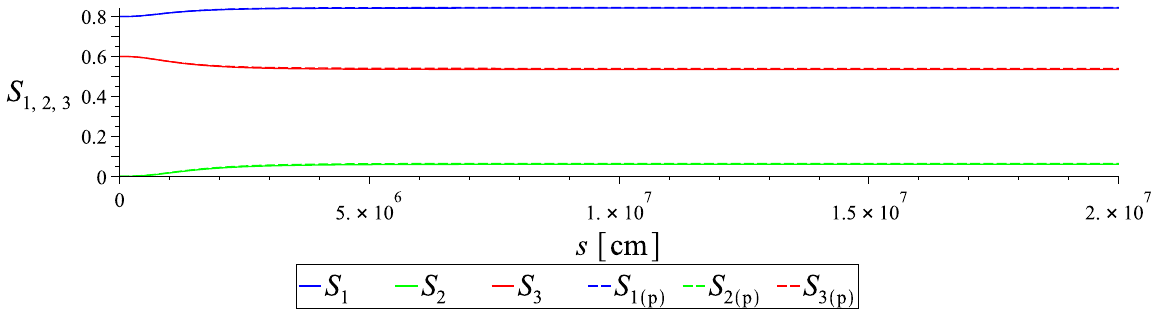}
\label{fig:subfig2}}

\caption{For Example (i): the evolution of the Stokes vector $\mathbf{S}\left(s\right)=
\left(S_{1}\left(s\right),S_{2}\left(s\right),S_{3}\left(s\right)\right)$, 
$0\le s \le 20r_{*}\left(\approx 2\times10^{7}\,\mathrm{cm}\right)$, for the X-ray emissions from the pulsar 
with $B_{\mathrm{max}}\approx10^{12}\,\mathrm{G}$ and $\Omega _{\mathrm{p}}\approx392.7\,\mathrm{Hz}$ ($P\approx0.016\,\mathrm{s}$); 
the subscript $\left(\mathrm{p}\right)$ stands for `perturbative'.}
\label{fig3}
\end{figure*}

\noindent
With $B_{\mathrm{max}}\approx10^{12}\,\mathrm{G}$ and $\Omega _{\mathrm{p}}\approx392.7\,\mathrm{Hz}$ ($P\approx0.016\,\mathrm{s}$), 
we obtain numerical solutions of Eqs. (\ref{ode1})-(\ref{ode3}) (in solid lines) or perturbative solutions 
by means of Eqs. (\ref{dS1})-(\ref{dS3}) (in dashed lines), as shown in Figs. \ref{fig:subfig1} and \ref{fig:subfig2}, given the initial Stokes vectors 
$\mathbf{S}\left(0\right)=\left(S_{1}\left(0\right),S_{2}\left(0\right),S_{3}\left(0\right)\right)=\left(1,0,0\right)$ and $\left(0.8,0,0.6\right)$, respectively.
The perturbative solutions agree well with numerical ones as $\left\vert k \hat{\Omega}_{1,2}\left( s\right)s \right\vert_{\mathrm{max}}\sim 
\left\vert k \hat{\Omega}_{1,2}\left( s_{\dagger 1,2} \right)s_{\dagger 1,2} \right\vert \sim 10^{-2} \ll 1$, where $s_{\dagger 1,2}$ 
is the extremum, i.e., $\mathrm{d}\hat{\Omega}_{1,2}\left( s_{\dagger 1,2} \right)/\mathrm{d}s=0$. On the Poincar\'{e} sphere, our solutions are 
represented by the magenta and light blue loci in Fig. \ref{fig:subfig7}, corresponding to Figs. \ref{fig:subfig1} and \ref{fig:subfig2}, respectively.
The loci imply a fraction of an oscillation for the polarization evolution, as is confirmed later by the approximate analytical solutions in Sect. \ref{fit}.\\

\noindent
\textbf{Example (ii)}

\begin{figure*}[!ht]
\centering

\subfloat[With $\mathbf{S}\left(0\right)=\left(1,0,0\right)$]
{\includegraphics[clip,width=0.86\textwidth]{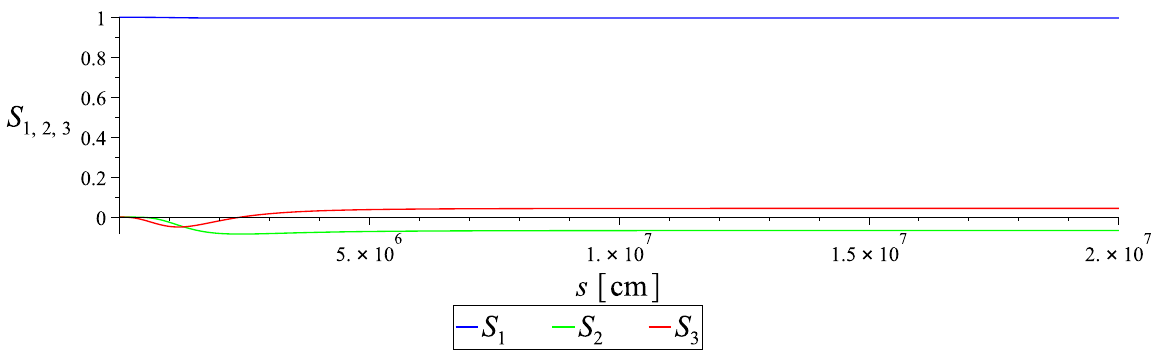}
\label{fig:subfig3}}

\subfloat[With $\mathbf{S}\left(0\right)=\left(0.8,0,0.6\right)$]
{\includegraphics[clip,width=0.87\textwidth]{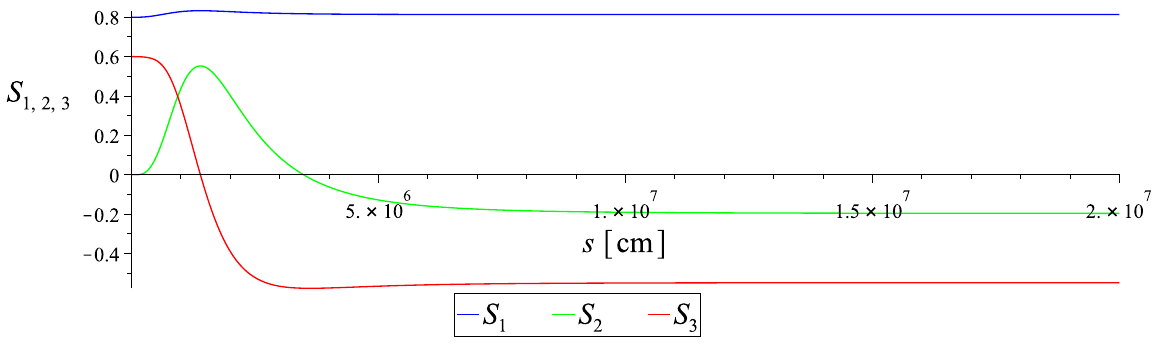}
\label{fig:subfig4}}

\caption{For Example (ii): the evolution of the Stokes vector $\mathbf{S}\left(s\right)=
\left(S_{1}\left(s\right),S_{2}\left(s\right),S_{3}\left(s\right)\right)$, 
$0\le s \le 20r_{*}\left(\approx 2\times10^{7}\,\mathrm{cm}\right)$, for the X-ray emissions from the pulsar 
with $B_{\mathrm{max}}\approx5.6\times10^{12}\,\mathrm{G}$ and $\Omega _{\mathrm{p}}\approx22.28\,\mathrm{Hz}$ 
($P\approx0.282\,\mathrm{s}$).}
\label{fig4}
\end{figure*}

\noindent
With $B_{\mathrm{max}}\approx5.6\times10^{12}\,\mathrm{G}$ and $\Omega _{\mathrm{p}}\approx22.28\,\mathrm{Hz}$ ($P\approx0.282\,\mathrm{s}$), 
we obtain numerical solutions of Eqs. (\ref{ode1})-(\ref{ode3}), as shown in Figs. \ref{fig:subfig3} and \ref{fig:subfig4}, given the initial Stokes vectors 
$\mathbf{S}\left(0\right)=\left(S_{1}\left(0\right),S_{2}\left(0\right),S_{3}\left(0\right)\right)=\left(1,0,0\right)$ and $\left(0.8,0,0.6\right)$, respectively. 
On the Poincar\'{e} sphere, our solutions are represented by the magenta and light blue loci in Fig. \ref{fig:subfig8}, corresponding to Figs. \ref{fig:subfig3} 
and \ref{fig:subfig4}, respectively. The loci imply about half an oscillation for the polarization evolution, as is confirmed later by the approximate analytical solutions in Sect. \ref{fit}.\\

\noindent
\textbf{Example (iii)}

\begin{figure*}[!ht]
\centering

\subfloat[With $\mathbf{S}\left(0\right)=\left(1,0,0\right)$]
{\includegraphics[clip,width=0.87\textwidth]{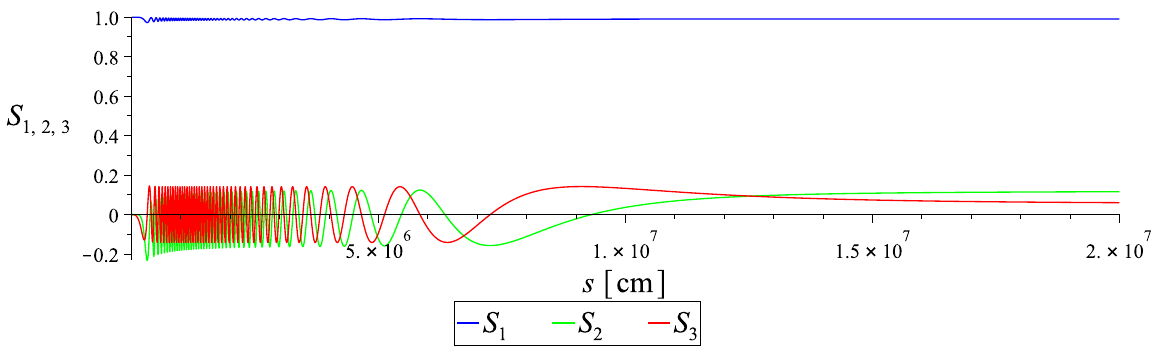}
\label{fig:subfig5}}

\subfloat[With $\mathbf{S}\left(0\right)=\left(0.8,0,0.6\right)$]
{\includegraphics[clip,width=0.87\textwidth]{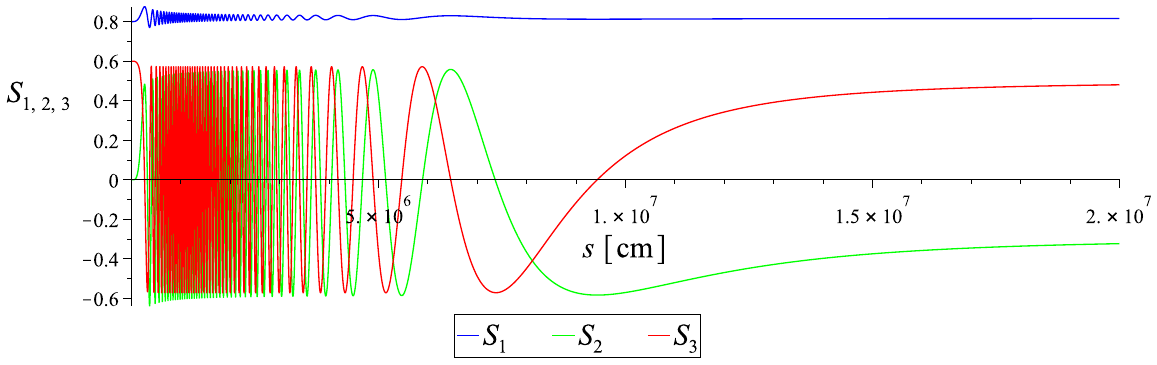}
\label{fig:subfig6}}

\caption{For Example (iii): the evolution of the Stokes vector $\mathbf{S}\left(s\right)=
\left(S_{1}\left(s\right),S_{2}\left(s\right),S_{3}\left(s\right)\right)$, 
$0\le s \le 20r_{*}\left(\approx 2\times10^{7}\,\mathrm{cm}\right)$, for the X-ray emissions from the pulsar 
with $B_{\mathrm{max}}\approx5.0\times10^{13}\,\mathrm{G}$ and 
$\Omega _{\mathrm{p}}\approx19.6\,\mathrm{Hz}$ ($P\approx0.32\,\mathrm{s}$).}
\label{fig5}
\end{figure*}

\begin{figure*}[!ht]
\centering
\def\twidth{0.33}
\centering\captionsetup{width=.33\textwidth}%
\subfloat[Example (i)]{%
		\includegraphics[angle=0, width=3.9cm]{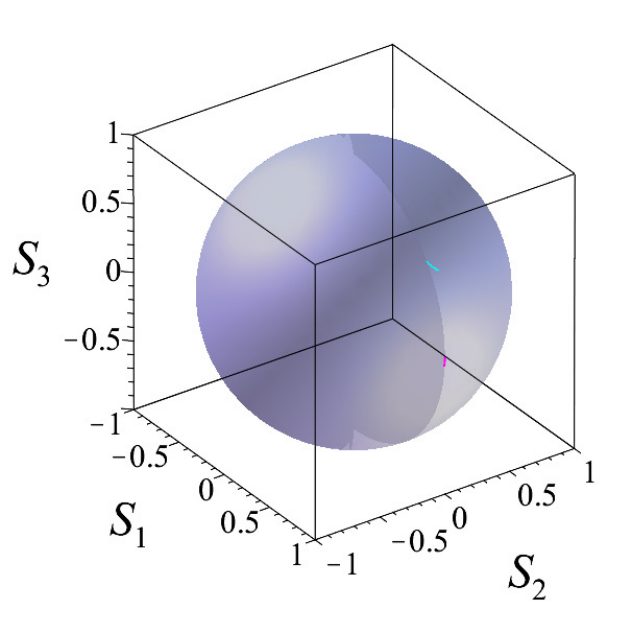}%
		\label{fig:subfig7}}\hfill
\centering\captionsetup{width=.33\textwidth}%
\subfloat[Example (ii)]{%
		\includegraphics[angle=0, width=3.9cm]{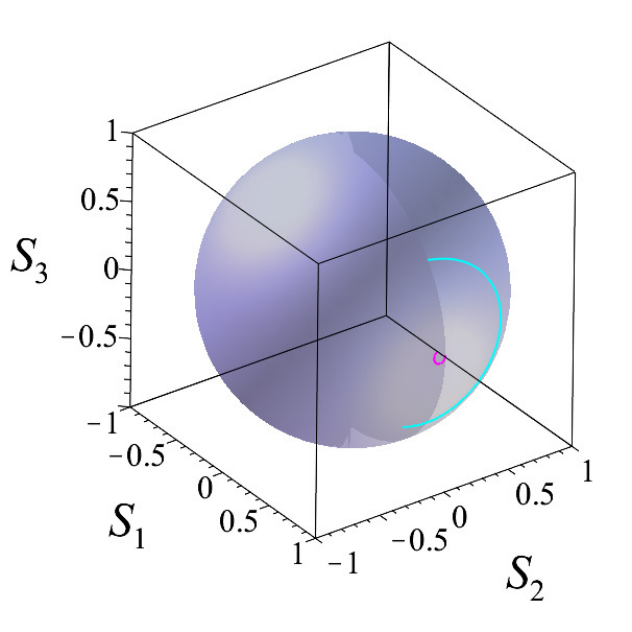}%
		\label{fig:subfig8}}\hfill
\centering\captionsetup{width=.33\textwidth}%
\subfloat[Example (iii)]{%
		\includegraphics[angle=0,width=3.9cm]{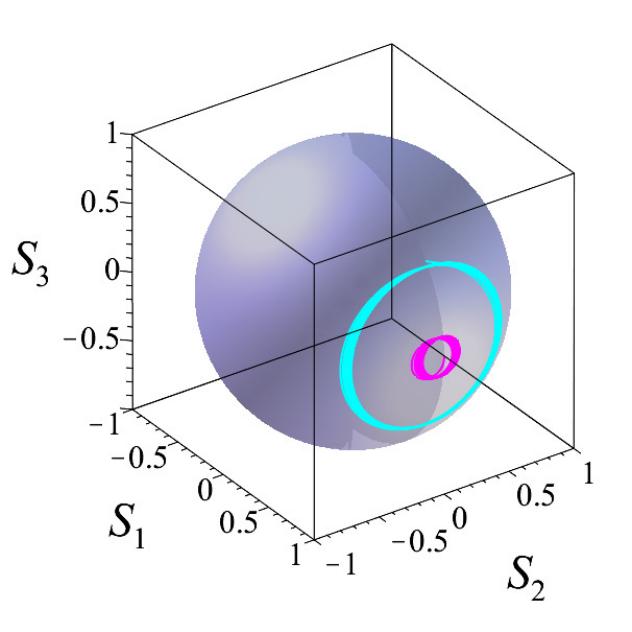}%
		\label{fig:subfig9}} 
\centering\captionsetup{width=\textwidth}%
\caption{Representations of the Stokes vectors from Examples (i)-(iii) on the Poincar\'{e} sphere. The loci imply patterns of the polarization evolution in terms of oscillation.}
\label{fig6}
\end{figure*}

\noindent
With $B_{\mathrm{max}}\approx5.0\times10^{13}\,\mathrm{G}$ and $\Omega _{\mathrm{p}}\approx19.6\,\mathrm{Hz}$ ($P\approx0.32\,\mathrm{s}$), 
we obtain numerical solutions of Eqs. (\ref{ode1})-(\ref{ode3}), as shown in Figs. \ref{fig:subfig5} and \ref{fig:subfig6}, given the initial Stokes vectors 
$\mathbf{S}\left(0\right)=\left(S_{1}\left(0\right),S_{2}\left(0\right),S_{3}\left(0\right)\right)=\left(1,0,0\right)$ and $\left(0.8,0,0.6\right)$, respectively. 
On the Poincar\'{e} sphere, our solutions are represented by the magenta and light blue loci in Fig. \ref{fig:subfig9}, corresponding to Figs. \ref{fig:subfig5} 
and \ref{fig:subfig6}, respectively. The loci imply multiple oscillations for the polarization evolution, as is confirmed later by the approximate analytical solutions in Sect. \ref{fit}.\\

With regard to the adiabatic evolution condition as mentioned in Refs. \cite{Heyl2000,Heyl2003highenergy,Heyl2018Strongly}, we carefully examine our results presented 
in Figs. \ref{fig3}-\ref{fig5} to see what interpretations the condition leads to. Solving the condition 
$\left\vert k\mathbf{\hat{\Omega}}\left( \mathrm{d}\ln \vert k\mathbf{\hat{\Omega}} \vert /\mathrm{d}s \right)^{-1} \right\vert \gtrsim 0.05$ 
for $s$ yields $s_{\mathrm{PL}1} \lesssim s \lesssim s_{\mathrm{PL}2}$,\footnote{Note that our $k\mathbf{\hat{\Omega}}$ is equivalent to the birefringent vector as defined 
in Refs. \cite{Heyl2000,Heyl2003highenergy,Heyl2018Strongly} and that we set the condition value to $0.05$ rather than $0.5$ as in the references.} 
where $s_{\mathrm{PL}1[2]}$ refers to the lower [upper] bound for the `polarization limiting' distance as measured from the emission point. Using this, one can check out the following: 
\\(1) for $6.2\times 10^{5}\,\mathrm{cm}\lesssim s \lesssim 1.8\times 10^{6}\,\mathrm{cm}$ in Fig. \ref{fig3},
\\(2) for $3.1\times 10^{5}\,\mathrm{cm}\lesssim s \lesssim 6.2\times 10^{6}\,\mathrm{cm}$ in Fig. \ref{fig4}, 
\\(3) for $1.1\times 10^{5}\,\mathrm{cm}\lesssim s \lesssim 1.9\times 10^{7}\,\mathrm{cm}$ in Fig. \ref{fig5},
\\our Stokes vector evolves evidently; otherwise, it freezes.

\subsubsection{Approximate analytical solutions}
\label{fit}
\begin{figure*}
\centering
\def\twidth{0.33}
\centering\captionsetup{width=.31\textwidth}%
\subfloat[For $B_{\mathrm{max}}\approx10^{12}\,\mathrm{G}$ and 
$\Omega _{\mathrm{p}}\approx392.7\,\mathrm{Hz}$ ($P\approx0.016\,\mathrm{s}$)]{%
		\includegraphics[angle=0, width=3.9cm]{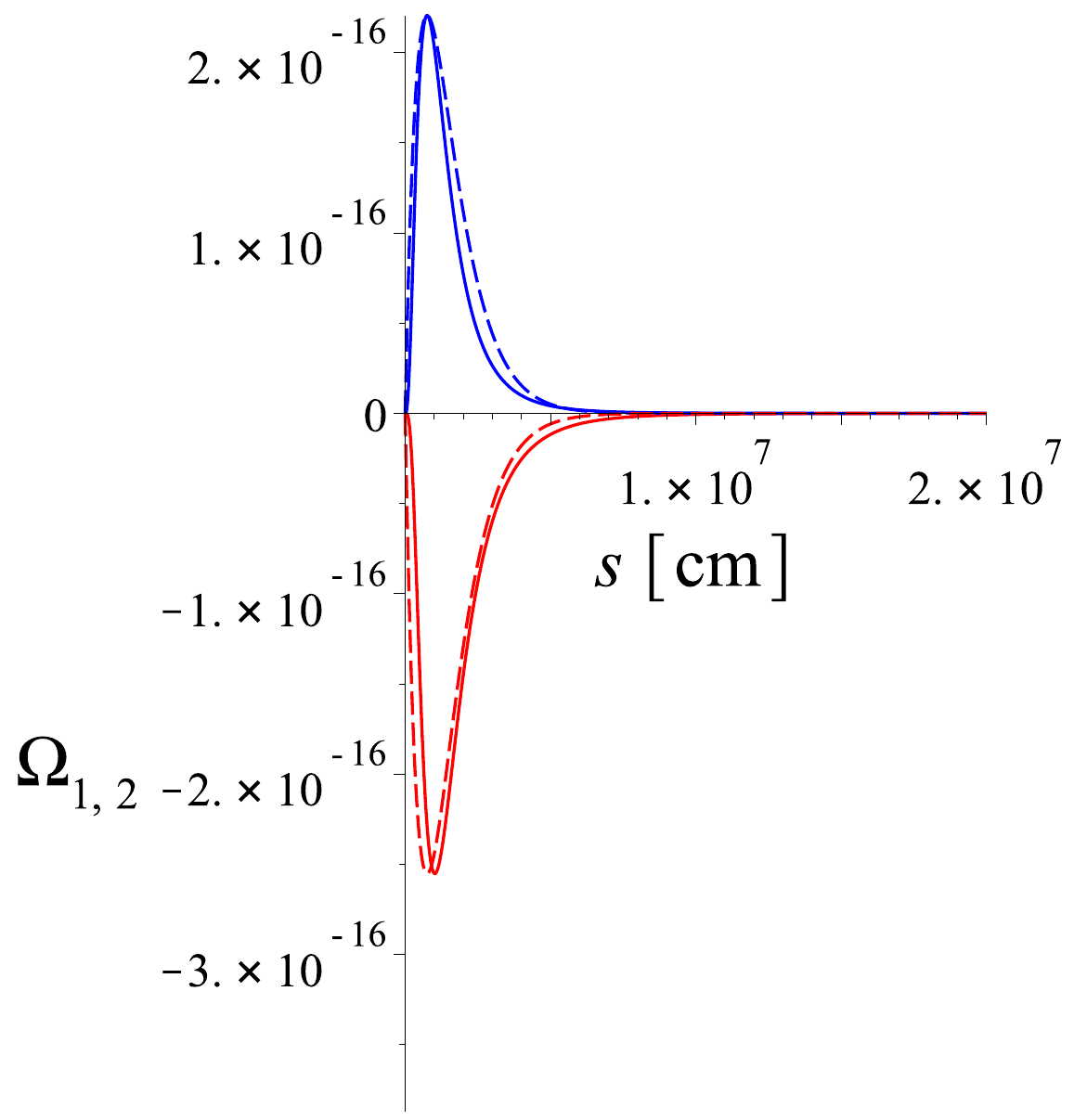}%
		\label{fig:subfig10}}\hfill
\centering\captionsetup{width=.31\textwidth}%
\subfloat[For $B_{\mathrm{max}}\approx5.6\times10^{12}\,\mathrm{G}$ and 
$\Omega _{\mathrm{p}}\approx22.28\,\mathrm{Hz}$ ($P\approx0.282\,\mathrm{s}$)]{%
		\includegraphics[angle=0, width=3.9cm]{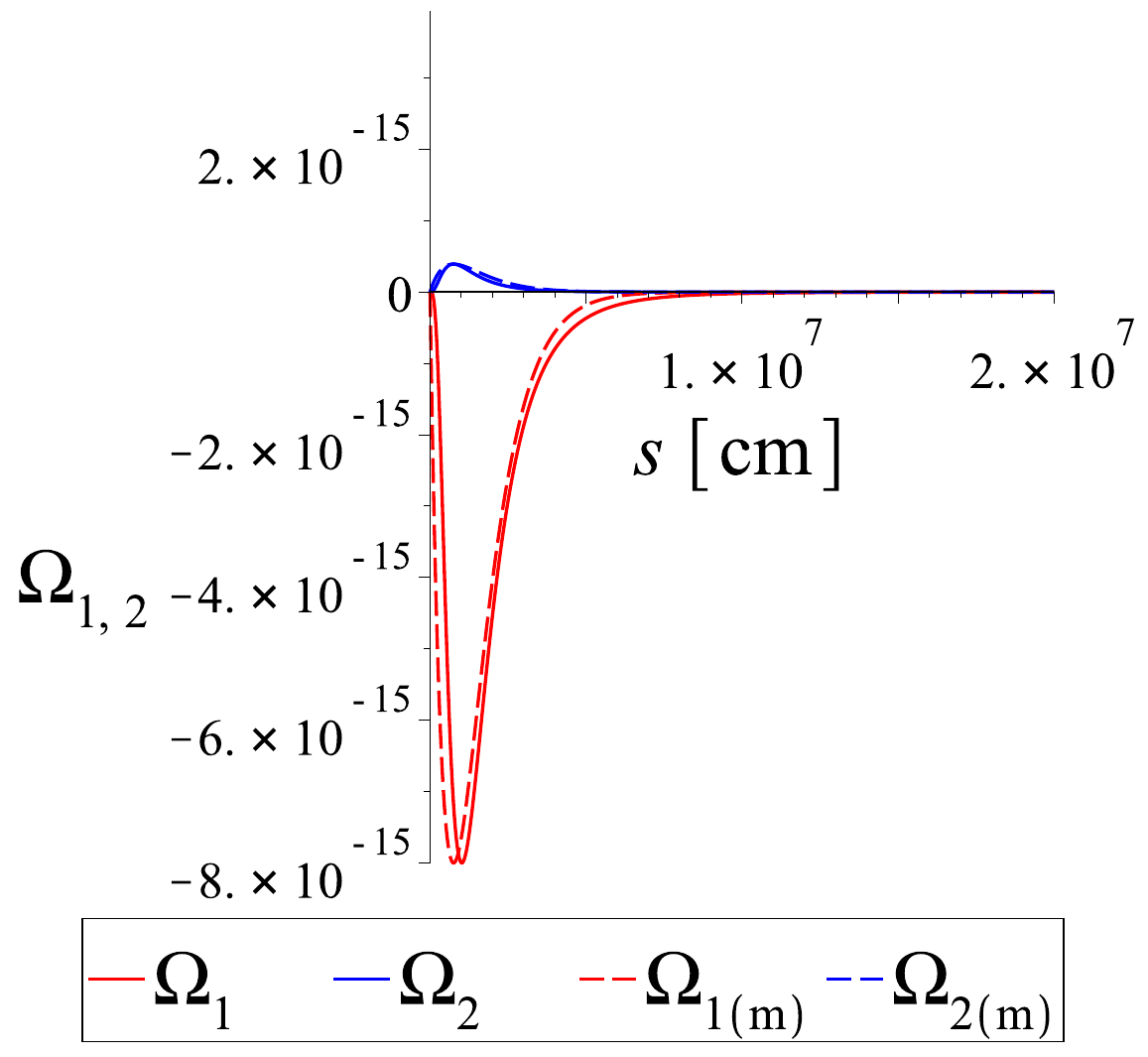}%
		\label{fig:subfig11}}\hfill 
\centering\captionsetup{width=.31\textwidth}%
\subfloat[For $B_{\mathrm{max}}\approx5.0\times10^{13}\,\mathrm{G}$ and 
$\Omega _{\mathrm{p}}\approx19.6\,\mathrm{Hz}$ ($P\approx0.32\,\mathrm{s}$)]{%
		\includegraphics[angle=0,width=3.9cm]{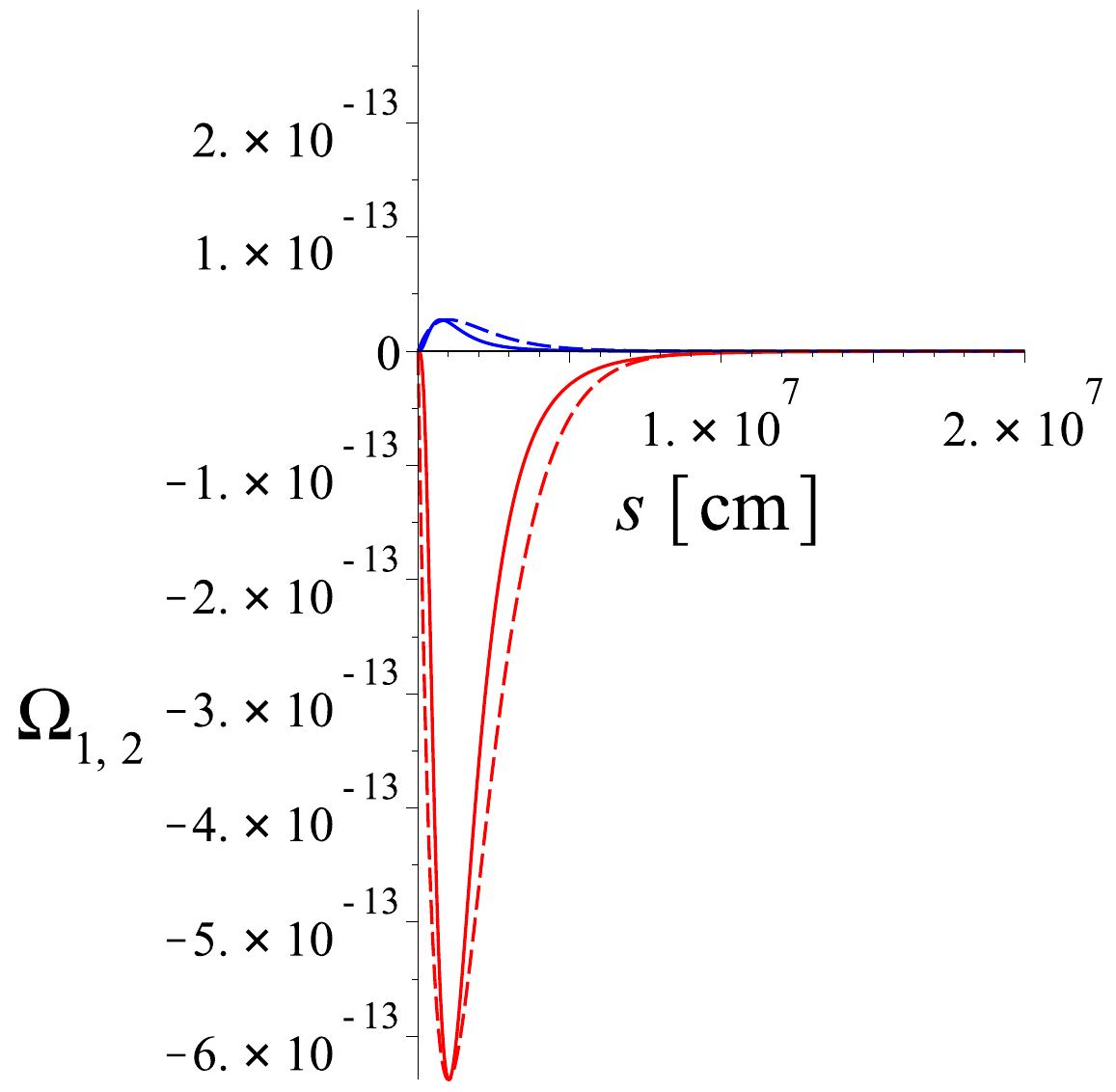}%
		\label{fig:subfig12}} 
\centering\captionsetup{width=\textwidth}%
\caption{Plots of the birefringent functions $\hat{\Omega}_{1}\left( s\right)$ and $\hat{\Omega}_{2}\left( s\right)$ and their approximate analytic models 
$\hat{\Omega}_{1\left(\mathrm{m}\right)}\left( s\right)$ and $\hat{\Omega}_{2\left(\mathrm{m}\right)}\left( s\right)$.}
\label{fig7}
\end{figure*}

Plotting the birefringent functions $\hat{\Omega}_{1}\left( s\right)$ and $\hat{\Omega}_{2}\left( s\right)$, 
as given by (\ref{br1}) and (\ref{br2}), respectively, one can observe that they feature distinctive patterns; 
they can be well approximated by some analytic models, whose curves resemble the original plots. In Fig. \ref{fig7} are plotted the birefringent functions for the three cases: 
(a) $B_{\mathrm{max}}\approx10^{12}\,\mathrm{G}$ and $\Omega _{\mathrm{p}}\approx392.7\,\mathrm{Hz}$ ($P\approx0.016\,\mathrm{s}$), 
(b) $B_{\mathrm{max}}\approx5.6\times10^{12}\,\mathrm{G}$ and $\Omega _{\mathrm{p}}\approx22.28\,\mathrm{Hz}$ ($P\approx0.282\,\mathrm{s}$), 
(c) $B_{\mathrm{max}}\approx5.0\times10^{13}\,\mathrm{G}$ and $\Omega _{\mathrm{p}}\approx19.6\,\mathrm{Hz}$ ($P\approx0.32\,\mathrm{s}$) 
with solid lines (see Figs. \ref{fig:subfig10}, \ref{fig:subfig11} and \ref{fig:subfig12}, respectively), where they have been evaluated with the same initial condition as assumed in Sect. \ref{ex}. 

In correspondence with the actual birefringent functions above, the following analytical models are also plotted with dashed lines in Fig. \ref{fig7}:
\begin{equation}
\hat{\Omega}_{1,2}\left( s\right) \approx -a_{1,2}s^{\frac{p+1}{p}}e^{-bs}
~~\text{for $0\le s\le 20r_{*}\left( \approx 2\times10^{7}\,\mathrm{cm} \right)$},  \label{Om} 
\end{equation}%
where $a_{1}>0$, $a_{2}<0$, $b>0$ and $p>0$ are free parameters; with suitable values chosen for these, our model functions can give rise to solutions of Eqs. (\ref{ode1})-(\ref{ode3}) 
that match fairly well the numerical results obtained in Sect. \ref{ex}. Here we express
\begin{equation}
a_{1,2}=-\hat{\Omega}_{1 \left(\mathrm{min} \right),2 \left(\mathrm{max} \right)}\left[\frac{e}{\left(1-q \right)s_{\dagger 1}+q s_{\dagger 2}} \right]^{\frac{p+1}{p}},
~~ b=\frac{p+1}{p\left[\left(1-q \right)s_{\dagger 1}+q s_{\dagger 2} \right]}, \label{ab} 
\end{equation}
where $\hat{\Omega}_{1 \left(\mathrm{min} \right),2 \left(\mathrm{max} \right)}=\hat{\Omega}_{1,2}\left( s_{\dagger 1,2} \right)$, evaluated from (\ref{br1}), (\ref{br2}), 
with $s_{\dagger 1,2}$ denoting the extremum. We have set $p=100$ (a sufficiently large number) for all three cases, and $q=0.99999$ for (a) and (b), 
and $q=0.0532$ for (c) in Fig. \ref{fig7}.\footnote{In particular, the values for $q$ have been chosen such that our solutions converge to the asymptotic limits 
that match well the numerical results given by Figs. \ref{fig:subfig2}, \ref{fig:subfig4} and \ref{fig:subfig6} in Sect. \ref{ex}, as $s$ tends to $\infty$.}

Plugging Eq. (\ref{Om}) into Eqs. (\ref{ode1})-(\ref{ode3}), we obtain analytical solutions as follows (for a complete derivation, see Appendix \ref{appB}):
\begin{align}
S_{1}\left( s\right)  &\approx \frac{a_{2}S_{\mathrm{o}}}{\sqrt{a_{1}^{2}+a_{2}^{2}}}
\cos \left( \Psi\left(s;p\right) +\delta \right) +\frac{a_{1}C}{\sqrt{a_{1}^{2}+a_{2}^{2}}}, \label{Sa1} \\
S_{2}\left( s\right)  &\approx -\frac{a_{1}S_{\mathrm{o}}}{\sqrt{a_{1}^{2}+a_{2}^{2}}}
\cos \left( \Psi\left(s;p\right) +\delta \right) +\frac{a_{2}C}{\sqrt{a_{1}^{2}+a_{2}^{2}}}, \label{Sa2} \\
S_{3}\left( s\right)  &\approx S_{\mathrm{o}}\sin \left( \Psi\left(s;p\right) +\delta \right), \label{Sa3}
\end{align}%
where 
\begin{equation}
\Psi\left(s;p\right) \equiv k\sqrt{a_{1}^{2}+a_{2}^{2}}b^{-\frac{4p+1}{2p}}s^{\frac{1}{2p}}e^{-\frac{1}{2}bs}
\left[ M_{\frac{1}{2p},\frac{p+1}{2p}}\left( bs\right) -\left(bs\right)^{\frac{2p+1}{2p}}e^{-\frac{1}{2}bs} \right], \label{Psi}
\end{equation}
and $M_{\kappa,\mu}\left( z\right)$ denotes a Whittaker function of the first kind,\footnote{The expression inside 
the square brackets in Eq. (\ref{Psi}) has been reduced from its original form as given by Eq. (\ref{Psio}) in Appendix \ref{appB}, 
using the identity $M_{\left(2p+1\right)/\left(2p\right),\left(p+1\right)/\left(2p\right)}\left( bs\right)
=\left(bs\right)^{\left(2p+1\right)/\left(2p\right)}e^{-bs/2}M\left(0,\left(2p+1\right)/p,bs \right)$, 
with the Kummer function $M\left(0,\left(2p+1\right)/p,bs \right)=1$ as a special case \cite{NIST2024}.} and $a_{1,2}$ and $b$ are given by (\ref{ab}). 
Here we determine $S_{\mathrm{o}}$, $C$ and $\delta$ by matching the initial value of the Stokes vector 
$\mathbf{S}\left(0\right)=\left( S_{1}\left( 0\right),S_{2}\left( 0\right),S_{3}\left( 0\right) \right)$ with Eqs. (\ref{Sa1})-(\ref{Sa3}) evaluated at $s=0$. 

\begin{figure*}
\centering

\subfloat[For $B_{\mathrm{max}}\approx10^{12}\,\mathrm{G}$ and 
$\Omega _{\mathrm{p}}\approx392.7\,\mathrm{Hz}$ ($P\approx0.016\,\mathrm{s}$)]
{\includegraphics[clip,width=0.859\textwidth]{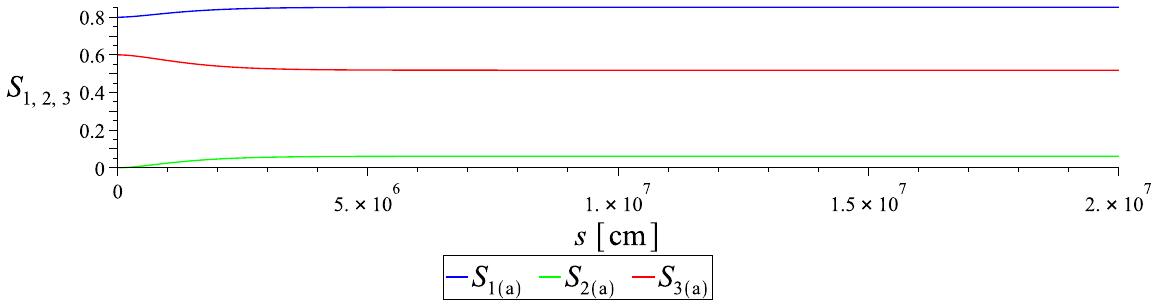}
\label{fig:subfig13}} 

\subfloat[For $B_{\mathrm{max}}\approx5.6\times10^{12}\,\mathrm{G}$ and 
$\Omega _{\mathrm{p}}\approx22.28\,\mathrm{Hz}$ ($P\approx0.282\,\mathrm{s}$)]
{\includegraphics[clip,width=0.869\textwidth]{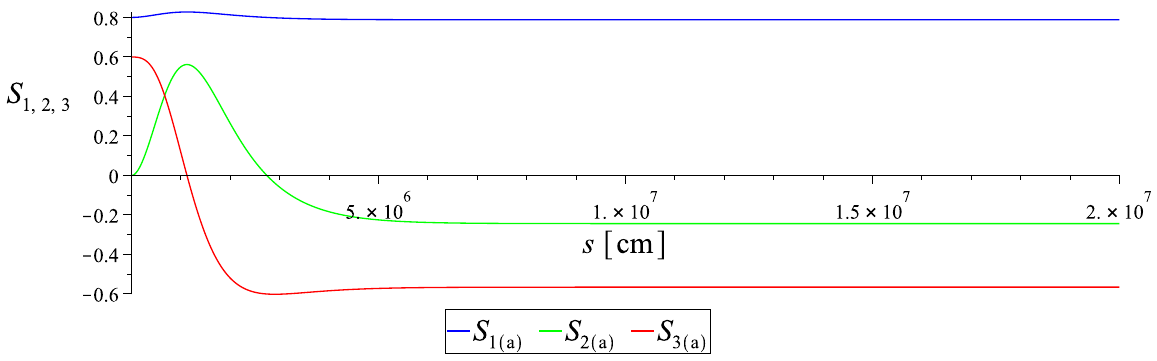}
\label{fig:subfig14}}

\subfloat[For $B_{\mathrm{max}}\approx5.0\times10^{13}\,\mathrm{G}$ and 
$\Omega _{\mathrm{p}}\approx19.6\,\mathrm{Hz}$ ($P\approx0.32\,\mathrm{s}$)]
{\includegraphics[clip,width=0.869\textwidth]{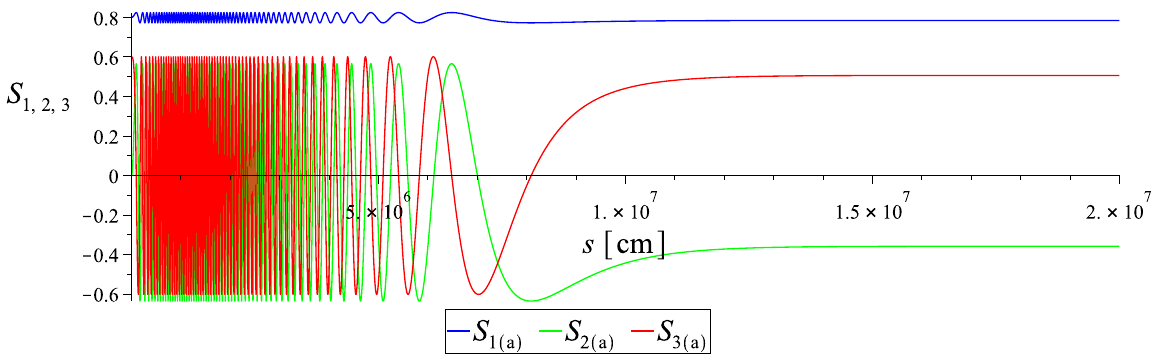}
\label{fig:subfig15}} 
\caption{Plots of the analytical solutions $\mathbf{S}_{\left(\mathrm{a}\right)}\left(s\right)=\left(S_{1\left(\mathrm{a}\right)}\left(s\right),S_{2\left(\mathrm{a}\right)}\left(s\right),
S_{3\left(\mathrm{a}\right)}\left(s\right)\right)$, given the initial Stokes vector $\mathbf{S}\left(0\right)=\left(0.8,0,0.6\right)$.}
\label{fig8}
\end{figure*}

In Fig. \ref{fig8} we plot the above solutions (\ref{Sa1})-(\ref{Sa3}) for the following cases, assuming $r_{\mathrm{o}}=2r_{*}\approx 2\times 10^{6}\,\mathrm{cm}$, $\theta _{\mathrm{o}}=60^{\circ }$, 
$\alpha=45^{\circ }$, $\eta\approx3.97\times 10^{-32}$ for $\hat{\Omega}_{1 \left(\mathrm{min} \right),2 \left(\mathrm{max} \right)}=\hat{\Omega}_{1,2}\left( s_{\dagger 1,2} \right)$ 
(evaluated via (\ref{br1}) and (\ref{br2})) and $\omega\approx2\pi\times 10^{18}\,\mathrm{Hz}$ ($k\approx2.0958\times10^{8}\,\mathrm{cm^{-1}}$) for the X-ray emissions, 
with the initial Stokes vector $\mathbf{S}\left(0\right)=\left(0.8,0,0.6\right)$: 
\\ \textbf{Case (a)} (see Fig. \ref{fig:subfig13});
\\for $B_{\mathrm{max}}\approx10^{12}\,\mathrm{G}$ and 
$\Omega _{\mathrm{p}}\approx392.7\,\mathrm{Hz}$ ($P\approx0.016\,\mathrm{s}$), with the parameters
\\$\hat{\Omega}_{1 \left(\mathrm{min} \right)}\approx -2.5511\times10^{-16}$, $\hat{\Omega}_{2 \left(\mathrm{max} \right)}\approx 2.2054\times10^{-16}$, 
$s_{\dagger 1}\approx1.0202\times10^{6}\,\mathrm{cm}$, $s_{\dagger 2}\approx7.5467\times10^{5}\,\mathrm{cm}$, $p=100$, $q=0.99999$ for $a_{1,2}$ and $b$ (evaluated via (\ref{ab})),  
and $S_{\mathrm{o}}\approx0.7961$, $C\approx0.6052$, $\delta\approx2.2879$,
\\ \textbf{Case (b)} (see Fig. \ref{fig:subfig14});
\\for $B_{\mathrm{max}}\approx5.6\times10^{12}\,\mathrm{G}$ and 
$\Omega _{\mathrm{p}}\approx22.28\,\mathrm{Hz}$ ($P\approx0.282\,\mathrm{s}$), with the parameters
\\$\hat{\Omega}_{1 \left(\mathrm{min} \right)}\approx -8.0001\times10^{-15}$, $\hat{\Omega}_{2 \left(\mathrm{max} \right)}\approx 3.9240\times10^{-16}$, 
$s_{\dagger 1}\approx1.0202\times10^{6}\,\mathrm{cm}$, $s_{\dagger 2}\approx7.5467\times10^{5}\,\mathrm{cm}$, $p=100$, $q=0.99999$ for $a_{1,2}$ and $b$ (evaluated via (\ref{ab})),
and $S_{\mathrm{o}}\approx0.6013$, $C\approx0.7990$, $\delta\approx1.6360$,
\\ \textbf{Case (c)} (see Fig. \ref{fig:subfig15});
\\for $B_{\mathrm{max}}\approx5.0\times10^{13}\,\mathrm{G}$ and 
$\Omega _{\mathrm{p}}\approx19.6\,\mathrm{Hz}$ ($P\approx0.32\,\mathrm{s}$), with the parameters
\\$\hat{\Omega}_{1 \left(\mathrm{min} \right)}\approx -6.3776\times10^{-13}$, $\hat{\Omega}_{2 \left(\mathrm{max} \right)}\approx 2.7567\times10^{-14}$, 
$s_{\dagger 1}\approx1.0202\times10^{6}\,\mathrm{cm}$, $s_{\dagger 2}\approx7.5467\times10^{5}\,\mathrm{cm}$, $p=100$, $q=0.0532$ for $a_{1,2}$ and $b$ (evaluated via (\ref{ab})),
and $S_{\mathrm{o}}\approx0.6010$, $C\approx0.7993$, $\delta\approx1.6283$. 
\\These plots compare with Figs. \ref{fig:subfig2} (for {Example (i)}), \ref{fig:subfig4} (for {Example (ii)}) and \ref{fig:subfig6} (for {Example (iii)}) in Sect. \ref{ex}, respectively. 

The analytical solutions (\ref{Sa1})-(\ref{Sa3}) provide a useful tool for understanding the different patterns of polarization evolution 
for the three cases above, as given by Figs. \ref{fig:subfig13}, \ref{fig:subfig14} and \ref{fig:subfig15}. 
Inspecting numerically the functional argument $\Psi\left(s;p \right)$ given by (\ref{Psi}), one can approximate it to a simpler form with the help of (\ref{Om}) and (\ref{ab}):
\\ For $0\le s \lesssim 2.5\times10^{6}\,\mathrm{cm}$,
\begin{align}
\Psi\left(s; p=100 \right) &= k\sqrt{a_{1}^{2}+a_{2}^{2}}b^{-\frac{401}{200}}s^{\frac{1}{200}}e^{-\frac{1}{2}bs} 
\left[ M_{\frac{1}{200},\frac{101}{200}}\left( bs\right) -\left(bs \right)^{\frac{201}{200}}e^{-\frac{1}{2}bs} \right] \notag \\
&\approx 0.24k\sqrt{\hat{\Omega}_{1 \left(\mathrm{min} \right)}^{2}+\hat{\Omega}_{2 \left(\mathrm{max} \right)}^{2}}\pi s. \label{arg} 
\end{align}%
Using this, we can estimate how much the oscillations for the three cases have progressed, for example, 
during $0\le s \lesssim 2.5\times10^{6}\,\mathrm{cm}$:
\begin{align}
\Psi\left(s\approx 2.5\times10^{6}\,\mathrm{cm}; p=100 \right) \approx \left\{
\begin{array}{ll}
0.04\pi \hspace{3pt} \text{(a fraction of an oscillation)} &\hspace{3pt} \text{for Case (a)}, \\
\pi \hspace{8pt} \text{(about half an oscillation)} &\hspace{3pt} \text{for Case (b)}, \\
80\pi \hspace{11pt} \text{(multiple oscillations)} &\hspace{3pt} \text{for Case (c)}, 
\end{array}
\right. \label{arg1}
\end{align}
each of which can be checked by comparison with Figs. \ref{fig:subfig13}, \ref{fig:subfig14} and \ref{fig:subfig15}, respectively. 

Recalling {Example (i)} from Sect. \ref{ex}, one finds that the condition for perturbation can be equivalently expressed with the help of (\ref{arg}): 
\begin{align}
\left\vert k \hat{\Omega}_{1,2}\left( s\right)s \right\vert_{\mathrm{max}}
\sim \left\vert k \hat{\Omega}_{1,2}\left( s_{\dagger 1,2} \right)s_{\dagger 1,2} \right\vert 
&\sim 0.24k\sqrt{\hat{\Omega}_{1 \left(\mathrm{min} \right)}^{2}+\hat{\Omega}_{2 \left(\mathrm{max} \right)}^{2}}\pi \times10^{6}\,\mathrm{cm} \notag \\
&\approx 0.017\pi \ll 1~~\text{for Example (i) or Case (a)}. \label{eqv}
\end{align}
Extending this argument, similarly to (\ref{arg1}) above we may state the following in reference to the patterns of polarization evolution:
\begin{align}
\left\vert k \hat{\Omega}_{1,2}\left( s\right)s \right\vert_{\mathrm{max}} \left\{
\begin{array}{ll}
\ll 1 \hspace{5pt} \text{(fractionally oscillatory - monotonic)} &\hspace{3pt} \text{for Example (i)}, \\
\sim \hspace{2pt} 1 \hspace{5pt} \text{(half oscillatory)} &\hspace{3pt} \text{for Example (ii)}, \\
\gg 1 \hspace{5pt} \text{(highly oscillatory)} &\hspace{3pt} \text{for Example (iii)}.
\end{array}
\right. \label{eqv1}
\end{align}
These features can also be checked by comparison with the loci on the Poincar\'{e} sphere, as given by Figs. \ref{fig:subfig7}, \ref{fig:subfig8} and \ref{fig:subfig9},  
which imply the three different patterns of the polarization evolution in terms of oscillation (by means of the number of turns of the circular loci).

Here the approximate analytical solutions serve our purpose well, in that they help us to understand the different patterns of polarization evolution, which depend largely on 
the major profiles of pulsar emission, such as the emission frequency, the magnetic field strength and the rotation frequency of the neutron stars, as implied from Eq. (\ref{arg}). 
On the other hand, it would be worthwhile to check how close the numerical and analytical solutions are to each other by evaluating the cross-correlations between them. For example, 
comparing Figs. \ref{fig:subfig6} and \ref{fig:subfig15}, over the entire region of polarization evolution 
($2\times10^{2}\,\mathrm{cm}\lesssim s \lesssim 2\times10^{7}\,\mathrm{cm}$), the cross-correlation coefficients between $S_{1}$ and $S_{1\left(\mathrm{a}\right)}$, 
between $S_{2}$ and $S_{2\left(\mathrm{a}\right)}$, and between $S_{3}$ and $S_{3\left(\mathrm{a}\right)}$ turn out to be approximately $0.9998$, $0.6757$, and $0.6521$, 
respectively. This shows that the two solutions are in decent agreement with each other; although the asymptotic values, $S_{1}\left(\infty\right)$ and 
$S_{1\left(\mathrm{a}\right)}\left(\infty\right)$, $S_{2}\left(\infty\right)$ and $S_{2\left(\mathrm{a}\right)}\left(\infty\right)$, and $S_{3}\left(\infty\right)$ and 
$S_{3\left(\mathrm{a}\right)}\left(\infty\right)$ are fairly closely matched to each other with less than $10\%$ differences, the correlations are decreased by the mismatched 
phases between the two solutions.

\section{Evolution of polarization states in strong magnetic field -- quadrudipole pulsars}
\label{evol2}

\subsection{Modified magnetic field geometry and evolution equations of Stokes vector}
\label{eveq2}

The pulsar magnetic field structure may not be assumed to be purely dipolar as given by (\ref{B}). This assumption is based on the behavior of the field 
in the far-field regime, where its high-order multipole ($\ell$) components decrease faster than low-order ones, like $r^{-\left(\ell+1\right)}$, which justifies the use of 
the dipole field as a good approximation \cite{Petri2016Theory}. However, taking into account the possible contributions from the multipolar components, 
especially in the vicinity of the neutron star, we need to extend our pulsar model by superposing the dipole and higher-order multipole fields. 
A simple extension can be implemented by considering a rotating off-centered dipole, and a number of studies have been carried out regarding a variety of astrophysical 
consequences of such extension models in pulsar astronomy (see \cite{Petri2016Theory} and references therein).\footnote{The multipole field structure extended in this way 
inevitably has the higher-order fields aligned with the dipolar axis. Although the aligned multipole fields might not accurately represent actual field geometries in nature 
(as illustrated in Ref. \cite{Kazmierczak2019}), the models would still be useful for estimating roughly the ‘beyond-dipole’ effects in pulsar emission in the near-field regime, 
as shown in Sect. \ref{ex2} later.} As the simplest model, one can consider `quadrudipole' fields, a superposition of dipole and quadrupole fields 
\cite{Gralla_2016,10.1093/mnras/stz2524,Kalapotharakos_2021}. 

The magnetic field of an oblique quadrudipole rotator can be written as
\begin{align}
& \mathbf{B}_{\mathrm{qd}}\left( r,\theta ,\phi \right)   \notag \\
& =\left[ \frac{2\mu _{\mathrm{d}}\left( \cos \alpha \cos \theta +\sin
\alpha \sin \theta \cos \phi \right) }{r^{3}}+\frac{\mu _{\mathrm{q}}\left(
3\left( \cos \alpha \cos \theta +\sin \alpha \sin \theta \cos \phi \right)
^{2}-1\right) }{r^{4}}\right] \mathbf{e}_{\hat{r}}  \notag \\
& \hspace{14pt} +\left[ \frac{\mu _{\mathrm{d}}\left( \cos \alpha \sin \theta -\sin \alpha
\cos \theta \cos \phi \right) }{r^{3}}\right.   \notag \\
& \hspace{31pt} \left. +\frac{2\mu _{\mathrm{q}}\left( \cos \alpha \sin \theta -\sin
\alpha \cos \theta \cos \phi \right) \left( \cos \alpha \cos \theta +\sin
\alpha \sin \theta \cos \phi \right) }{r^{4}}\right] \mathbf{e}_{\hat{\theta}%
}  \notag \\
& \hspace{14pt} +\left[ \frac{\mu _{\mathrm{d}}\sin \alpha \sin \phi }{r^{3}}+\frac{2\mu
_{\mathrm{q}}\sin \alpha \sin \phi \left( \cos \alpha \cos \theta +\sin \alpha \sin
\theta \cos \phi \right) }{r^{4}}\right] \mathbf{e}_{\hat{\phi}}.
\label{Bqd}
\end{align}
Here the subscript `qd' on the left-hand side stands for `quadrudipole' (hereafter, this will be attached to notations for any quantities affected by the quadrudipole field), 
and $\mu _{\mathrm{d}}$ and $\mu _{\mathrm{q}}$ denote the magnetic dipole and quadrupole moments, respectively.

\begin{figure*}[!ht]
\centering
\includegraphics[width=\textwidth]{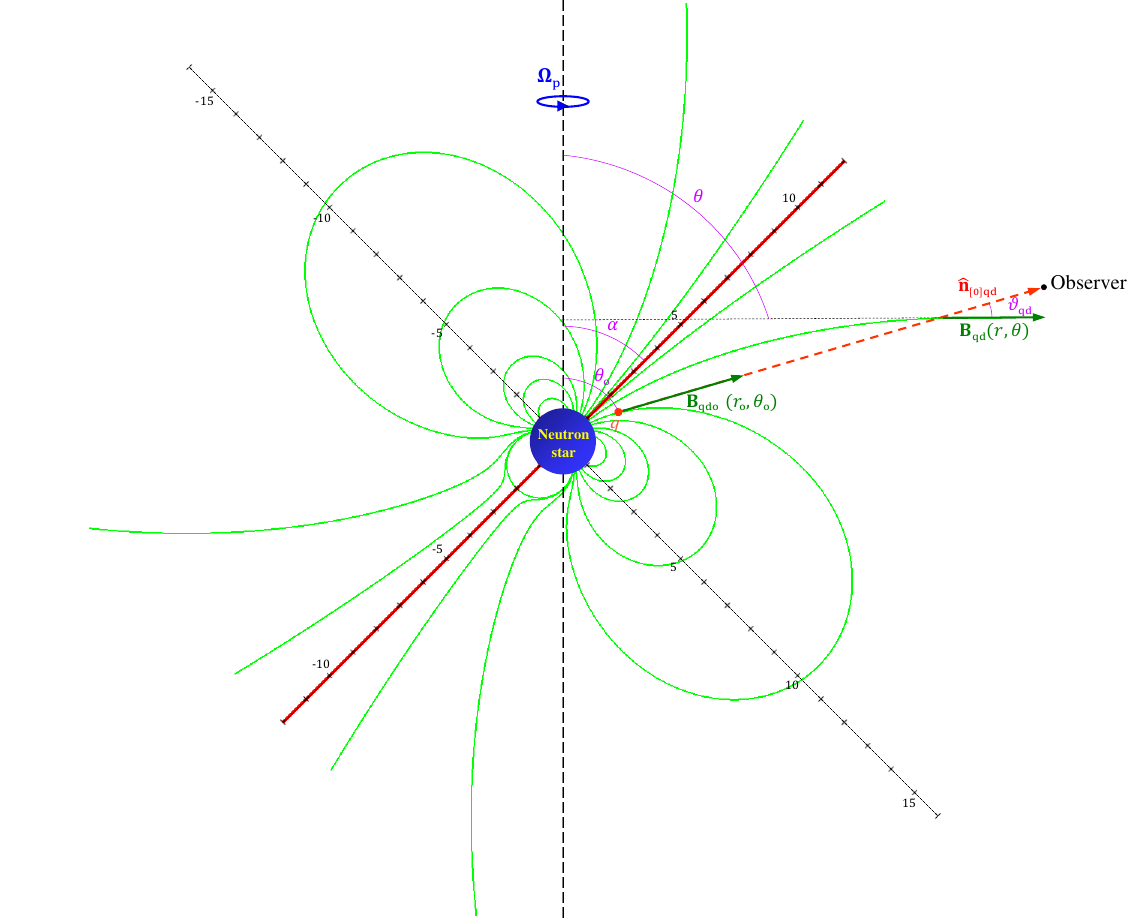}
\caption{A cross-sectional view of a pulsar magnetosphere with the quadrudipole (dipole $+$ quadrupole) magnetic field lines (green) around a neutron star, 
where the relation between the magnetic dipole and quadrupole moments is set by $\mu _{\mathrm{q}}=2r_{\ast}\mu _{\mathrm{d}}$ for the total field (\ref{Bqd}). 
The vertical dashed line (black) and the inclined solid line (red) represent the rotation axis and the magnetic axis, respectively. $\alpha$ between these axes denotes 
the inclination angle. The scale of the unity in this graph is equivalent to the neutron star radius $\sim 10^{6}\,\mathrm{cm}$. The red dashed line represents 
the trajectory curve of the light ray traced by the propagation vector $\mathbf{\hat{n}}_{\left[ 0\right]\mathrm{qd}}$ as projected onto the $xz$-plane.}
\label{fig9}
\end{figure*}

The magnetic dipole moment can be expressed as $\mu _{\mathrm{d}}=\pi r_{\ast}^{2}I$, as produced by a static loop current $I$ of radius of the neutron star $r_{\ast}$, 
encircling its equator. Similarly, one can express the magnetic quadrupole moment as $\mu _{\mathrm{q}}=2\chi\pi r_{\ast}^{3}I=2\chi r_{\ast}\mu _{\mathrm{d}}$, 
as produced by two identical magnetic dipole loops carrying opposing equal currents $I$, each of radius $r_{\ast}$, separated by distance $\chi r_{\ast}$ (i.e., anti-Helmholtz coils), 
where $\chi>0$ is a free parameter to determine the ratio between the dipole and quadrupole moments. For example, the magnetic field lines of an oblique quadrudipole rotator, 
with $\chi=1$, that is, $\mu _{\mathrm{q}}=2r_{\ast}\mu _{\mathrm{d}}$, set for Eq. (\ref{Bqd}) are illustrated in Fig. \ref{fig9}.

As the magnetic field geometry changes from (\ref{B}) to (\ref{Bqd}), the classical propagation vector shall be modified from (\ref{mhd}) to 
\begin{equation}
\mathbf{\hat{n}}_{\left[ 0\right] \mathrm{qd}}=\beta_{\mathrm{qd}} \mathbf{\hat{B}}_\mathrm{qd} + \frac{\mathbf{\Omega }_{\mathrm{p}}\times \mathbf{r}}{c}, 
\label{mhd2}
\end{equation}%
where $\mathbf{\hat{B}}_\mathrm{qd} \equiv \mathbf{B}_{\mathrm{qd}}/\left\vert \mathbf{B}_{\mathrm{qd}} \right\vert$ and 
\begin{equation}
\beta_{\mathrm{qd}} \equiv \left[ 1-\left( \frac{\Omega _{\mathrm{p}}r}{c}\right) ^{2}\sin
^{2}\theta \left( 1-\frac{\sin ^{2}\alpha \sin ^{2}\phi }{f \left( \theta ^{\prime}; \chi \right)}\right) \right] ^{1/2}
-\frac{\Omega _{\mathrm{p}}r}{c}\frac{\sin \alpha \sin \theta \sin \phi }{\left[f \left( \theta ^{\prime}; \chi \right) \right] ^{1/2}},  \label{beta2}
\end{equation}%
with 
\begin{equation}
f \left( \theta ^{\prime}; \chi \right) \equiv 3\cos ^{2}\theta ^{\prime
}+1 +16\chi \frac{r_{\ast }}{r}\cos ^{3}\theta ^{\prime }+4\chi^{2}\left( \frac{r_{\ast }}{r}\right) ^{2}
\left( 5\cos ^{4}\theta ^{\prime}-2\cos ^{2}\theta ^{\prime }+1 \right) \label{fth}
\end{equation}%
and $\cos \theta ^{\prime }\equiv \cos \alpha \cos \theta +\sin \alpha \sin \theta \cos \phi $. 

Similarly to Sect. \ref{eveq}, taking only the leading order expansions of $\mathbf{\hat{B}}_{\mathrm{qd}} \left( r_{\mathrm{o}},\theta _{\mathrm{o}},\phi \right) $
and $\beta_{\mathrm{qd}} \left( r_{\mathrm{o}},\theta _{\mathrm{o}},\phi \right)$ in $\phi$ from Eqs. (\ref{Bqd}) and (\ref{beta2}), respectively,  
one can write down the classical propagation vector out of Eq. (\ref{mhd2}) in Cartesian coordinates as
\begin{equation}
\mathbf{\hat{n}}_{\left[ 0\right] \mathrm{qd}}=\hat{n}_{x\left[ 0\right] \mathrm{qd}}\mathbf{e}_{x}
+\hat{n}_{y\left[ 0\right] \mathrm{qd}}\mathbf{e}_{y}+\hat{n}_{z\left[ 0\right] \mathrm{qd}}\mathbf{e}_{z} \label{no2} \\
\end{equation}%
where
\begin{align}
\hat{n}_{x\left[ 0\right] \mathrm{qd}}\approx & \left[ f\left( \theta _{\mathrm{o}%
}-\alpha ;\chi \right) \right] ^{-1/2}\left\{ 2\cos \left( \theta _{\mathrm{o%
}}-\alpha \right) \sin \theta _{\mathrm{o}}+\sin \left( \theta _{\mathrm{o}%
}-\alpha \right) \cos \theta _{\mathrm{o}}~_{~_{~_{~_{~_{}}}}}^{~^{~^{~^{~^{}}}}}\right. 
\notag \\
& \left. +\hspace{1pt} 2\chi \frac{r_{\ast }}{r_{\mathrm{o}}}\left[ \left( 3\cos
^{2}\left( \theta _{\mathrm{o}}-\alpha \right) -1\right) \sin \theta _{%
\mathrm{o}}+2\cos \left( \theta _{\mathrm{o}}-\alpha \right) \sin \left(
\theta _{\mathrm{o}}-\alpha \right) \cos \theta _{\mathrm{o}}\right]
\right\} \notag \\ 
&+\mathcal{O}\left( \phi^{2},\left( \Omega _{\mathrm{p}}r_{\mathrm{o}}/c\right) ^{2},
\phi \left(\Omega _{\mathrm{p}}r_{\mathrm{o}}/c\right) \right),  \label{nox2} 
\end{align}%
\begin{align}
\hat{n}_{z\left[ 0\right] \mathrm{qd}}\approx & \left[ f\left( \theta _{\mathrm{o}%
}-\alpha ;\chi \right) \right] ^{-1/2}\left\{ 2\cos \left( \theta _{\mathrm{o%
}}-\alpha \right) \cos \theta _{\mathrm{o}}-\sin \left( \theta _{\mathrm{o}%
}-\alpha \right) \sin \theta _{\mathrm{o}}~_{~_{~_{~_{~_{}}}}}^{~^{~^{~^{~^{}}}}}\right. 
\notag \\
& \left. +\hspace{1pt} 2\chi \frac{r_{\ast }}{r_{\mathrm{o}}}\left[ \left( 3\cos
^{2}\left( \theta _{\mathrm{o}}-\alpha \right) -1\right) \cos \theta _{%
\mathrm{o}}-2\cos \left( \theta _{\mathrm{o}}-\alpha \right) \sin \left(
\theta _{\mathrm{o}}-\alpha \right) \sin \theta _{\mathrm{o}}\right]
\right\} \notag \\  
&+\mathcal{O}\left( \phi^{2},\left( \Omega _{\mathrm{p}}r_{\mathrm{o}}/c\right) ^{2},
\phi \left(\Omega _{\mathrm{p}}r_{\mathrm{o}}/c\right) \right),  \label{noz2}
\end{align}%
and 
\begin{align}
\hat{n}_{y\left[ 0\right] \mathrm{qd}}\approx & \frac{\Omega _{\mathrm{p}}}{c}\left\{ %
\left[ f\left( \theta _{\mathrm{o}}-\alpha ;\chi \right) \right] ^{-1/2}\sin
\alpha \left( 1+4\chi \frac{r_{\ast }}{r_{\mathrm{o}}}\cos \left( \theta _{%
\mathrm{o}}-\alpha \right) \right) s+r_{\mathrm{o}}\sin \theta _{\mathrm{o}%
}\right\} \notag \\ 
&+\mathcal{O}\left( \phi^{2},\left( \Omega _{\mathrm{p}}r_{\mathrm{o}}/c\right) ^{2},
\phi \left(\Omega _{\mathrm{p}}r_{\mathrm{o}}/c\right) \right),   \label{noy2}
\end{align}%
with $f\left( \theta _{\mathrm{o}}-\alpha ;\chi \right)$ referring to Eq. (\ref{fth}) for $\theta=\theta _{\mathrm{o}}$ and $\phi=0$. 
In association with $\mathbf{\hat{n}}_{\left[ 0\right] \mathrm{qd}}$, the orthogonal pair of classical mode polarization vectors are determined as 
\begin{align}
\boldsymbol{\varepsilon }_{\mathrm{I}\left[ 0\right] \mathrm{qd}}&=\hat{n}_{z\left[ 0\right] \mathrm{qd}}\mathbf{e}_{x}
+\hat{n}_{y\left[ 0\right] \mathrm{qd}}\mathbf{e}_{y}-\hat{n}_{x\left[ 0\right] \mathrm{qd}}\mathbf{e}_{z},  \label{ei02} \\
\boldsymbol{\varepsilon }_{\mathrm{II}\left[ 0\right] \mathrm{qd}}&=-\left( \hat{n}_{x\left[0\right] \mathrm{qd}}+\hat{n}_{z\left[ 0\right] \mathrm{qd}}\right) 
\hat{n}_{y\left[ 0\right] \mathrm{qd}}\mathbf{e}_{x}+\mathbf{e}_{y}+\left( \hat{n}_{x\left[ 0\right] \mathrm{qd}}-\hat{n}_{z\left[ 0\right] \mathrm{qd}}\right) 
\hat{n}_{y\left[ 0\right] \mathrm{qd}}\mathbf{e}_{z}, \label{eii02}
\end{align}%
such that the three vectors, $\mathbf{\hat{n}}_{\left[ 0\right] \mathrm{qd}}$, $\boldsymbol{\varepsilon }_{\mathrm{I}\left[ 0\right] \mathrm{qd}}$ 
and $\boldsymbol{\varepsilon }_{\mathrm{II}\left[ 0\right] \mathrm{qd}}$ form an orthonormal basis. 

In addition, due to (\ref{Bqd}) and (\ref{no2}), the angle between the photon trajectory and the local magnetic field line, as defined by Eq. (\ref{th}) shall be modified. Then we have 
\begin{align}
\cos \vartheta_{\mathrm{qd}} &\approx \left[ f\left( \theta _{\mathrm{o}}-\alpha ;\chi \right) \right] ^{-1/2}
\left[ f\left( \theta -\alpha ;\chi \right) \right] ^{-1/2}  \notag \\
&\hspace{12pt} \times \left[ 4\cos \left( \theta _{\mathrm{o}}-\alpha \right) \cos \left(\theta -\alpha \right) 
+\sin \left( \theta _{\mathrm{o}}-\alpha \right) \sin \left( \theta -\alpha \right) +\chi g_{1}\left( r,\theta \right) +\chi^{2}g_{2}\left( r,\theta \right) \right] \notag \\
&\hspace{12pt}+\mathcal{O}\left( \phi^{2},\left( \Omega _{\mathrm{p}}r_{\mathrm{o}}/c\right) ^{2},\phi \left(\Omega _{\mathrm{p}}r_{\mathrm{o}}/c\right) \right) ,  \label{cos2} 
\end{align}%
where 
\begin{align}
g_{1}\left( r,\theta \right)  &=4\frac{r_{\ast }}{r_{\mathrm{o}}}\left[\left( 3\cos^{2}\left( \theta _{\mathrm{o}}-\alpha \right) -1\right) \cos \left( \theta -\alpha \right) 
+\cos \left( \theta _{\mathrm{o}}-\alpha\right) \sin \left( \theta _{\mathrm{o}}-\alpha \right) \sin \left( \theta-\alpha \right) \right]   \notag \\
&\hspace{12pt} +4\frac{r_{\ast }}{r}\left[ \left( 3\cos^{2}\left( \theta -\alpha \right) -1\right) \cos \left( \theta _{\mathrm{o}}-\alpha \right) 
+\cos \left( \theta -\alpha \right) \sin \left( \theta -\alpha \right) \sin \left( \theta_{\mathrm{o}}-\alpha \right) \right],   \label{g1} \\
g_{2}\left( r,\theta \right)  &=4\frac{r_{\ast }^{2}}{r_{\mathrm{o}}r}\left[\left(3\cos^{2}\left( \theta _{\mathrm{o}}-\alpha \right) -1\right) 
\left(3\cos ^{2}\left( \theta -\alpha \right) -1\right) \right.   \notag \\
&\hspace{6pt} \left. _{\,_{{}}}^{\,^{{}}}+4\cos \left( \theta _{\mathrm{o}}-\alpha \right) \sin \left( \theta _{\mathrm{o}}-\alpha \right) 
\cos \left( \theta-\alpha \right) \sin \left( \theta -\alpha \right) \right].  \label{g2} 
\end{align}%

Now, having $\mathbf{B}_{\mathrm{qd}}$, $\mathbf{\hat{n}}_{\left[ 0\right] \mathrm{qd}}$, $\boldsymbol{\varepsilon }_{\mathrm{I}\left[ 0\right] \mathrm{qd}}$, 
$\boldsymbol{\varepsilon }_{\mathrm{II}\left[ 0\right] \mathrm{qd}}$ and $\cos \vartheta_{\mathrm{qd}}$ at hand,
as given by Eqs. (\ref{Bqd}), (\ref{no2}), (\ref{ei02}), (\ref{eii02}) and (\ref{cos2}) above, respectively, we modify the evolution equations of the Stokes vector (\ref{ev}) to 
\begin{equation}
\frac{\mathrm{d}\mathbf{S}}{\mathrm{d}s}=k\mathbf{\hat{\Omega}}_{\mathrm{qd}}\times \mathbf{S},  \label{ev2}
\end{equation}%
where
\begin{equation}
\mathbf{\hat{\Omega}}_{\mathrm{qd}}\equiv \frac{\alpha_{\mathrm{e}}}{30\pi }\left( B_{\mathrm{qd}}/B_{\mathrm{c}}\right) ^{2}\sin ^{2}\vartheta_{\mathrm{qd}} 
\left( \mathcal{E}_{\mathrm{I} \mathrm{qd}}^{2}-\mathcal{E}_{\mathrm{II} \mathrm{qd}}^{2},
2\mathcal{E}_{\mathrm{I} \mathrm{qd}}\mathcal{E}_{\mathrm{II \mathrm{qd}}},0\right) . \label{br3}
\end{equation}%
Here $\sin \vartheta_{\mathrm{qd}}$ is defined via (\ref{cos2}), and
\begin{align}
&\mathcal{E}_{\mathrm{I} \mathrm{qd}}=-\mathbf{\hat{B}_{\mathrm{qd}}\cdot }
\left( \mathbf{\hat{n}}_{\left[ 0\right] \mathrm{qd}}\times \boldsymbol{\varepsilon }_{\mathrm{I}\left[ 0\right] \mathrm{qd}}\right)   \notag \\
&\approx \left[ f\left( \theta _{\mathrm{o}}-\alpha ;\chi \right) \right] ^{-1/2} 
\left[ f\left( \theta -\alpha ;\chi \right) \right] ^{-1/2}  \notag \\
&\hspace{8pt} \times \left\{ \left[ 2\cos \left( \theta _{\mathrm{o}}-\alpha \right)
+2\chi \frac{r_{\ast }}{r_{\mathrm{o}}}\left( 3\cos ^{2}\left( \theta _{\mathrm{o}}-\alpha \right) 
-1\right) \right] \right.  \notag \\
&\hspace{24pt} \times \left[ 2\cos \left( \theta-\alpha \right) +2\chi \frac{r_{\ast }}{r}
\left( 3\cos ^{2}\left( \theta-\alpha \right) -1\right) \right]   \notag \\
&\hspace{24pt} \left. +\sin \left( \theta _{\mathrm{o}}-\alpha \right) \left[ 1+4\chi 
\frac{r_{\ast }}{r_{\mathrm{o}}}\cos \left( \theta _{\mathrm{o}}-\alpha
\right) \right] \sin \left( \theta -\alpha \right) \left[ 1+4\chi \frac{r_{\ast }}{r}
\cos \left( \theta -\alpha \right) \right] \right\}\hat{n}_{y\left[ 0\right] \mathrm{qd}}   \notag \\
&\hspace{8pt} -\mathcal{E}_{\mathrm{II} \mathrm{qd}}\hat{n}_{y\left[ 0\right] \mathrm{qd}} - \frac{\Omega _{\mathrm{p}}\sin \alpha}{c} \left[ f\left( \theta -\alpha ;\chi \right) \right] ^{-1/2}
\left[ 1+4\chi \frac{r_{\ast }}{r}\cos \left( \theta -\alpha \right) \right]s  \notag  \\
&\hspace{8pt}+\mathcal{O}\left( \phi ^{2},\left( \Omega_{\mathrm{p}}r_{\mathrm{o}}/c\right) ^{2},\phi \left( \Omega _{\mathrm{p}}r_{\mathrm{o}}/c\right) \right), \label{EI2} 
\end{align}
\begin{align}
&\mathcal{E}_{\mathrm{II} \mathrm{qd}}=-\mathbf{\hat{B}_{\mathrm{qd}}\cdot }
\left( \mathbf{\hat{n}}_{\left[ 0\right] \mathrm{qd}}\times \boldsymbol{\varepsilon }_{\mathrm{II}\left[ 0\right] \mathrm{qd}}\right)   \notag \\
&\approx \left[ f\left( \theta _{\mathrm{o}}-\alpha ;\chi \right) \right] ^{-1/2} 
\left[ f\left( \theta -\alpha ;\chi \right) \right] ^{-1/2}  \notag \\
&\hspace{8pt} \times \left\{ \left[ 2\cos \left( \theta -\alpha \right) +2\chi \frac{r_{\ast }}{r} 
\left( 3\cos ^{2}\left( \theta -\alpha \right) -1\right) \right]
\sin \left( \theta _{\mathrm{o}}-\alpha \right) \left[ 1+4\chi \frac{r_{\ast}}{r_{\mathrm{o}}}
\cos \left( \theta _{\mathrm{o}}-\alpha \right) \right] \right.   \notag \\
&\hspace{24pt} \left. -\left[ 2\cos \left( \theta _{\mathrm{o}}-\alpha \right) +2\chi 
\frac{r_{\ast }}{r_{\mathrm{o}}}\left( 3\cos ^{2}\left( \theta _{\mathrm{o}}-\alpha \right) -1\right) 
\right] \sin \left( \theta -\alpha \right) \left[1+4\chi \frac{r_{\ast }}{r}\cos \left( \theta -\alpha \right) \right] \right\}  \notag \\
&\hspace{8pt}+\mathcal{O}\left( \phi ^{2},\left( \Omega_{\mathrm{p}}r_{\mathrm{o}}/c\right) ^{2},\phi \left( \Omega _{\mathrm{p}}r_{\mathrm{o}}/c\right) \right),  \label{EII2}
\end{align}%
with $f\left( \theta -\alpha ;\chi \right)$ referring to Eq. (\ref{fth}) for $\phi=0$. 

\subsection{Solving the evolution equations}
\label{soleveq2}

From Eq. (\ref{br3}) one can write out the non-zero components of the birefringent vector:
\begin{align}
\hat{\Omega}_{1 \mathrm{qd}} &\approx -\eta B^{2}_{\mathrm{qd}} \sin^{2}\vartheta_{\mathrm{qd}} \mathcal{E}^2_{\mathrm{II} \mathrm{qd}} 
+\mathcal{O}\left( \phi ^{2},\left( \Omega _{\mathrm{p}}r_{\mathrm{o}}/c\right) ^{2},\phi \left( \Omega _{\mathrm{p}}r_{\mathrm{o}}/c\right) \right) ,  \label{Om1} \\
\hat{\Omega}_{2 \mathrm{qd}} &\approx 2\eta B^{2}_{\mathrm{qd}} \sin^{2}\vartheta_{\mathrm{qd}} \mathcal{E}_{\mathrm{I} \mathrm{qd}}\mathcal{E}_{\mathrm{II} \mathrm{qd}} 
+\mathcal{O}\left( \phi ^{2},\left( \Omega _{\mathrm{p}}r_{\mathrm{o}}/c\right) ^{2},\phi \left( \Omega _{\mathrm{p}}r_{\mathrm{o}}/c\right) \right) ,  \label{Om2} 
\end{align}%
where $\eta \equiv \alpha_\mathrm{e} /\left( 30\pi B_{\mathrm{c}}^{2}\right) $, and $\mathcal{E}_{\mathrm{I} \mathrm{qd}}$ 
and $\mathcal{E}_{\mathrm{II} \mathrm{qd}}$ refer to Eqs. (\ref{EI2}) and (\ref{EII2}), respectively, and $\sin \vartheta_{\mathrm{qd}}$ is defined via Eq. (\ref{cos2}), while
\begin{equation}
B_{\mathrm{qd}}=\frac{B_{\mathrm{max}}r_{*}^{3}\, \left[ f\left( \theta -\alpha ;\chi \right) \right]^{1/2}}{2\left( 1+2\chi \right) \left( x^{2}+z^{2}\right) ^{3/2}} \label{Bqd1}
\end{equation}
due to Eq. (\ref{Bqd}), with $B_{\mathrm{max}}$ being the maximum magnetic field intensity at the polar cap, $r_{*}$ being the neutron star radius and 
$f\left( \theta -\alpha ;\chi \right)$ referring to Eq. (\ref{fth}) for $\phi=0$. Having Eqs. (\ref{Om1}) and (\ref{Om2}) at hand, we solve a system of differential equations, i.e.,
the evolution equations of the Stokes vector, written out in component form from (\ref{ev2}):
\begin{align}
\dot{S}_{1}\left( s\right)  &=k\hat{\Omega}_{2 \mathrm{qd}}\left( s\right) S_{3}\left(
s\right) ,  \label{qdode1} \\
\dot{S}_{2}\left( s\right)  &=-k\hat{\Omega}_{1 \mathrm{qd}}\left( s\right) S_{3}\left(
s\right) ,  \label{qdode2} \\
\dot{S}_{3}\left( s\right)  &=k\left[ \hat{\Omega}_{1 \mathrm{qd}}\left( s\right)
S_{2}\left( s\right) -\hat{\Omega}_{2 \mathrm{qd}}\left( s\right) S_{1}\left( s\right) %
\right] .  \label{qdode3}
\end{align}%

\subsubsection{Examples}
\label{ex2}
We consider again X-ray emissions from the same three RPPs as in Sect. \ref{ex}: ($\mathrm{i}^\prime$) one with $B_{\mathrm{max}}\approx10^{12}\,\mathrm{G}$ 
and $\Omega _{\mathrm{p}}\approx392.7\,\mathrm{Hz}$ ($P\approx0.016\,\mathrm{s}$), 
($\mathrm{ii}^\prime$) another with $B_{\mathrm{max}}\approx5.6\times10^{12}\,\mathrm{G}$ and $\Omega _{\mathrm{p}}\approx22.28\,\mathrm{Hz}$ 
($P\approx0.282\,\mathrm{s}$), ($\mathrm{iii}^\prime$) the third with $B_{\mathrm{max}}\approx5.0\times10^{13}\,\mathrm{G}$ and 
$\Omega _{\mathrm{p}}\approx19.6\,\mathrm{Hz}$ ($P\approx0.32\,\mathrm{s}$). Again, for all three, we assume
$r_{\mathrm{o}}=2r_{*}\approx 2\times 10^{6}\,\mathrm{cm}$, $\theta _{\mathrm{o}}=60^{\circ }$, $\alpha=45^{\circ }$, 
$\eta\approx3.97\times 10^{-32}$ and $\omega\approx2\pi\times 10^{18}\,\mathrm{Hz}$ ($k\approx2.0958\times10^{8}\,\mathrm{cm^{-1}}$).  
However, unlike the dipole pulsars as in Sect. \ref{ex}, quadrudipole pulsars can be modeled by setting the value of an arbitrary parameter $\chi$ to determine 
the ratio between the magnetic dipole and quadrupole moments; we choose two values, $\chi=1.585$ and $0.85$ to model each RPP in our analysis here.\footnote
{The values $\chi=1.585$ and $0.85$ have been determined such that they optimize $\hat{\Omega}_{1 \mathrm{qd}} \left( s;\chi \right) \le 0$ and 
$\hat{\Omega}_{2 \mathrm{qd}} \left( s;\chi \right) \ge 0$, respectively. That is to say, the optimal values of $\hat{\Omega}_{1 \mathrm{qd}\left(\mathrm{min} \right)}$ and 
$\hat{\Omega}_{2 \mathrm{qd}\left(\mathrm{max} \right)}$ are found at $\left(s\approx8.2444\times 10^5\,\mathrm{cm};\chi\approx1.585 \right)$ and 
$\left(s\approx6.3588\times 10^5\,\mathrm{cm};\chi\approx0.85 \right)$, respectively, which are determined from
$\partial \hat{\Omega}_{1 \mathrm{qd}} \left( s; \chi \right) / \partial s = \partial \hat{\Omega}_{1 \mathrm{qd}} \left( s; \chi \right) / \partial \chi =0$ and 
$\partial \hat{\Omega}_{2 \mathrm{qd}} \left( s; \chi \right) / \partial s = \partial \hat{\Omega}_{2 \mathrm{qd}} \left( s; \chi \right) / \partial \chi =0$, respectively. 
In consideration of Eqs. (\ref{qdode1})-(\ref{qdode3}), these values will maximize the effects of birefringence on our evolution system.}

\begin{figure*}[!ht]
\centering

\subfloat[With $\mathbf{S}\left(0\right)=\left(1,0,0\right)$ and $\chi=1.585$]
{\includegraphics[clip,width=0.86\textwidth]{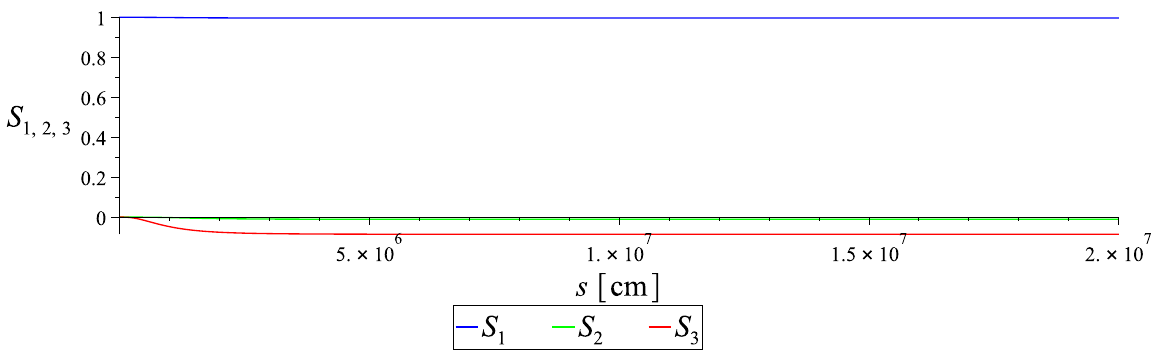}
\label{fig:subfig16}}

\subfloat[With $\mathbf{S}\left(0\right)=\left(1,0,0\right)$ and $\chi=0.85$]
{\includegraphics[clip,width=0.86\textwidth]{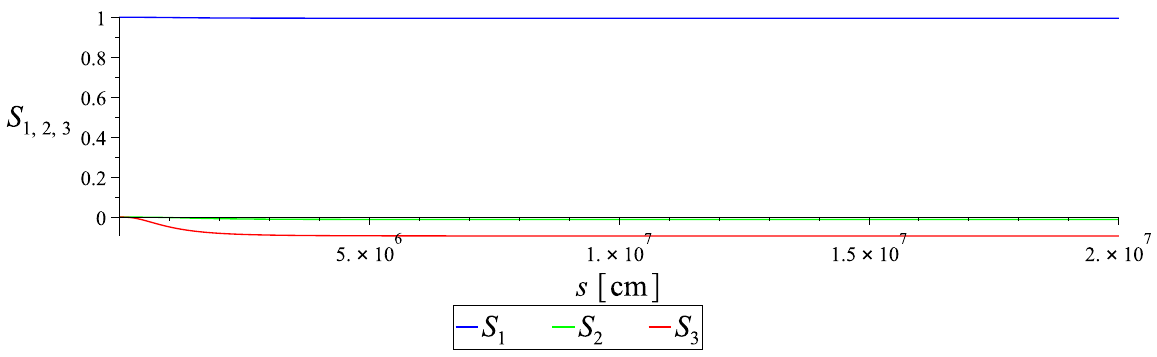}
\label{fig:subfig17}}

\subfloat[With $\mathbf{S}\left(0\right)=\left(0.8,0,0.6\right)$ and $\chi=1.585$]
{\includegraphics[clip,width=0.86\textwidth]{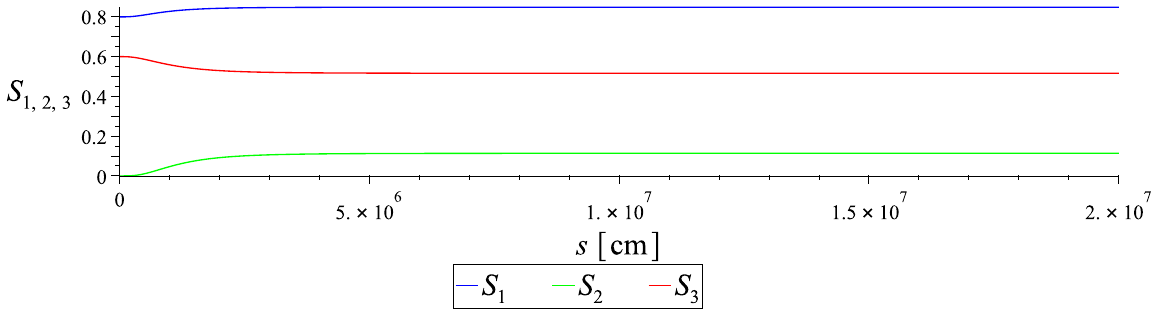}
\label{fig:subfig18}}

\subfloat[With $\mathbf{S}\left(0\right)=\left(0.8,0,0.6\right)$ and $\chi=0.85$]
{\includegraphics[clip,width=0.86\textwidth]{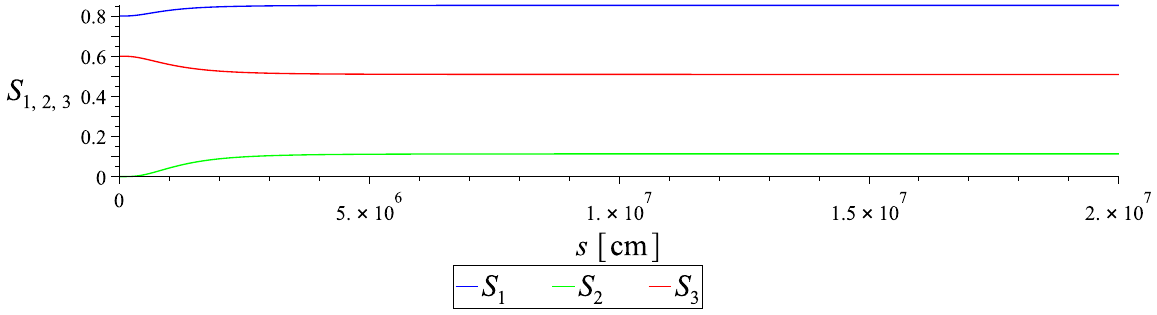}
\label{fig:subfig19}}

\caption{For Example ($\mathrm{i}^\prime$): the evolution of the Stokes vector $\mathbf{S}\left(s\right)=
\left(S_{1}\left(s\right),S_{2}\left(s\right),S_{3}\left(s\right)\right)$, 
$0\le s \le 20r_{*}\left(\approx 2\times10^{7}\,\mathrm{cm}\right)$, for the X-ray emissions from the quadrudipole pulsar 
with $B_{\mathrm{max}}\approx10^{12}\,\mathrm{G}$ and $\Omega _{\mathrm{p}}\approx392.7\,\mathrm{Hz}$ ($P\approx0.016\,\mathrm{s}$).}
\label{fig10}
\end{figure*}

\begin{figure*}[!ht]
\centering

\subfloat[With $\mathbf{S}\left(0\right)=\left(1,0,0\right)$ and $\chi=1.585$]
{\includegraphics[clip,width=0.86\textwidth]{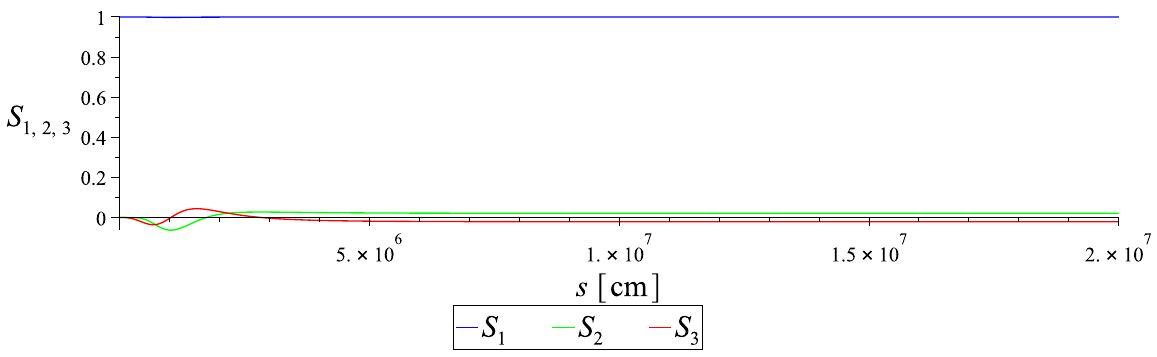}
\label{fig:subfig20}}

\subfloat[With $\mathbf{S}\left(0\right)=\left(1,0,0\right)$ and $\chi=0.85$]
{\includegraphics[clip,width=0.86\textwidth]{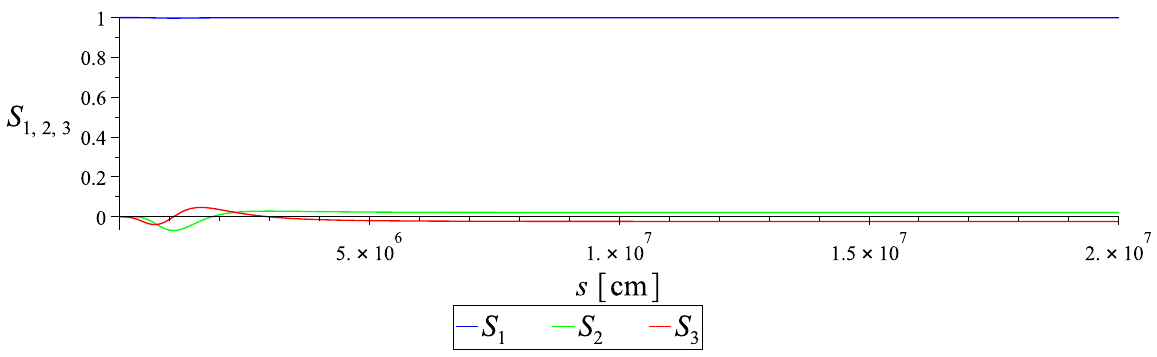}
\label{fig:subfig21}}

\subfloat[With $\mathbf{S}\left(0\right)=\left(0.8,0,0.6\right)$ and $\chi=1.585$]
{\includegraphics[clip,width=0.87\textwidth]{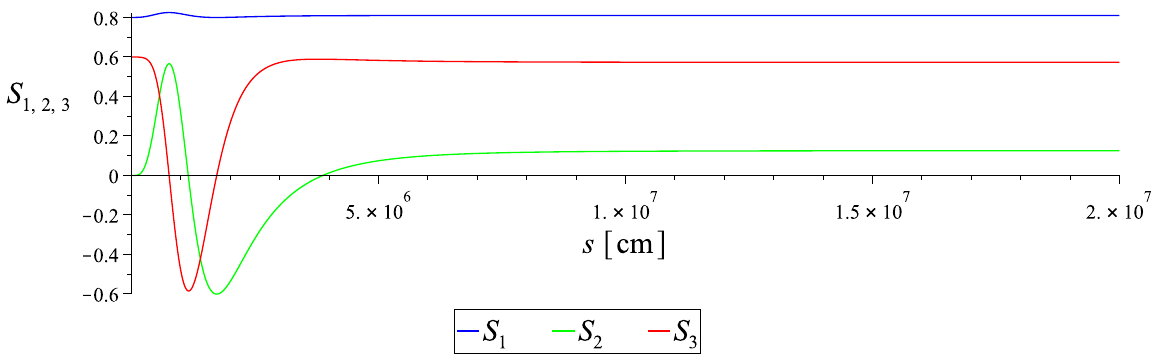}
\label{fig:subfig22}}

\subfloat[With $\mathbf{S}\left(0\right)=\left(0.8,0,0.6\right)$ and $\chi=0.85$]
{\includegraphics[clip,width=0.87\textwidth]{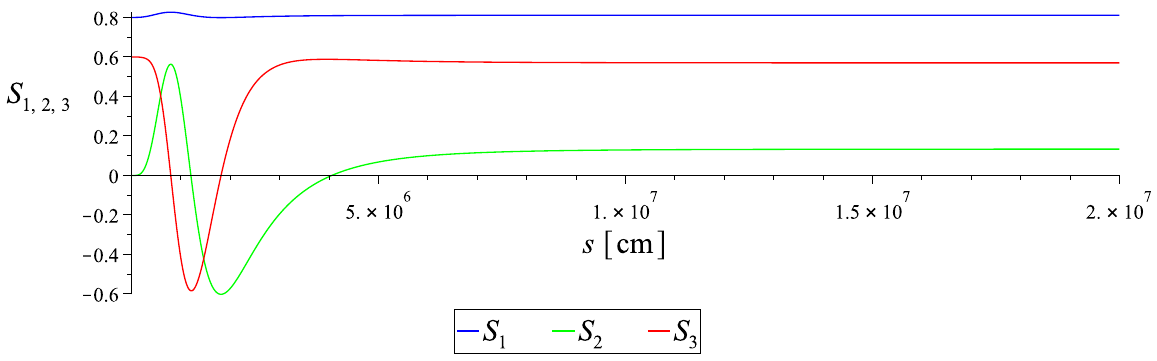}
\label{fig:subfig23}}

\caption{For Example ($\mathrm{ii}^\prime$): the evolution of the Stokes vector $\mathbf{S}\left(s\right)=
\left(S_{1}\left(s\right),S_{2}\left(s\right),S_{3}\left(s\right)\right)$, 
$0\le s \le 20r_{*}\left(\approx 2\times10^{7}\,\mathrm{cm}\right)$, for the X-ray emissions from the quadrudipole pulsar 
with $B_{\mathrm{max}}\approx5.6\times10^{12}\,\mathrm{G}$ and $\Omega _{\mathrm{p}}\approx22.28\,\mathrm{Hz}$ 
($P\approx0.282\,\mathrm{s}$).}
\label{fig11}
\end{figure*}

\begin{figure*}[!ht]
\centering

\subfloat[With $\mathbf{S}\left(0\right)=\left(1,0,0\right)$ and $\chi=1.585$]
{\includegraphics[clip,width=0.86\textwidth]{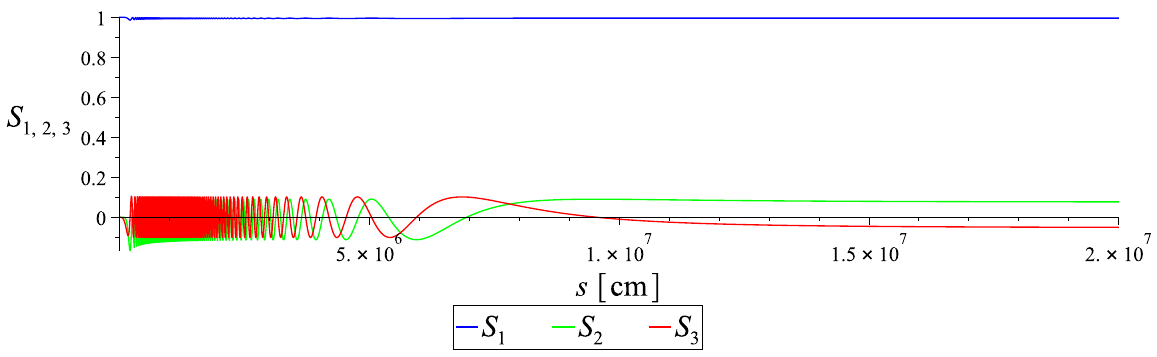}
\label{fig:subfig24}}

\subfloat[With $\mathbf{S}\left(0\right)=\left(1,0,0\right)$ and $\chi=0.85$]
{\includegraphics[clip,width=0.86\textwidth]{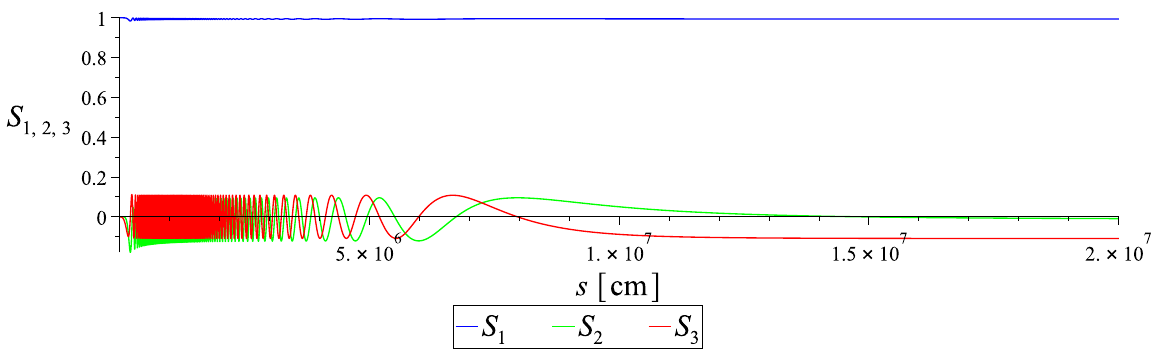}
\label{fig:subfig25}}

\subfloat[With $\mathbf{S}\left(0\right)=\left(0.8,0,0.6\right)$ and $\chi=1.585$]
{\includegraphics[clip,width=0.87\textwidth]{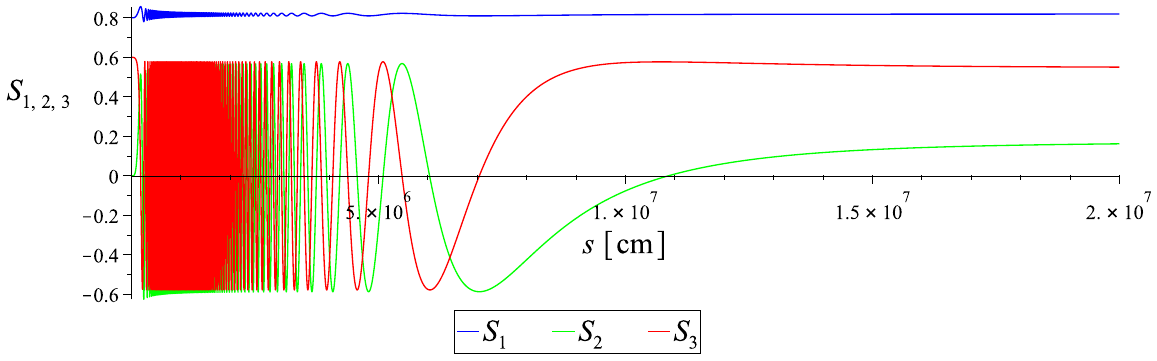}
\label{fig:subfig26}}

\subfloat[With $\mathbf{S}\left(0\right)=\left(0.8,0,0.6\right)$ and $\chi=0.85$]
{\includegraphics[clip,width=0.87\textwidth]{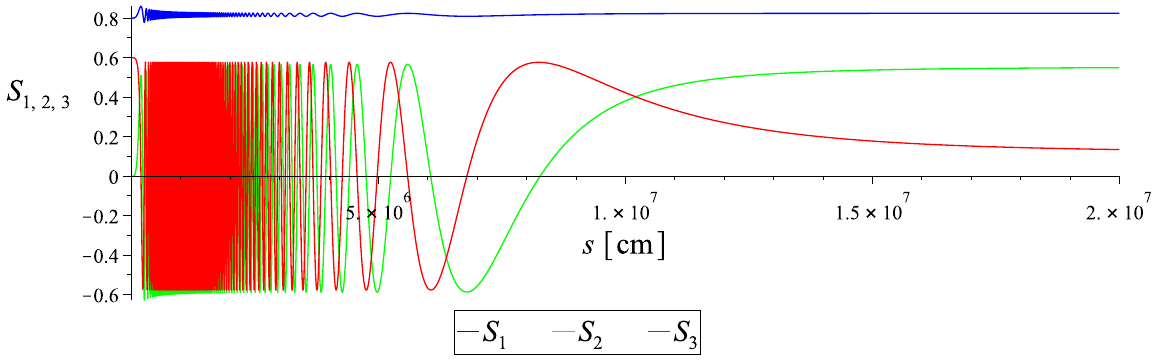}
\label{fig:subfig27}}

\caption{For Example ($\mathrm{iii}^\prime$): the evolution of the Stokes vector $\mathbf{S}\left(s\right)=
\left(S_{1}\left(s\right),S_{2}\left(s\right),S_{3}\left(s\right)\right)$, 
$0\le s \le 20r_{*}\left(\approx 2\times10^{7}\,\mathrm{cm}\right)$, for the X-ray emissions from the quadrudipole pulsar 
with $B_{\mathrm{max}}\approx5.0\times10^{13}\,\mathrm{G}$ and 
$\Omega _{\mathrm{p}}\approx19.6\,\mathrm{Hz}$ ($P\approx0.32\,\mathrm{s}$).}
\label{fig12}
\end{figure*}

\begin{figure*}[!t]
\centering
\def\twidth{0.33}
\centering\captionsetup{width=.33\textwidth}%
\subfloat[Example ($\mathrm{i}^\prime$)]{%
		\includegraphics[angle=0, width=3.9cm]{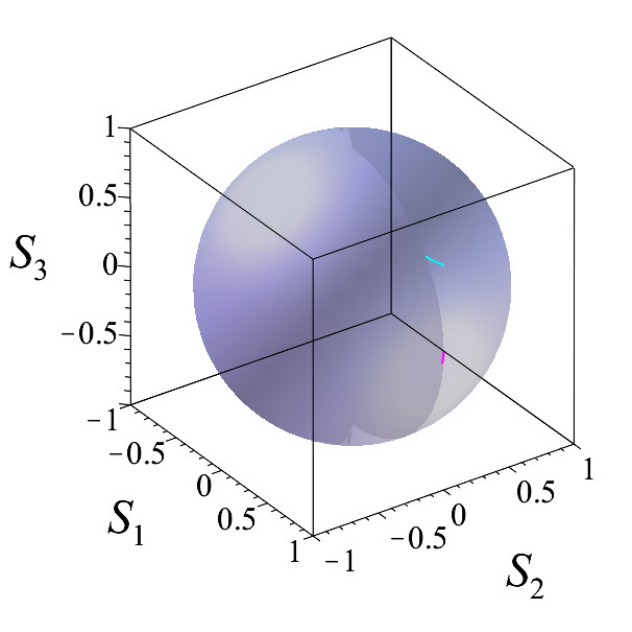}%
		\label{fig:subfig28}}\hfill
\centering\captionsetup{width=.33\textwidth}%
\subfloat[Example ($\mathrm{ii}^\prime$)]{%
		\includegraphics[angle=0, width=3.9cm]{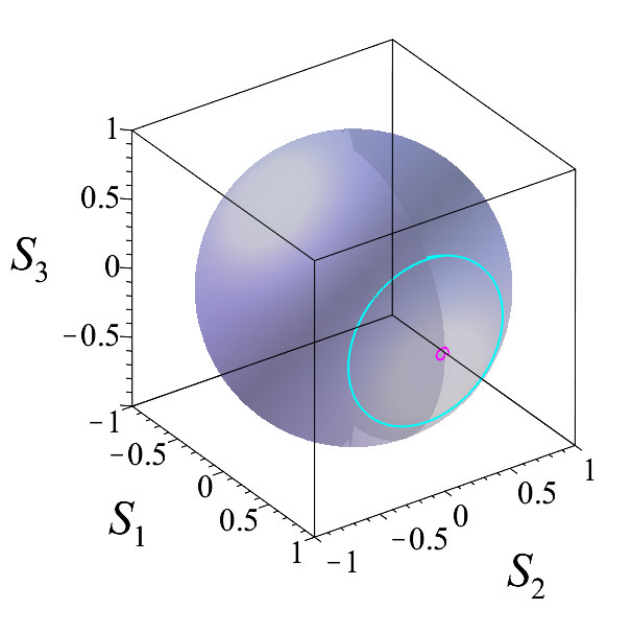}%
		\label{fig:subfig29}}\hfill
\centering\captionsetup{width=.33\textwidth}%
\subfloat[Example ($\mathrm{iii}^\prime$)]{%
		\includegraphics[angle=0,width=3.9cm]{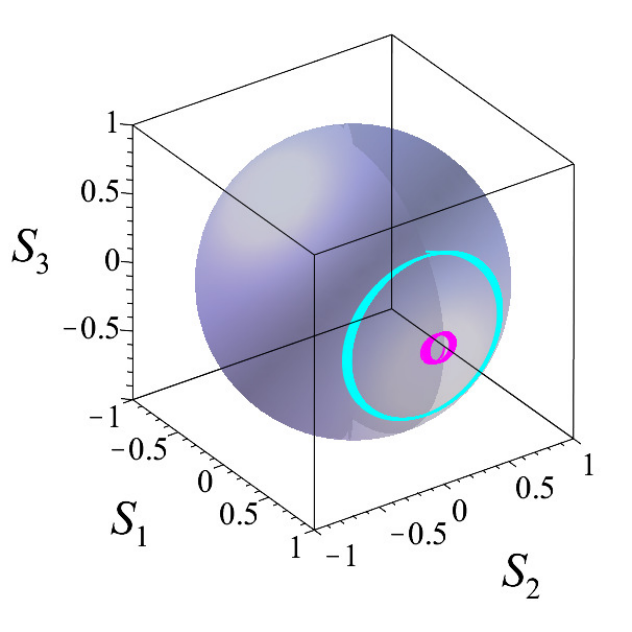}%
		\label{fig:subfig30}} 
\centering\captionsetup{width=\textwidth}%
\caption{Representations of the Stokes vectors from Examples ($\mathrm{i}^\prime$)-($\mathrm{iii}^\prime$) for $\chi=1.585$ on the Poincar\'{e} sphere. 
The loci imply patterns of the polarization evolution in terms of oscillation.}
\label{fig13}
\end{figure*}

In Figs. \ref{fig10}, \ref{fig11} and \ref{fig12} are plotted our numerical solutions of Eqs. (\ref{qdode1})-(\ref{qdode3}) for the Stokes vectors in Examples ($\mathrm{i}^\prime$), 
($\mathrm{ii}^\prime$) and ($\mathrm{iii}^\prime$), respectively. Also, in Fig. \ref{fig13} we present the solutions as represented on the Poincar\'{e} sphere for $\chi=1.585$ only; 
there is little difference in the representations of solutions between $\chi=1.585$ and $0.85$ cases. The solutions represented by the three magenta loci and the three light blue loci 
in Fig. \ref{fig13} correspond to Figs. \ref{fig:subfig16}, \ref{fig:subfig20}, \ref{fig:subfig24} and Figs. \ref{fig:subfig18}, \ref{fig:subfig22}, \ref{fig:subfig26}, respectively. 
In Fig. \ref{fig10} for Example ($\mathrm{i}^\prime$), where $B_{\mathrm{max}}$ is relatively weak among the three RPPs, general patterns of polarization evolution are shown 
to be nearly the same for $\chi=1.585$ and $0.85$, and to be almost the same even to the dipole case as given by Fig. \ref{fig3} for Example (i) in Sect. \ref{ex}; it is also 
confirmed by comparing the representations in Figs. \ref{fig:subfig7} and \ref{fig:subfig28}. In contrast, in Fig. \ref{fig11} for Example ($\mathrm{ii}^\prime$), where 
$B_{\mathrm{max}}$ is intermediate among the three RPPs, general patterns of polarization evolution are shown to be nearly the same for $\chi=1.585$ and $0.85$, 
but to be noticeably different from the dipole case as given by Fig. \ref{fig4} for Example (ii) in Sect. \ref{ex}; comparing the representations in Figs. \ref{fig:subfig8} 
and \ref{fig:subfig29}, the number of cycles appears to become nearly doubled (from half a cycle to one cycle). However, in Fig. \ref{fig12} for Example ($\mathrm{iii}^\prime$), 
where $B_{\mathrm{max}}$ is the strongest among the three RPPs, general patterns of polarization evolution are shown to be similar for $\chi=1.585$ and $0.85$, 
with noticeable differences in phase, but to be significantly different from the dipole case as given by Fig. \ref{fig5} for Example (iii) in Sect. \ref{ex}; the plots 
appear to be much denser in the early part of evolution in Fig. \ref{fig12} than in Fig. \ref{fig5}, as the polarization states oscillate much more frequently in the quadrudipole field 
than in the dipole field, which can also be confirmed by comparing the representations in Figs. \ref{fig:subfig9} and \ref{fig:subfig30}.

\section{Conclusions and discussion}
\label{conclusions}
We have investigated the evolution of polarization states of pulsar emission under the quantum refraction effects, combined with the dependence on the emission frequency, 
for both dipole and quadrudipole pulsar models. To this end, we have solved a system of evolution equations of the Stokes vector given by (\ref{ev}) (or by (\ref{ode1})-(\ref{ode3})) 
and (\ref{ev2}) (or by (\ref{qdode1})-(\ref{qdode3})) in the dipole and quadrudipole cases, respectively, for three examples of RPPs at a fixed frequency for specific emissions 
(e.g., X-rays as in sections \ref{ex}, \ref{fit} and \ref{ex2}). Our main results are presented by Figs. \ref{fig3}-\ref{fig5} and \ref{fig10}-\ref{fig12} in the dipole and quadrudipole cases, 
respectively, from numerical solutions and in part from perturbative solutions. Also, we have replaced the birefringent vector with some approximate models as in Fig. \ref{fig7} 
to solve the evolution equations analytically in the dipole case, and obtained the results as presented by Fig. \ref{fig8}.
It is noteworthy that at a fixed frequency of emission the evolution of the Stokes vector largely exhibits three different patterns, depending on the magnitudes of the birefringent vector, 
in which the magnetic field strength is a dominant factor: (i) fractionally oscillatory - monotonic, or (ii) half-oscillatory, or (iii) highly oscillatory behaviors. 
These features are shown by the numerical solutions in Figs. \ref{fig3}-\ref{fig6}, and also confirmed by the approximate analytical solutions in Fig. \ref{fig8}. 
In addition, we have examined how the aforementioned features change in the quadrudipole case.

This study is centered on solving the evolution equations for polarization states (\ref{ev}), wherein the birefringent vector that contains all information about the quantum refraction effects, 
coupled to the frequency of pulsar emission, acts on the Stokes vector; the evolution results from the combination of the quantum refraction effects and the frequency dependence 
of the emission. This is a major difference from our previous work \cite{10.1093/mnras/stae1304}, wherein the same effects have no connection to the emission frequency; the work solely 
focuses on the quantum refraction effects on the propagation and polarization vectors in pulsar emission, with no reference to other properties, such as the emission frequency. 
In this regard, it is worthwhile to draw comparison between the two quantities, the polarization vector and the Stokes vector, both of which are used to describe polarization states. 
The polarization vector is defined directly from the radiative electric field vector (i.e., the unit electric field vector), and it is parallel-transported along the the propagation vector; 
usually, we consider such two vectors orthogonal to each other and to the propagation vector to define an orthonormal basis consisting of the three vectors. 
In contrast, the Stokes vector is defined from Stokes parameters which are built out of the radiative electric field vector \cite{Rybicki:847173}. The representation of the Stokes vector is abstract 
in the sense that it is a vector defined on the Poincar\'{e} sphere. The Stokes vector is not parallel-transported along the the propagation vector, but can still be defined along the propagation vector 
as the two polarization vectors move along it; hence, it can be parameterized by $s$ to represent polarization states along the photon trajectory. However, the Stokes vector has a crucial advantage 
over the polarization vector in representing polarization states in some astrophysical studies like this: it can be directly estimated from polarimetric measurements and accommodate depolarization 
effects due to incomplete coherence and random processes during the photon propagation \cite{Scully1997}. 

Our results in this study may be of some significance for the currently operating and planned X-ray space telescopes: Imaging X-ray Polarimetry Explorer (IXPE) \cite{IXPE}, X-ray Polarimeter Satellite (XPoSat) 
\cite{XPoSat}, the enhanced X-ray Timing and Polarimetry mission (eXTP) \cite{Santangelo_2019} and the Compton Telescope project \cite{Wadiasingh2019Magnetars}. 
These telescopes measure the polarization of the X-rays from energetic compact objects such as magnetars and black holes to unveil their geometry and physical environment in detail. 
Several magnetars observed by IXPE have been estimated to have overcritical field strengths \cite{Taverna2022Polarized, Turolla2024IXPE}. Furthermore, a recent measurement of the X-ray 
polarization of the magnetar 4U 0142+61 has shown that the polarization degree and angle change as a function of X-ray energy, the interpretation of which has led to two competing scenarios about 
the X-ray emission of the magnetar \cite{Taverna2022Polarized,Lai2023IXPE}. In fact, a full analysis of the polarimetry data would require a physical model that comprehensively incorporates the properties 
regarding the polarization of surface emission, the photon propagation through magnetized plasmas, birefringence due to a magnetized quantum vacuum, and gravitational effects on photon propagation 
\cite{Taverna2015Polarization,Caiazzo2022Probing}. 

In this study, we have focused on vacuum birefringence as it is one of the most significant phenomenological issues to be tested by the X-ray polarimetry in practice. When the magnetic field is 
sufficiently strong and slowly varying, the polarization states evolve due to vacuum birefringence; that is, the Stokes vector components change during the photon propagation within the so-called 
polarization-limiting radius \cite{Heyl2000,Heyl2003highenergy}, which can be several to a couple dozen neutron-star radii, depending on the magnetic field strength at the surface of a neutron star 
and the emission frequency. However, beyond the polarization-limiting radius, the polarization states `freeze’, remaining the same until finally being observed through polarimetry 
\cite{Heyl2003highenergy,Caiazzo2022Probing}. The new features of the polarization evolution presented in our study, such as the three different oscillatory patterns of the Stokes vectors and the effects 
of the possible contributions from the multipolar components may all closely concern the observation through the X-ray polarimetry, and therefore should be taken into proper consideration for a more 
accurate model for pulsar emission.

Effects of gravitation have not been considered in this study. However, close to the neutron star, where gravitation due to the neutron star mass may not be negligible, 
its effects must be taken into account in our analysis. Then, basically, the following shall be redefined in curved spacetime:
(1) the QED one-loop effective Lagrangian, 
(2) the refractive index for the photon propagation, 
(3) the magnetic field geometry in the magnetosphere, 
(4) the radiative electric field due to a charge moving along a magnetic field line,  
(5) the photon trajectory.
All these have not been rigorously dealt with in previous studies. In this regard, inclusion of the gravitational effects will involve non-trivial and immense analyses, 
and therefore shall be conducted for a long-term plan in our future studies.

\backmatter

\bmhead{Acknowledgments}
D.-H.K was supported by the Basic Science Research Program through the National Research Foundation
of Korea (NRF) funded by the Ministry of Education (NRF-2021R1I1A1A01054781). C.M.K. was supported by Ultrashort Quantum
Beam Facility operation program (140011) through APRI, GIST and GIST Research Institute (GRI) grant funded by GIST. 
S.P.K. was also in part supported by National Research Foundation of Korea (NRF) funded by the Ministry of Education (NRF-2019R1I1A3A01063183).

\begin{appendices}

\section{The classical Stokes vector}
\label{appA}

Consider a particle with a charge $q$ moving along a curved trajectory (a magnetic field line). 
Then the curvature radiation due to this can be expressed by the electric field: 
\begin{equation}
\mathbf{E}\left( t\right) =\frac{q}{c\left\vert \mathbf{r}-\boldsymbol{\xi }%
\left( t_{\mathrm{ret}}\right) \right\vert }\frac{\mathbf{n\times }\left[
\left( \mathbf{n-}\frac{\dot{\boldsymbol{\xi}}\left( t_{\mathrm{ret}}\right) }{c}%
\right) \times \frac{\ddot{\boldsymbol{\xi}}\left( t_{\mathrm{ret}}\right) }{c}%
\right] }{\left( 1-\frac{\dot{\boldsymbol{\xi}}\left( t_{\mathrm{ret}}\right) }{c%
}\cdot \mathbf{n}\right) ^{3}}, \label{cre}
\end{equation}
where $t_{\mathrm{ret}}\equiv t-r/c$ is the retarded time, $\boldsymbol{\xi}$ represents the particle's trajectory, 
$\mathbf{n}$ is the propagation direction of the radiation, and an over-dot ${\dot{}}$ denotes differentiation with respect to $t$.    
In a suitably chosen Cartesian frame, by setting $\boldsymbol{\xi}\left( t_{\mathrm{ret}}\right)
=\rho\left(\sin\left(\beta c t_{\mathrm{ret}}/\rho \right),0,\cos\left(\beta c t_{\mathrm{ret}}/\rho \right) \right)$, 
with $\rho$ being the radius of curvature of the particle's trajectory, and $\mathbf{n}=\left(\cos\varphi,\sin\varphi,0 \right)$, 
with $\varphi$ being the angle measured from the $x$-axis to the plane of the particle's motion, 
we can construct a simple toy model for pulse profiles of pulsar curvature emission as described below \cite{Kim2021Generala}.

One can express Stokes parameters out of the radiation field (\ref{cre}), which describe its polarization properties \cite{Gil1990Curvature}: 
\begin{align}
&I =\tilde{E}_{\parallel }^{\ast }\tilde{E}_{\parallel }+\tilde{E}_{\perp
}^{\ast }\tilde{E}_{\perp }  \notag \\
& =\mathcal{E}_{\mathrm{o}}^{2}\omega ^{2}\left[ \left( \delta ^{2}+\varphi
^{2}\right) ^{2}\mathrm{K}_{2/3}^{2}\left( \frac{\rho \omega }{3\beta c}%
\left( \delta ^{2}+\varphi ^{2}\right) ^{3/2}\right) +\,\varphi ^{2}\left(
\delta ^{2}+\varphi ^{2}\right) \mathrm{K}_{1/3}^{2}\left( \frac{\rho \omega 
}{3\beta c}\left( \delta ^{2}+\varphi ^{2}\right) ^{3/2}\right) \right]\!,
\label{pp1} \\
&Q =\tilde{E}_{\parallel }^{\ast }\tilde{E}_{\parallel }-\tilde{E}_{\perp
}^{\ast }\tilde{E}_{\perp }  \notag \\
& =\mathcal{E}_{\mathrm{o}}^{2}\omega ^{2}\left[ \left( \delta ^{2}+\varphi
^{2}\right) ^{2}\mathrm{K}_{2/3}^{2}\left( \frac{\rho \omega }{3\beta c}%
\left( \delta ^{2}+\varphi ^{2}\right) ^{3/2}\right) -\,\varphi ^{2}\left(
\delta ^{2}+\varphi ^{2}\right) \mathrm{K}_{1/3}^{2}\left( \frac{\rho \omega 
}{3\beta c}\left( \delta ^{2}+\varphi ^{2}\right) ^{3/2}\right) \right]\!,
\label{pp2} \\
&U =\tilde{E}_{\parallel }^{\ast }\tilde{E}_{\perp }+\tilde{E}_{\parallel }%
\tilde{E}_{\perp }^{\ast }=0,  \label{pp3} \\
&V =-\mathrm{i}\left( \tilde{E}_{\parallel }^{\ast }\tilde{E}_{\perp }-%
\tilde{E}_{\parallel }\tilde{E}_{\perp }^{\ast }\right)   \notag \\
& =-2\mathcal{E}_{\mathrm{o}}^{2}\omega ^{2}\varphi \left( \delta
^{2}+\varphi ^{2}\right) ^{3/2}\mathrm{K}_{2/3}\left( \frac{\rho \omega }{%
3\beta c}\left( \delta ^{2}+\varphi ^{2}\right) ^{3/2}\right) \mathrm{K}%
_{1/3}\left( \frac{\rho \omega }{3\beta c}\left( \delta ^{2}+\varphi
^{2}\right) ^{3/2}\right)\!,  \label{pp4}
\end{align}%
where $\tilde{E}_{\parallel }$ and $\tilde{E}_{\perp }$ denote the components of the Fourier
transform $\mathbf{\tilde{E}}\left( \omega \right) =\tilde{E}_{\parallel }\left( \omega \right) \mathbf{e}_{z}
+\tilde{E}_{\perp }\left( \omega \right) \,\mathbf{e}_{y} =\int_{-\infty }^{\infty }\mathbf{E}\left( t\right) \exp \left( \mathrm{i}\omega t\right) \,\mathrm{d}t$,
expressed in the Cartesian frame, and $^{\ast }$ means the complex conjugate, and $\mathcal{E}_{\mathrm{o}}
=q\beta/\left( 2\sqrt{3}\pi ^{2}r \rho \right) $, and $\delta \equiv \gamma ^{-1}=\left( 1-\beta^{2}\right) ^{1/2}\ll 1$ 
is the half-angle of the beam emission, and $\mathrm{K}_{1/3}$ and $\mathrm{K}_{2/3}$ denote the modified Bessel
functions of the second kind. With regard to the polarization state of the radiation field, $I$ is a measure of the total intensity, 
$Q$ and $U$ jointly describe the linear polarization, and $V$ describes the circular polarization. These parameters can be plotted 
as functions of the phase angle $\varphi$, where $\varphi\le\delta\ll5^{\circ}$ usually, to simulate the pulse profiles of pulsar emission theoretically. 

\begin{figure*}
\centering
\includegraphics[width=12cm]{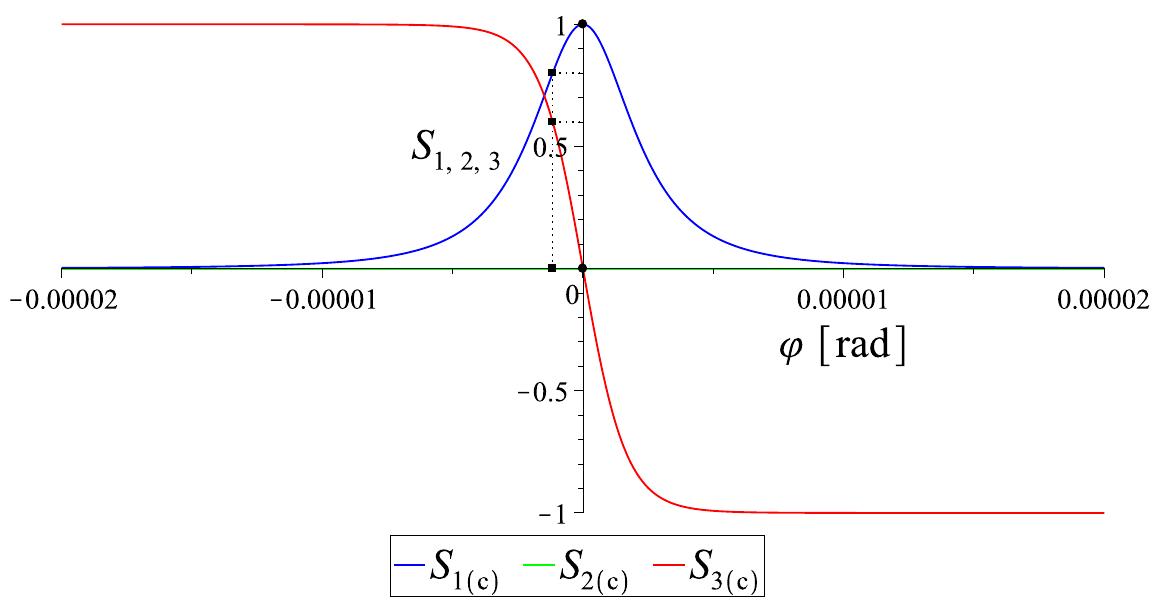}
\caption{The classical Stokes vector $\mathbf{S}_{\left(\mathrm{c}\right)}\left(\varphi\right)=\left(S_{1\left(\mathrm{c}\right)}\left(\varphi\right),
S_{2\left(\mathrm{c}\right)}\left(\varphi\right),S_{3\left(\mathrm{c}\right)}\left(\varphi\right)\right)$ 
plotted against the phase angle $\varphi$. Its initial values $\mathbf{S}\left(\varphi=0\right)=\left(1,0,0\right)$ and 
$\mathbf{S}\left(\varphi\approx-1.16\times10^{-6}\, \mathrm{rad}\right)=\left(0.8,0,0.6\right)$ are marked by solid circles and solid boxes, respectively.}
\label{figA1}
\end{figure*}

Out of the Stokes parameters, one can define the Stokes vector $\mathbf{S}=\left( S_{1},S_{2},S_{3}\right) \allowbreak \equiv \left(Q/I,U/I,V/I\right)$ 
and express it using (\ref{pp1})-(\ref{pp4}):
\begin{align}
S_{1} &=\frac{\left( \delta ^{2}+\varphi ^{2}\right) \mathrm{K}%
_{2/3}^{2}\left( \frac{\rho \omega }{3\beta c}\left( \delta ^{2}+\varphi
^{2}\right) ^{3/2}\right) -\,\varphi ^{2}\mathrm{K}_{1/3}^{2}\left( \frac{%
\rho \omega }{3\beta c}\left( \delta ^{2}+\varphi ^{2}\right) ^{3/2}\right) 
}{\left( \delta ^{2}+\varphi ^{2}\right) \mathrm{K}_{2/3}^{2}\left( \frac{%
\rho \omega }{3\beta c}\left( \delta ^{2}+\varphi ^{2}\right) ^{3/2}\right)
+\,\varphi ^{2}\mathrm{K}_{1/3}^{2}\left( \frac{\rho \omega }{3\beta c}%
\left( \delta ^{2}+\varphi ^{2}\right) ^{3/2}\right) },  \label{S1} \\
S_{2} &=0,  \label{S2} \\
S_{3} &=-\frac{2\varphi \left( \delta ^{2}+\varphi ^{2}\right)
^{1/2}\mathrm{K}_{2/3}\left( \frac{\rho \omega }{3\beta c}\left( \delta
^{2}+\varphi ^{2}\right) ^{3/2}\right) \mathrm{K}_{1/3}\left( \frac{\rho
\omega }{3\beta c}\left( \delta ^{2}+\varphi ^{2}\right) ^{3/2}\right) }{%
\left( \delta ^{2}+\varphi ^{2}\right) \mathrm{K}_{2/3}^{2}\left( \frac{\rho
\omega }{3\beta c}\left( \delta ^{2}+\varphi ^{2}\right) ^{3/2}\right)
+\,\varphi ^{2}\mathrm{K}_{1/3}^{2}\left( \frac{\rho \omega }{3\beta c}%
\left( \delta ^{2}+\varphi ^{2}\right) ^{3/2}\right) }.  \label{S3}
\end{align}%
In Fig. \ref{figA1} is plotted the classical Stokes vector against the phase angle $\varphi$, 
where we have set, for example, $\delta\approx10^{-8}$, $\rho\approx10^{8}\,\mathrm{cm}$, $\beta\approx1$ 
and $\omega\approx2\pi\times 10^{18}\,\mathrm{Hz}$ to model pulse profiles of X-ray pulsar emission. Here the initial values for the Stokes vector, 
$\mathbf{S}\left(\varphi=0\right)=\left(1,0,0\right)$ and $\mathbf{S}\left(\varphi\approx-1.16\times10^{-6}\, \mathrm{rad}\right)=\left(0.8,0,0.6\right)$, 
as in the examples given in Sect. \ref{ex}, are marked by solid circles and solid boxes, respectively. 

\section{Approximate analytical solutions to evolution equations}
\label{appB}

Substituting Eq. (\ref{Om}) into Eqs. (\ref{ode1})-(\ref{ode3}), the evolution equations can be reduced as follows:
\\ For $0\le s\le 20r_{*}$,
\begin{align}
&\dot{S}_{1}\left( s\right) \approx -ka_{2}s^{\frac{p+1}{p}}e^{-bs}S_{3}\left( s\right), \label{S1a} \\
&\dot{S}_{2}\left( s\right) \approx ka_{1}s^{\frac{p+1}{p}}e^{-bs}S_{3}\left( s\right), \label{S2a} \\
&\ddot{S}_{3}\left( s\right) -\left( \frac{p+1}{ps}-b\right) \dot{S}_{3}\left( s\right) 
+k^{2}\left( a_{1}^{2}+a_{2}^{2}\right) \left( s^{\frac{p+1}{p}}e^{-bs}\right) ^{2}S_{3}\left( s\right) \approx 0. \label{S3a}
\end{align}

First, we solve Eq. (\ref{S3a}) for $S_{3}\left( s\right)$, and then using this solution, obtain $S_{1}\left( s\right)$ and $S_{2}\left( s\right)$, 
by integrating Eqs. (\ref{S1a}) and (\ref{S2a}), respectively:
\begin{align}
S_{3}\left( s\right)  &\approx S_{\mathrm{o}}\sin \left( \Psi \left(s;p \right) +\delta \right), \label{S3b} \\
S_{1}\left( s\right)  &\approx \frac{a_{2}S_{\mathrm{o}}}{\sqrt{a_{1}^{2}+a_{2}^{2}}}\cos \left( \Psi \left(s;p \right) +\delta \right) +S_{1\mathrm{o}}, \label{S1b} \\
S_{2}\left( s\right)  &\approx -\frac{a_{1}S_{\mathrm{o}}}{\sqrt{a_{1}^{2}+a_{2}^{2}}}\cos \left( \Psi \left(s;p \right)+\delta \right) +S_{2\mathrm{o}}, \label{S2b} 
\end{align}
where
\begin{equation}
\Psi \left(s;p \right) \equiv k\sqrt{a_{1}^{2}+a_{2}^{2}}b^{-\frac{4p+1}{2p}}s^{\frac{1}{2p}}e^{-\frac{1}{2}bs}\left[ M_{\frac{1}{2p},\frac{p+1}{2p}}\left( bs\right) 
-M_{\frac{2p+1}{2p},\frac{p+1}{2p}}\left( bs\right) \right], \label{Psio}
\end{equation}
and $M_{\kappa,\mu}\left( z\right)$ denotes a Whittaker function of the first kind. Here employing the identity 
$S_{1}^{2}\left( s\right)+S_{2}^{2}\left( s\right)+S_{3}^{2}\left( s\right)=1$ (conservation of the degree of polarization), 
one can specify $S_{1\mathrm{o}}$ and $S_{2\mathrm{o}}$ in terms of $a_{1}$, $a_{2}$ and a constant $C$, and establish a relation 
between $S_{\mathrm{o}}$ and $C$: 
\begin{equation}
S_{1\mathrm{o}} =\frac{Ca_{1}}{\sqrt{a_{1}^{2}+a_{2}^{2}}},~~
S_{2\mathrm{o}} =\frac{Ca_{2}}{\sqrt{a_{1}^{2}+a_{2}^{2}}},~~
S_{\mathrm{o}}^{2}+C^{2}=1. \label{So12}
\end{equation}%
Then $S_{\mathrm{o}}$, $C$ and $\delta$ are determined by matching the initial value of the Stokes vector 
$\mathbf{S}\left(0\right)=\left( S_{1}\left( 0\right),S_{2}\left( 0\right),S_{3}\left( 0\right) \right)$ with Eqs. (\ref{S3b})-(\ref{S2b}) evaluated at $s=0$.

\end{appendices}

\bibliography{refs_VB_master}


\begin{thebibliography}{52}
\ifx \bisbn   \undefined \def \bisbn  #1{ISBN #1}\fi
\ifx \binits  \undefined \def \binits#1{#1}\fi
\ifx \bauthor  \undefined \def \bauthor#1{#1}\fi
\ifx \batitle  \undefined \def \batitle#1{#1}\fi
\ifx \bjtitle  \undefined \def \bjtitle#1{#1}\fi
\ifx \bvolume  \undefined \def \bvolume#1{\textbf{#1}}\fi
\ifx \byear  \undefined \def \byear#1{#1}\fi
\ifx \bissue  \undefined \def \bissue#1{#1}\fi
\ifx \bfpage  \undefined \def \bfpage#1{#1}\fi
\ifx \blpage  \undefined \def \blpage #1{#1}\fi
\ifx \burl  \undefined \def \burl#1{\textsf{#1}}\fi
\ifx \doiurl  \undefined \def \doiurl#1{\url{https://doi.org/#1}}\fi
\ifx \betal  \undefined \def \betal{\textit{et al.}}\fi
\ifx \binstitute  \undefined \def \binstitute#1{#1}\fi
\ifx \binstitutionaled  \undefined \def \binstitutionaled#1{#1}\fi
\ifx \bctitle  \undefined \def \bctitle#1{#1}\fi
\ifx \beditor  \undefined \def \beditor#1{#1}\fi
\ifx \bpublisher  \undefined \def \bpublisher#1{#1}\fi
\ifx \bbtitle  \undefined \def \bbtitle#1{#1}\fi
\ifx \bedition  \undefined \def \bedition#1{#1}\fi
\ifx \bseriesno  \undefined \def \bseriesno#1{#1}\fi
\ifx \blocation  \undefined \def \blocation#1{#1}\fi
\ifx \bsertitle  \undefined \def \bsertitle#1{#1}\fi
\ifx \bsnm \undefined \def \bsnm#1{#1}\fi
\ifx \bsuffix \undefined \def \bsuffix#1{#1}\fi
\ifx \bparticle \undefined \def \bparticle#1{#1}\fi
\ifx \barticle \undefined \def \barticle#1{#1}\fi
\bibcommenthead
\ifx \bconfdate \undefined \def \bconfdate #1{#1}\fi
\ifx \botherref \undefined \def \botherref #1{#1}\fi
\ifx \url \undefined \def \url#1{\textsf{#1}}\fi
\ifx \bchapter \undefined \def \bchapter#1{#1}\fi
\ifx \bbook \undefined \def \bbook#1{#1}\fi
\ifx \bcomment \undefined \def \bcomment#1{#1}\fi
\ifx \oauthor \undefined \def \oauthor#1{#1}\fi
\ifx \citeauthoryear \undefined \def \citeauthoryear#1{#1}\fi
\ifx \endbibitem  \undefined \def \endbibitem {}\fi
\ifx \bconflocation  \undefined \def \bconflocation#1{#1}\fi
\ifx \arxivurl  \undefined \def \arxivurl#1{\textsf{#1}}\fi
\csname PreBibitemsHook\endcsname

\bibitem{Heisenberg1936Folgerungen}
\begin{barticle}
\bauthor{\bsnm{Heisenberg}, \binits{W.}},
\bauthor{\bsnm{Euler}, \binits{H.}}:
\batitle{Folgerungen aus der {Diracschen} {Theorie} des {Positrons}}.
\bjtitle{Zeitschr. Phys}
\bvolume{98}(\bissue{11-12}),
\bfpage{714}
(\byear{1936}).
\doiurl{10.1007/BF01343663}
\end{barticle}
\endbibitem

\bibitem{Schwinger1951Gauge}
\begin{barticle}
\bauthor{\bsnm{Schwinger}, \binits{J.}}:
\batitle{On {Gauge} {Invariance} and {Vacuum} {Polarization}}.
\bjtitle{Phys. Rev.}
\bvolume{82}(\bissue{5}),
\bfpage{664}--\blpage{679}
(\byear{1951}).
\doiurl{10.1103/PhysRev.82.664}
\end{barticle}
\endbibitem

\bibitem{Meszaros1992High}
\begin{bbook}
\bauthor{\bsnm{M{\'e}sz{\'a}ros}, \binits{P.}}:
\bbtitle{High-Energy Radiation from Magnetized Neutron Stars}.
\bsertitle{Astrophysics/Physics}.
\bpublisher{University of Chicago Press},
\blocation{Chicago}
(\byear{1992}).
\burl{https://books.google.co.kr/books?id=BVY7gEs11IYC}
\end{bbook}
\endbibitem

\bibitem{Kim2023Vacuum}
\begin{barticle}
\bauthor{\bsnm{Kim}, \binits{C.M.}},
\bauthor{\bsnm{Kim}, \binits{S.P.}}:
\batitle{Vacuum birefringence at one-loop in a supercritical magnetic field
  superposed with a weak electric field and application to pulsar
  magnetosphere}.
\bjtitle{Eur. Phys. J. C}
\bvolume{83}(\bissue{2}),
\bfpage{104}
(\byear{2023}).
\doiurl{10.1140/epjc/s10052-023-11243-1}
\end{barticle}
\endbibitem

\bibitem{Ruffini_2010}
\begin{barticle}
\bauthor{\bsnm{Ruffini}, \binits{R.}},
\bauthor{\bsnm{Vereshchagin}, \binits{G.}},
\bauthor{\bsnm{Xue}, \binits{S.-S.}}:
\batitle{Electron–positron pairs in physics and astrophysics: From heavy
  nuclei to black holes}.
\bjtitle{Physics Reports}
\bvolume{487}(\bissue{1–4}),
\bfpage{1}--\blpage{140}
(\byear{2010}).
\doiurl{10.1016/j.physrep.2009.10.004}
\end{barticle}
\endbibitem

\bibitem{Fedotov:2022ely}
\begin{barticle}
\bauthor{\bsnm{Fedotov}, \binits{A.}},
\bauthor{\bsnm{Ilderton}, \binits{A.}},
\bauthor{\bsnm{Karbstein}, \binits{F.}},
\bauthor{\bsnm{King}, \binits{B.}},
\bauthor{\bsnm{Seipt}, \binits{D.}},
\bauthor{\bsnm{Taya}, \binits{H.}},
\bauthor{\bsnm{Torgrimsson}, \binits{G.}}:
\batitle{{Advances in QED with intense background fields}}.
\bjtitle{Phys. Rept.}
\bvolume{1010},
\bfpage{1}--\blpage{138}
(\byear{2023}).
\doiurl{10.1016/j.physrep.2023.01.003}
\end{barticle}
\endbibitem

\bibitem{Hattori:2023egw}
\begin{barticle}
\bauthor{\bsnm{Hattori}, \binits{K.}},
\bauthor{\bsnm{Itakura}, \binits{K.}},
\bauthor{\bsnm{Ozaki}, \binits{S.}}:
\batitle{{Strong-field physics in QED and QCD: From fundamentals to
  applications}}.
\bjtitle{Prog. Part. Nucl. Phys.}
\bvolume{133},
\bfpage{104068}
(\byear{2023}).
\doiurl{10.1016/j.ppnp.2023.104068}
\end{barticle}
\endbibitem

\bibitem{Ejlli2020PVLASa}
\begin{barticle}
\bauthor{\bsnm{{Ejlli}}, \binits{A.}},
\bauthor{\bsnm{{Della Valle}}, \binits{F.}},
\bauthor{\bsnm{{Gastaldi}}, \binits{U.}},
\bauthor{\bsnm{{Messineo}}, \binits{G.}},
\bauthor{\bsnm{{Pengo}}, \binits{R.}},
\bauthor{\bsnm{{Ruoso}}, \binits{G.}},
\bauthor{\bsnm{{Zavattini}}, \binits{G.}}:
\batitle{{The PVLAS experiment: A 25 year effort to measure vacuum magnetic
  birefringence}}.
\bjtitle{Phys. Rept.}
\bvolume{871},
\bfpage{1}--\blpage{74}
(\byear{2020}).
\doiurl{10.1016/j.physrep.2020.06.001}
\end{barticle}
\endbibitem

\bibitem{Karbstein2021Vacuum}
\begin{barticle}
\bauthor{\bsnm{Karbstein}, \binits{F.}},
\bauthor{\bsnm{Sundqvist}, \binits{C.}},
\bauthor{\bsnm{Schulze}, \binits{K.S.}},
\bauthor{\bsnm{Uschmann}, \binits{I.}},
\bauthor{\bsnm{Gies}, \binits{H.}},
\bauthor{\bsnm{Paulus}, \binits{G.G.}}:
\batitle{Vacuum birefringence at x-ray free-electron lasers}.
\bjtitle{New J. Phys.}
\bvolume{23}(\bissue{9}),
\bfpage{095001}
(\byear{2021}).
\doiurl{10.1088/1367-2630/ac1df4}
\end{barticle}
\endbibitem

\bibitem{Shen2018Exploring}
\begin{barticle}
\bauthor{\bsnm{Shen}, \binits{B.}},
\bauthor{\bsnm{Bu}, \binits{Z.}},
\bauthor{\bsnm{Xu}, \binits{J.}},
\bauthor{\bsnm{Xu}, \binits{T.}},
\bauthor{\bsnm{Ji}, \binits{L.}},
\bauthor{\bsnm{Li}, \binits{R.}},
\bauthor{\bsnm{Xu}, \binits{Z.}}:
\batitle{Exploring vacuum birefringence based on a 100 {PW} laser and an x-ray
  free electron laser beam}.
\bjtitle{Plasma Phys. Control. Fusion}
\bvolume{60}(\bissue{4}),
\bfpage{044002}
(\byear{2018}).
\doiurl{10.1088/1361-6587/aaa7fb}
\end{barticle}
\endbibitem

\bibitem{Yu2023X}
\begin{barticle}
\bauthor{\bsnm{Yu}, \binits{Q.}},
\bauthor{\bsnm{Xu}, \binits{D.}},
\bauthor{\bsnm{Shen}, \binits{B.}},
\bauthor{\bsnm{Cowan}, \binits{T.E.}},
\bauthor{\bsnm{Schlenvoigt}, \binits{H.-P.}}:
\batitle{X-ray polarimetry and its application to strong-field quantum
  electrodynamics}.
\bjtitle{High Power Laser Sci. Eng.}
\bvolume{11},
\bfpage{71}
(\byear{2023}).
\doiurl{10.1017/hpl.2023.45}
\end{barticle}
\endbibitem

\bibitem{Danson2019Petawatt}
\begin{barticle}
\bauthor{\bsnm{Danson}, \binits{C.N.}},
\bauthor{\bsnm{Haefner}, \binits{C.}},
\bauthor{\bsnm{Bromage}, \binits{J.}},
\bauthor{\bsnm{Butcher}, \binits{T.}},
\bauthor{\bsnm{Chanteloup}, \binits{J.-C.F.}},
\bauthor{\bsnm{Chowdhury}, \binits{E.A.}},
\bauthor{\bsnm{Galvanauskas}, \binits{A.}},
\bauthor{\bsnm{Gizzi}, \binits{L.A.}},
\bauthor{\bsnm{Hein}, \binits{J.}},
\bauthor{\bsnm{Hillier}, \binits{D.I.}},
\bauthor{\bparticle{et} \bsnm{al.}}:
\batitle{Petawatt and exawatt class lasers worldwide}.
\bjtitle{High Power Laser Sci. Eng.}
\bvolume{7},
\bfpage{54}
(\byear{2019}).
\doiurl{10.1017/hpl.2019.36}
\end{barticle}
\endbibitem

\bibitem{Harding2006Physics}
\begin{barticle}
\bauthor{\bsnm{Harding}, \binits{A.K.}},
\bauthor{\bsnm{Lai}, \binits{D.}}:
\batitle{Physics of strongly magnetized neutron stars}.
\bjtitle{Rep. Prog. Phys.}
\bvolume{69}(\bissue{9}),
\bfpage{2631}--\blpage{2708}
(\byear{2006}).
\doiurl{10.1088/0034-4885/69/9/R03}
\end{barticle}
\endbibitem

\bibitem{IXPE}
\begin{botherref}
{{IXPE}}.
\url{https://ixpe.msfc.nasa.gov/index.html}
\end{botherref}
\endbibitem

\bibitem{XPoSat}
\begin{botherref}
{{XPoSat}}.
\url{https://www.isro.gov.in/XPoSat.html}
\end{botherref}
\endbibitem

\bibitem{Santangelo_2019}
\begin{botherref}
\oauthor{\bsnm{Santangelo}, \binits{A.}},
\oauthor{\bsnm{Zane}, \binits{S.}},
\oauthor{\bsnm{Feng}, \binits{H.}},
\oauthor{\bsnm{Xu}, \binits{R.}},
\oauthor{\bsnm{Doroshenko}, \binits{V.}},
\oauthor{\bsnm{Bozzo}, \binits{E.}},
\oauthor{\bsnm{Caiazzo}, \binits{I.}},
\oauthor{\bsnm{Zelati}, \binits{F.C.}},
\oauthor{\bsnm{Esposito}, \binits{P.}},
\oauthor{\bsnm{González-Caniulef}, \binits{D.}},
\oauthor{\bsnm{Heyl}, \binits{J.}},
\oauthor{\bsnm{Huppenkothen}, \binits{D.}},
\oauthor{\bsnm{Israel}, \binits{G.}},
\oauthor{\bsnm{Li}, \binits{Z.}},
\oauthor{\bsnm{Lin}, \binits{L.}},
\oauthor{\bsnm{Mignani}, \binits{R.}},
\oauthor{\bsnm{Rea}, \binits{N.}},
\oauthor{\bsnm{Orlandini}, \binits{M.}},
\oauthor{\bsnm{Taverna}, \binits{R.}},
\oauthor{\bsnm{Tong}, \binits{H.}},
\oauthor{\bsnm{Turolla}, \binits{R.}},
\oauthor{\bsnm{Baglio}, \binits{C.}},
\oauthor{\bsnm{Bernardini}, \binits{F.}},
\oauthor{\bsnm{Bucciantini}, \binits{N.}},
\oauthor{\bsnm{Feroci}, \binits{M.}},
\oauthor{\bsnm{Fürst}, \binits{F.}},
\oauthor{\bsnm{Göğüş}, \binits{E.}},
\oauthor{\bsnm{Güngör}, \binits{C.}},
\oauthor{\bsnm{Ji}, \binits{L.}},
\oauthor{\bsnm{Lu}, \binits{F.}},
\oauthor{\bsnm{Manousakis}, \binits{A.}},
\oauthor{\bsnm{Mereghetti}, \binits{S.}},
\oauthor{\bsnm{Mikusincova}, \binits{R.}},
\oauthor{\bsnm{Paul}, \binits{B.}},
\oauthor{\bsnm{Prescod-Weinstein}, \binits{C.}},
\oauthor{\bsnm{Younes}, \binits{G.}},
\oauthor{\bsnm{Tiengo}, \binits{A.}},
\oauthor{\bsnm{Xu}, \binits{Y.}},
\oauthor{\bsnm{Watts}, \binits{A.}},
\oauthor{\bsnm{Zhang}, \binits{S.}},
\oauthor{\bsnm{Zhan}, \binits{S.-N.}}:
Physics and astrophysics of strong magnetic field systems with extp.
Science China Physics, Mechanics \& Astronomy
\textbf{62}(2)
(2019).
\doiurl{10.1007/s11433-018-9234-3}
\end{botherref}
\endbibitem

\bibitem{Wadiasingh2019Magnetars}
\begin{botherref}
\oauthor{\bsnm{Wadiasingh}, \binits{Z.}},
\oauthor{\bsnm{Younes}, \binits{G.}},
\oauthor{\bsnm{Baring}, \binits{M.G.}},
\oauthor{\bsnm{Harding}, \binits{A.K.}},
\oauthor{\bsnm{Gonthier}, \binits{P.L.}},
\oauthor{\bsnm{Hu}, \binits{K.}},
\oauthor{\bsnm{Horst}, \binits{A.v.d.}},
\oauthor{\bsnm{Zane}, \binits{S.}},
\oauthor{\bsnm{Kouveliotou}, \binits{C.}},
\oauthor{\bsnm{Beloborodov}, \binits{A.M.}},
\oauthor{\bsnm{Prescod-Weinstein}, \binits{C.}},
\oauthor{\bsnm{Chattopadhyay}, \binits{T.}},
\oauthor{\bsnm{Chandra}, \binits{S.}},
\oauthor{\bsnm{Kalapotharakos}, \binits{C.}},
\oauthor{\bsnm{Parfrey}, \binits{K.}},
\oauthor{\bsnm{Kazanas}, \binits{D.}}:
Magnetars as {Astrophysical} {Laboratories} of {Extreme} {Quantum}
  {Electrodynamics}: The {Case} for a {Compton} {Telescope}.
Bulletin of the American Astronomical Society
\textbf{51}(3)
(2019).
\url{https://baas.aas.org/pub/2020n3i292}
\end{botherref}
\endbibitem

\bibitem{Raffelt1996Stars}
\begin{bbook}
\bauthor{\bsnm{Raffelt}, \binits{G.G.}}:
\bbtitle{Stars as Laboratories for Fundamental Physics: the Astrophysics of
  Neutrinos, Axions, and Other Weakly Interacting Particles}.
\bpublisher{The University of Chicago Press},
\blocation{Chicago}
(\byear{1996}).
\burl{https://ui.adsabs.harvard.edu/abs/1996slfp.book.....R}
\end{bbook}
\endbibitem

\bibitem{Kim2024Magnetars}
\begin{barticle}
\bauthor{\bsnm{Kim}, \binits{C.M.}},
\bauthor{\bsnm{Kim}, \binits{S.P.}}:
\batitle{{Magnetars as laboratories for strong field QED}}.
\bjtitle{AIP Conference Proceedings}
\bvolume{2874}(\bissue{1}),
\bfpage{020013}
(\byear{2024}).
\doiurl{10.1063/5.0215939}
\end{barticle}
\endbibitem

\bibitem{Adler1971Photon}
\begin{barticle}
\bauthor{\bsnm{Adler}, \binits{S.L.}}:
\batitle{Photon splitting and photon dispersion in a strong magnetic field}.
\bjtitle{Ann. Phys.}
\bvolume{67}(\bissue{2}),
\bfpage{599}--\blpage{647}
(\byear{1971}).
\doiurl{10.1016/0003-4916(71)90154-0}
\end{barticle}
\endbibitem

\bibitem{Ni2013Foundations}
\begin{barticle}
\bauthor{\bsnm{Ni}, \binits{W.-T.}},
\bauthor{\bsnm{Mei}, \binits{H.-H.}},
\bauthor{\bsnm{Wu}, \binits{S.-J.}}:
\batitle{Foundations of classical electrodynamics, equivalence principle and
  cosmic interactions: A short exposition and an update}.
\bjtitle{Mod. Phys. Lett. A}
\bvolume{28}(\bissue{03}),
\bfpage{1340013}
(\byear{2013}).
\doiurl{10.1142/S0217732313400130}
\end{barticle}
\endbibitem

\bibitem{Denisov2016Pulsar}
\begin{barticle}
\bauthor{\bsnm{Denisov}, \binits{V.I.}},
\bauthor{\bsnm{Shvilkin}, \binits{B.N.}},
\bauthor{\bsnm{Sokolov}, \binits{V.A.}},
\bauthor{\bsnm{Vasili'ev}, \binits{M.I.}}:
\batitle{Pulsar radiation in post-maxwellian vacuum nonlinear electrodynamics}.
\bjtitle{Phys. Rev. D}
\bvolume{94}(\bissue{4}),
\bfpage{045021}
(\byear{2016}).
\doiurl{10.1103/PhysRevD.94.045021}
\end{barticle}
\endbibitem

\bibitem{10.1093/mnras/stae1304}
\begin{barticle}
\bauthor{\bsnm{Kim}, \binits{D.-H.}},
\bauthor{\bsnm{Kim}, \binits{C.M.}},
\bauthor{\bsnm{Kim}, \binits{S.P.}}:
\batitle{{Quantum refraction effects in pulsar emission}}.
\bjtitle{Monthly Notices of the Royal Astronomical Society}
\bvolume{531}(\bissue{1}),
\bfpage{2148}--\blpage{2161}
(\byear{2024}).
\doiurl{10.1093/mnras/stae1304}
\end{barticle}
\endbibitem

\bibitem{Gralla_2016}
\begin{barticle}
\bauthor{\bsnm{Gralla}, \binits{S.E.}},
\bauthor{\bsnm{Lupsasca}, \binits{A.}},
\bauthor{\bsnm{Philippov}, \binits{A.}}:
\batitle{Pulsar magnetospheres: Beyond the flat spacetime dipole}.
\bjtitle{The Astrophysical Journal}
\bvolume{833}(\bissue{2}),
\bfpage{258}
(\byear{2016}).
\doiurl{10.3847/1538-4357/833/2/258}
\end{barticle}
\endbibitem

\bibitem{10.1093/mnras/stz2524}
\begin{barticle}
\bauthor{\bsnm{Lockhart}, \binits{W.}},
\bauthor{\bsnm{Gralla}, \binits{S.E.}},
\bauthor{\bsnm{Özel}, \binits{F.}},
\bauthor{\bsnm{Psaltis}, \binits{D.}}:
\batitle{{X-ray light curves from realistic polar cap models: inclined pulsar
  magnetospheres and multipole fields}}.
\bjtitle{Monthly Notices of the Royal Astronomical Society}
\bvolume{490}(\bissue{2}),
\bfpage{1774}--\blpage{1783}
(\byear{2019}).
\doiurl{10.1093/mnras/stz2524}
\end{barticle}
\endbibitem

\bibitem{Kazmierczak2019}
\begin{botherref}
\oauthor{\bsnm{Kazmierczak}, \binits{J.}}:
{NASA’s NICER Delivers Best-ever Pulsar Measurements, 1st Surface Map}
(2019).
\url{https://www.nasa.gov/universe/nasas-nicer-delivers-best-ever-pulsar-measurements-1st-surface-map/}
\end{botherref}
\endbibitem

\bibitem{Kalapotharakos_2021}
\begin{barticle}
\bauthor{\bsnm{Kalapotharakos}, \binits{C.}},
\bauthor{\bsnm{Wadiasingh}, \binits{Z.}},
\bauthor{\bsnm{Harding}, \binits{A.K.}},
\bauthor{\bsnm{Kazanas}, \binits{D.}}:
\batitle{The multipolar magnetic field of the millisecond pulsar psr
  j0030+0451}.
\bjtitle{The Astrophysical Journal}
\bvolume{907}(\bissue{2}),
\bfpage{63}
(\byear{2021}).
\doiurl{10.3847/1538-4357/abcec0}
\end{barticle}
\endbibitem

\bibitem{Petri2016Theory}
\begin{botherref}
\oauthor{\bsnm{Petri}, \binits{J.}}:
Theory of pulsar magnetosphere and wind.
Journal of Plasma Physics
\textbf{82}(5)
(2016).
\doiurl{10.1017/S0022377816000763}
\end{botherref}
\endbibitem

\bibitem{Heyl2000}
\begin{barticle}
\bauthor{\bsnm{Heyl}, \binits{J.S.}},
\bauthor{\bsnm{Shaviv}, \binits{N.J.}}:
\batitle{{Polarization evolution in strong magnetic fields}}.
\bjtitle{Monthly Notices of the Royal Astronomical Society}
\bvolume{311}(\bissue{3}),
\bfpage{555}--\blpage{564}
(\byear{2000}).
\doiurl{10.1046/j.1365-8711.2000.03076.x}
\end{barticle}
\endbibitem

\bibitem{Wang2007Wave}
\begin{barticle}
\bauthor{\bsnm{Wang}, \binits{C.}},
\bauthor{\bsnm{Lai}, \binits{D.}}:
\batitle{Wave modes in the magnetospheres of pulsars and magnetars}.
\bjtitle{Mon. Not. R. Astron Soc.}
\bvolume{377}(\bissue{3}),
\bfpage{1095}--\blpage{1112}
(\byear{2007}).
\doiurl{10.1111/j.1365-2966.2007.11531.x}
\end{barticle}
\endbibitem

\bibitem{Kubo1981}
\begin{barticle}
\bauthor{\bsnm{Kubo}, \binits{H.}},
\bauthor{\bsnm{Nagata}, \binits{R.}}:
\batitle{{Determination of dielectric tensor fields in weakly inhomogeneous
  anisotropic media. II}}.
\bjtitle{J. Opt. Soc. Am.}
\bvolume{71}(\bissue{3}),
\bfpage{327}--\blpage{333}
(\byear{1981}).
\doiurl{10.1364/JOSA.71.000327}
\end{barticle}
\endbibitem

\bibitem{Kubo1983}
\begin{barticle}
\bauthor{\bsnm{Kubo}, \binits{H.}},
\bauthor{\bsnm{Nagata}, \binits{R.}}:
\batitle{Vector representation of behavior of polarized light in a weakly
  inhomogeneous medium with birefringence and dichroism}.
\bjtitle{J. Opt. Soc. Am.}
\bvolume{73}(\bissue{12}),
\bfpage{1719}--\blpage{1724}
(\byear{1983}).
\doiurl{10.1364/JOSA.73.001719}
\end{barticle}
\endbibitem

\bibitem{Kubo1985}
\begin{barticle}
\bauthor{\bsnm{Kubo}, \binits{H.}},
\bauthor{\bsnm{Nagata}, \binits{R.}}:
\batitle{{Vector representation of behavior of polarized light in a weakly
  inhomogeneous medium with birefringence and dichroism. II. Evolution of
  polarization states}}.
\bjtitle{J. Opt. Soc. Am. A}
\bvolume{2}(\bissue{1}),
\bfpage{30}--\blpage{34}
(\byear{1985}).
\doiurl{10.1364/JOSAA.2.000030}
\end{barticle}
\endbibitem

\bibitem{Heyl2003highenergy}
\begin{barticle}
\bauthor{\bsnm{Heyl}, \binits{J.S.}},
\bauthor{\bsnm{Shaviv}, \binits{N.J.}},
\bauthor{\bsnm{Lloyd}, \binits{D.}}:
\batitle{{The high-energy polarization-limiting radius of neutron star
  magnetospheres — I. Slowly rotating neutron stars}}.
\bjtitle{Monthly Notices of the Royal Astronomical Society}
\bvolume{342}(\bissue{1}),
\bfpage{134}--\blpage{144}
(\byear{2003}).
\doiurl{10.1046/j.1365-8711.2003.06521.x}
\end{barticle}
\endbibitem

\bibitem{Heyl2018Strongly}
\begin{barticle}
\bauthor{\bsnm{Heyl}, \binits{J.}},
\bauthor{\bsnm{Caiazzo}, \binits{I.}}:
\batitle{{Strongly Magnetized Sources: QED and X-ray Polarization}}.
\bjtitle{Galaxies}
\bvolume{6}(\bissue{3}),
\bfpage{76}
(\byear{2018}).
\doiurl{10.3390/galaxies6030076}
\end{barticle}
\endbibitem

\bibitem{Novak2018}
\begin{barticle}
\bauthor{\bsnm{Novak}, \binits{O.}},
\bauthor{\bsnm{Diachenko}, \binits{M.}},
\bauthor{\bsnm{Padusenko}, \binits{E.}},
\bauthor{\bsnm{Kholodov}, \binits{R.}}:
\batitle{{Vacuum Birefringence in the Fields of a Current Coil and a Guided
  Electromagnetic Wave}}.
\bjtitle{Ukrainian Journal of Physics}
\bvolume{63}(\bissue{11}),
\bfpage{979}
(\byear{2018}).
\doiurl{10.15407/ujpe63.11.979}
\end{barticle}
\endbibitem

\bibitem{Kim2021Generala}
\begin{botherref}
\oauthor{\bsnm{Kim}, \binits{D.-H.}},
\oauthor{\bsnm{Trippe}, \binits{S.}}:
{General relativistic effects on pulsar radiation}
(2021).
\url{https://arxiv.org/abs/2109.13387}
\end{botherref}
\endbibitem

\bibitem{Blandford1982}
\begin{barticle}
\bauthor{\bsnm{Blandford}, \binits{R.D.}},
\bauthor{\bsnm{Payne}, \binits{D.G.}}:
\batitle{{Hydromagnetic flows from accretion discs and the production of radio
  jets}}.
\bjtitle{Monthly Notices of the Royal Astronomical Society}
\bvolume{199}(\bissue{4}),
\bfpage{883}--\blpage{903}
(\byear{1982}).
\doiurl{10.1093/mnras/199.4.883}
\end{barticle}
\endbibitem

\bibitem{Gangadhara2005}
\begin{barticle}
\bauthor{\bsnm{Gangadhara}, \binits{R.T.}}:
\batitle{{On the method of estimating emission altitude from relativistic phase
  shift in pulsars}}.
\bjtitle{The Astrophysical Journal}
\bvolume{628},
\bfpage{923}--\blpage{930}
(\byear{2005}).
\doiurl{10.1086/431138}
\end{barticle}
\endbibitem

\bibitem{Euler1935Ueber}
\begin{barticle}
\bauthor{\bsnm{Euler}, \binits{H.}},
\bauthor{\bsnm{Kockel}, \binits{B.}}:
\batitle{Ueber die streuung von licht an licht nach der diracschen theorie}.
\bjtitle{Die Naturwissenschaften}
\bvolume{23}(\bissue{15}),
\bfpage{246}--\blpage{247}
(\byear{1935}).
\doiurl{10.1007/BF01493898}
\end{barticle}
\endbibitem

\bibitem{Pavlov2013}
\begin{botherref}
\oauthor{\bsnm{Pavlov}, \binits{G.}},
\oauthor{\bsnm{Kargaltsev}, \binits{O.}},
\oauthor{\bsnm{Durant}, \binits{B.} \bsuffix{M.~Posselt}}:
{X-ray Observations of Rotation Powered Pulsars}
(2013).
\url{https://www.cosmos.esa.int/documents/332006/943890/GPavlov\_t.pdf}
\end{botherref}
\endbibitem

\bibitem{ANTF2024}
\begin{botherref}
\oauthor{\bsnm{Hobbs}, \binits{G.}},
\oauthor{\bsnm{Manchester}, \binits{R.N.}},
\oauthor{\bsnm{Toomey}, \binits{L.}},
\oauthor{\bsnm{Kapur}, \binits{A.}}:
{ATNF Pulsar Catalogue}
(2024).
\url{https://www.atnf.csiro.au/research/pulsar/psrcat}
\end{botherref}
\endbibitem

\bibitem{Smith_2023}
\begin{barticle}
\bauthor{\bsnm{Smith}, \binits{D.A.}},
\bauthor{\bsnm{Abdollahi}, \binits{S.}},
\bauthor{\bsnm{Ajello}, \binits{M.}},
\bauthor{\bsnm{Bailes}, \binits{M.}},
\bauthor{\bsnm{Baldini}, \binits{L.}},
\bauthor{\bsnm{Ballet}, \binits{J.}},
\bauthor{\bsnm{Baring}, \binits{M.G.}},
\bauthor{\bsnm{Bassa}, \binits{C.}},
\bauthor{\bsnm{Gonzalez}, \binits{J.B.}},
\bauthor{\bsnm{Bellazzini}, \binits{R.}},
\bauthor{\bsnm{Berretta}, \binits{A.}},
\bauthor{\bsnm{Bhattacharyya}, \binits{B.}},
\bauthor{\bsnm{Bissaldi}, \binits{E.}},
\bauthor{\bsnm{Bonino}, \binits{R.}},
\bauthor{\bsnm{Bottacini}, \binits{E.}},
\bauthor{\bsnm{Bregeon}, \binits{J.}},
\bauthor{\bsnm{Bruel}, \binits{P.}},
\bauthor{\bsnm{Burgay}, \binits{M.}},
\bauthor{\bsnm{Burnett}, \binits{T.H.}},
\bauthor{\bsnm{Cameron}, \binits{R.A.}},
\bauthor{\bsnm{Camilo}, \binits{F.}},
\bauthor{\bsnm{Caputo}, \binits{R.}},
\bauthor{\bsnm{Caraveo}, \binits{P.A.}},
\bauthor{\bsnm{Cavazzuti}, \binits{E.}},
\bauthor{\bsnm{Chiaro}, \binits{G.}},
\bauthor{\bsnm{Ciprini}, \binits{S.}},
\bauthor{\bsnm{Clark}, \binits{C.J.}},
\bauthor{\bsnm{Cognard}, \binits{I.}},
\bauthor{\bsnm{Corongiu}, \binits{A.}},
\bauthor{\bsnm{Orestano}, \binits{P.C.}},
\bauthor{\bsnm{Crnogorcevic}, \binits{M.}},
\bauthor{\bsnm{Cuoco}, \binits{A.}},
\bauthor{\bsnm{Cutini}, \binits{S.}},
\bauthor{\bsnm{D’Ammando}, \binits{F.}},
\bauthor{\bparticle{de} \bsnm{Angelis}, \binits{A.}},
\bauthor{\bsnm{DeCesar}, \binits{M.E.}},
\bauthor{\bsnm{Gaetano}, \binits{S.D.}},
\bauthor{\bparticle{de} \bsnm{Menezes}, \binits{R.}},
\bauthor{\bsnm{Deneva}, \binits{J.}},
\bauthor{\bparticle{de} \bsnm{Palma}, \binits{F.}},
\bauthor{\bsnm{Lalla}, \binits{N.D.}},
\bauthor{\bsnm{Dirirsa}, \binits{F.}},
\bauthor{\bsnm{Venere}, \binits{L.D.}},
\bauthor{\bsnm{Domínguez}, \binits{A.}},
\bauthor{\bsnm{Dumora}, \binits{D.}},
\bauthor{\bsnm{Fegan}, \binits{S.J.}},
\bauthor{\bsnm{Ferrara}, \binits{E.C.}},
\bauthor{\bsnm{Fiori}, \binits{A.}},
\bauthor{\bsnm{Fleischhack}, \binits{H.}},
\bauthor{\bsnm{Flynn}, \binits{C.}},
\bauthor{\bsnm{Franckowiak}, \binits{A.}},
\bauthor{\bsnm{Freire}, \binits{P.C.C.}},
\bauthor{\bsnm{Fukazawa}, \binits{Y.}},
\bauthor{\bsnm{Fusco}, \binits{P.}},
\bauthor{\bsnm{Galanti}, \binits{G.}},
\bauthor{\bsnm{Gammaldi}, \binits{V.}},
\bauthor{\bsnm{Gargano}, \binits{F.}},
\bauthor{\bsnm{Gasparrini}, \binits{D.}},
\bauthor{\bsnm{Giacchino}, \binits{F.}},
\bauthor{\bsnm{Giglietto}, \binits{N.}},
\bauthor{\bsnm{Giordano}, \binits{F.}},
\bauthor{\bsnm{Giroletti}, \binits{M.}},
\bauthor{\bsnm{Green}, \binits{D.}},
\bauthor{\bsnm{Grenier}, \binits{I.A.}},
\bauthor{\bsnm{Guillemot}, \binits{L.}},
\bauthor{\bsnm{Guiriec}, \binits{S.}},
\bauthor{\bsnm{Gustafsson}, \binits{M.}},
\bauthor{\bsnm{Harding}, \binits{A.K.}},
\bauthor{\bsnm{Hays}, \binits{E.}},
\bauthor{\bsnm{Hewitt}, \binits{J.W.}},
\bauthor{\bsnm{Horan}, \binits{D.}},
\bauthor{\bsnm{Hou}, \binits{X.}},
\bauthor{\bsnm{Jankowski}, \binits{F.}},
\bauthor{\bsnm{Johnson}, \binits{R.P.}},
\bauthor{\bsnm{Johnson}, \binits{T.J.}},
\bauthor{\bsnm{Johnston}, \binits{S.}},
\bauthor{\bsnm{Kataoka}, \binits{J.}},
\bauthor{\bsnm{Keith}, \binits{M.J.}},
\bauthor{\bsnm{Kerr}, \binits{M.}},
\bauthor{\bsnm{Kramer}, \binits{M.}},
\bauthor{\bsnm{Kuss}, \binits{M.}},
\bauthor{\bsnm{Latronico}, \binits{L.}},
\bauthor{\bsnm{Lee}, \binits{S.-H.}},
\bauthor{\bsnm{Li}, \binits{D.}},
\bauthor{\bsnm{Li}, \binits{J.}},
\bauthor{\bsnm{Limyansky}, \binits{B.}},
\bauthor{\bsnm{Longo}, \binits{F.}},
\bauthor{\bsnm{Loparco}, \binits{F.}},
\bauthor{\bsnm{Lorusso}, \binits{L.}},
\bauthor{\bsnm{Lovellette}, \binits{M.N.}},
\bauthor{\bsnm{Lower}, \binits{M.}},
\bauthor{\bsnm{Lubrano}, \binits{P.}},
\bauthor{\bsnm{Lyne}, \binits{A.G.}},
\bauthor{\bsnm{Maan}, \binits{Y.}},
\bauthor{\bsnm{Maldera}, \binits{S.}},
\bauthor{\bsnm{Manchester}, \binits{R.N.}},
\bauthor{\bsnm{Manfreda}, \binits{A.}},
\bauthor{\bsnm{Marelli}, \binits{M.}},
\bauthor{\bsnm{Martí-Devesa}, \binits{G.}},
\bauthor{\bsnm{Mazziotta}, \binits{M.N.}},
\bauthor{\bsnm{McEnery}, \binits{J.E.}},
\bauthor{\bsnm{Mereu}, \binits{I.}},
\bauthor{\bsnm{Michelson}, \binits{P.F.}},
\bauthor{\bsnm{Mickaliger}, \binits{M.}},
\bauthor{\bsnm{Mitthumsiri}, \binits{W.}},
\bauthor{\bsnm{Mizuno}, \binits{T.}},
\bauthor{\bsnm{Moiseev}, \binits{A.A.}},
\bauthor{\bsnm{Monzani}, \binits{M.E.}},
\bauthor{\bsnm{Morselli}, \binits{A.}},
\bauthor{\bsnm{Negro}, \binits{M.}},
\bauthor{\bsnm{Nemmen}, \binits{R.}},
\bauthor{\bsnm{Nieder}, \binits{L.}},
\bauthor{\bsnm{Nuss}, \binits{E.}},
\bauthor{\bsnm{Omodei}, \binits{N.}},
\bauthor{\bsnm{Orienti}, \binits{M.}},
\bauthor{\bsnm{Orlando}, \binits{E.}},
\bauthor{\bsnm{Ormes}, \binits{J.F.}},
\bauthor{\bsnm{Palatiello}, \binits{M.}},
\bauthor{\bsnm{Paneque}, \binits{D.}},
\bauthor{\bsnm{Panzarini}, \binits{G.}},
\bauthor{\bsnm{Parthasarathy}, \binits{A.}},
\bauthor{\bsnm{Persic}, \binits{M.}},
\bauthor{\bsnm{Pesce-Rollins}, \binits{M.}},
\bauthor{\bsnm{Pillera}, \binits{R.}},
\bauthor{\bsnm{Poon}, \binits{H.}},
\bauthor{\bsnm{Porter}, \binits{T.A.}},
\bauthor{\bsnm{Possenti}, \binits{A.}},
\bauthor{\bsnm{Principe}, \binits{G.}},
\bauthor{\bsnm{Rainò}, \binits{S.}},
\bauthor{\bsnm{Rando}, \binits{R.}},
\bauthor{\bsnm{Ransom}, \binits{S.M.}},
\bauthor{\bsnm{Ray}, \binits{P.S.}},
\bauthor{\bsnm{Razzano}, \binits{M.}},
\bauthor{\bsnm{Razzaque}, \binits{S.}},
\bauthor{\bsnm{Reimer}, \binits{A.}},
\bauthor{\bsnm{Reimer}, \binits{O.}},
\bauthor{\bsnm{Renault-Tinacci}, \binits{N.}},
\bauthor{\bsnm{Romani}, \binits{R.W.}},
\bauthor{\bsnm{Sánchez-Conde}, \binits{M.}},
\bauthor{\bsnm{Parkinson}, \binits{P.M.S.}},
\bauthor{\bsnm{Scotton}, \binits{L.}},
\bauthor{\bsnm{Serini}, \binits{D.}},
\bauthor{\bsnm{Sgrò}, \binits{C.}},
\bauthor{\bsnm{Shannon}, \binits{R.}},
\bauthor{\bsnm{Sharma}, \binits{V.}},
\bauthor{\bsnm{Shen}, \binits{Z.}},
\bauthor{\bsnm{Siskind}, \binits{E.J.}},
\bauthor{\bsnm{Spandre}, \binits{G.}},
\bauthor{\bsnm{Spinelli}, \binits{P.}},
\bauthor{\bsnm{Stappers}, \binits{B.W.}},
\bauthor{\bsnm{Stephens}, \binits{T.E.}},
\bauthor{\bsnm{Suson}, \binits{D.J.}},
\bauthor{\bsnm{Tabassum}, \binits{S.}},
\bauthor{\bsnm{Tajima}, \binits{H.}},
\bauthor{\bsnm{Tak}, \binits{D.}},
\bauthor{\bsnm{Theureau}, \binits{G.}},
\bauthor{\bsnm{Thompson}, \binits{D.J.}},
\bauthor{\bsnm{Tibolla}, \binits{O.}},
\bauthor{\bsnm{Torres}, \binits{D.F.}},
\bauthor{\bsnm{Valverde}, \binits{J.}},
\bauthor{\bsnm{Venter}, \binits{C.}},
\bauthor{\bsnm{Wadiasingh}, \binits{Z.}},
\bauthor{\bsnm{Wang}, \binits{N.}},
\bauthor{\bsnm{Wang}, \binits{N.}},
\bauthor{\bsnm{Wang}, \binits{P.}},
\bauthor{\bsnm{Weltevrede}, \binits{P.}},
\bauthor{\bsnm{Wood}, \binits{K.}},
\bauthor{\bsnm{Yan}, \binits{J.}},
\bauthor{\bsnm{Zaharijas}, \binits{G.}},
\bauthor{\bsnm{Zhang}, \binits{C.}},
\bauthor{\bsnm{Zhu}, \binits{W.}}:
\batitle{The third fermi large area telescope catalog of gamma-ray pulsars}.
\bjtitle{The Astrophysical Journal}
\bvolume{958}(\bissue{2}),
\bfpage{191}
(\byear{2023}).
\doiurl{10.3847/1538-4357/acee67}
\end{barticle}
\endbibitem

\bibitem{NIST2024}
\begin{botherref}
\oauthor{\bsnm{{NIST}}}:
{Digital Library of Mathematical Functions, National Institute of Standards and
  Technology, U.S. Department of Commerce}
(2024).
\url{https://dlmf.nist.gov}
\end{botherref}
\endbibitem

\bibitem{Rybicki:847173}
\begin{bbook}
\bauthor{\bsnm{Rybicki}, \binits{G.B.}},
\bauthor{\bsnm{Lightman}, \binits{A.P.}}:
\bbtitle{Radiative Processes in Astrophysics}.
\bpublisher{Wiley},
\blocation{New York, NY}
(\byear{1985}).
\doiurl{10.1002/9783527618170}.
\burl{https://cds.cern.ch/record/847173}
\end{bbook}
\endbibitem

\bibitem{Scully1997}
\begin{bbook}
\bauthor{\bsnm{Scully}, \binits{M.O.}},
\bauthor{\bsnm{Zubairy}, \binits{M.S.}}:
\bbtitle{Quantum Optics}.
\bpublisher{Cambridge University Press},
\blocation{Cambridge}
(\byear{1997}).
\doiurl{10.1017/CBO9780511813993}.
\burl{http://ebooks.cambridge.org/ref/id/CBO9780511813993}
\end{bbook}
\endbibitem

\bibitem{Taverna2022Polarized}
\begin{barticle}
\bauthor{\bsnm{Taverna}, \binits{R.}}, \betal:
\batitle{Polarized x-rays from a magnetar}.
\bjtitle{Science}
\bvolume{378}(\bissue{6620}),
\bfpage{646}--\blpage{650}
(\byear{2022}).
\doiurl{10.1126/science.add0080}
\end{barticle}
\endbibitem

\bibitem{Turolla2024IXPE}
\begin{barticle}
\bauthor{\bsnm{Turolla}, \binits{R.}},
\bauthor{\bsnm{Taverna}, \binits{R.}},
\bauthor{\bsnm{Zane}, \binits{S.}},
\bauthor{\bsnm{Heyl}, \binits{J.}}:
\batitle{{{IXPE Observations}} of {{Magnetar Sources}}}.
\bjtitle{Galaxies}
\bvolume{12}(\bissue{5}),
\bfpage{53}
(\byear{2024}).
\doiurl{10.3390/galaxies12050053}
\end{barticle}
\endbibitem

\bibitem{Lai2023IXPE}
\begin{botherref}
\oauthor{\bsnm{Lai}, \binits{D.}}:
Ixpe detection of polarized x-rays from magnetars and photon mode conversion at
  qed vacuum resonance.
Proc. Natl. Acad. Sci.
\textbf{120}(17)
(2023).
\doiurl{10.1073/pnas.2216534120}
\end{botherref}
\endbibitem

\bibitem{Taverna2015Polarization}
\begin{barticle}
\bauthor{\bsnm{Taverna}, \binits{R.}},
\bauthor{\bsnm{Turolla}, \binits{R.}},
\bauthor{\bsnm{Gonzalez~Caniulef}, \binits{D.}},
\bauthor{\bsnm{Zane}, \binits{S.}},
\bauthor{\bsnm{Muleri}, \binits{F.}},
\bauthor{\bsnm{Soffitta}, \binits{P.}}:
\batitle{Polarization of neutron star surface emission: A systematic analysis}.
\bjtitle{Monthly Notices of the Royal Astronomical Society}
\bvolume{454}(\bissue{3}),
\bfpage{3254}--\blpage{3266}
(\byear{2015}).
\doiurl{10.1093/mnras/stv2168}
\end{barticle}
\endbibitem

\bibitem{Caiazzo2022Probing}
\begin{barticle}
\bauthor{\bsnm{Caiazzo}, \binits{I.}},
\bauthor{\bsnm{Gonz{\'{a}}lez-Caniulef}, \binits{D.}},
\bauthor{\bsnm{Heyl}, \binits{J.}},
\bauthor{\bsnm{Fern{\'{a}}ndez}, \binits{R.}}:
\batitle{Probing magnetar emission mechanisms with x-ray spectropolarimetry}.
\bjtitle{Monthly Notices of the Royal Astronomical Society}
\bvolume{514}(\bissue{4}),
\bfpage{5024}--\blpage{5034}
(\byear{2022}).
\doiurl{10.1093/mnras/stac1571}
\end{barticle}
\endbibitem

\bibitem{Gil1990Curvature}
\begin{barticle}
\bauthor{\bsnm{Gil}, \binits{J.A.}},
\bauthor{\bsnm{Snakowski}, \binits{J.K.}}:
\batitle{{Curvature radiation and the core emission of pulsars.}}
\bjtitle{Astronomy and Astrophysics}
\bvolume{234},
\bfpage{237}--\blpage{242}
(\byear{1990}).
\bcomment{\url{https://ui.adsabs.harvard.edu/abs/1990A\&A...234..237G}}
\end{barticle}
\endbibitem

\end{thebibliography}



\end{document}